\documentclass[preprint2]{aastex}
\usepackage{amsmath}
\usepackage{natbib}
\usepackage{graphicx}
\usepackage{txfonts}
\usepackage[pagebackref,breaklinks,colorlinks,citecolor=blue]{hyperref}
\usepackage[all]{hypcap}

\usepackage[nameinlink]{cleveref}
\usepackage{todonotes}
\usepackage{epsfig,color}

\newcommand{\eq}[1]{\begin{equation}  #1 \end{equation}}   
\newcommand{\eqa}[1]{\begin{align}  #1 \end{align}} 
\newcommand{\br}[1]{\left( #1 \right)}   
\newcommand{\bc}[1]{\left\{ #1 \right\}}   
\newcommand{\bb}[1]{\left[ #1 \right]}   
\newcommand{\ba}[1]{\left\langle #1 \right\rangle}   
\newcommand{\nn}{\nonumber}

\newcommand{\dd}{{\rm d}}    
\newcommand{\expo}[1]{~{\rm e}^{ #1 }}  
\newcommand{\vek}[1]{\mbox{\boldmath $#1$}}  
\newcommand{\ic}{{\rm i}}  
\newcommand{\permapj}[1]{\textit{\copyright\ AAS. Reproduced with permission from} \citet{#1}.}  
\newcommand{\permaa}[1]{\textit{Reproduced with permission from} \citet{#1} \textit{\copyright\ ESO.}}  
\newcommand{\permmn}[1]{\textit{Reproduced with permission from} \citet{#1}.}  

\onecolumn

\begin{document}
\title{Galaxy alignments: An overview}

\author{Benjamin Joachimi\altaffilmark{1}, Marcello Cacciato\altaffilmark{2}, Thomas D. Kitching\altaffilmark{3}, Adrienne Leonard\altaffilmark{1}, Rachel Mandelbaum\altaffilmark{4}, Bj{\"o}rn Malte Sch{\"a}fer\altaffilmark{5}, Crist{\'o}bal Sif{\'o}n\altaffilmark{2}, Henk Hoekstra\altaffilmark{2}, Alina Kiessling\altaffilmark{6}, Donnacha Kirk\altaffilmark{1}, Anais Rassat\altaffilmark{7}}
\email{b.joachimi@ucl.ac.uk}

\altaffiltext{1}{Department of Physics and Astronomy, University College London, Gower Street, London WC1E 6BT, UK}
\altaffiltext{2}{Leiden Observatory, Leiden University, PO Box 9513, 2300 RA, Leiden, the Netherlands}
\altaffiltext{3}{Mullard Space Science Laboratory, University College London, Holmbury St Mary, Dorking, Surrey RH5 6NT, UK}
\altaffiltext{4}{McWilliams Center for Cosmology, Department of Physics, Carnegie Mellon University, Pittsburgh, PA 15213, USA}
\altaffiltext{5}{Zentrum f{\"u}r Astronomie der Universit{\"a}t Heidelberg, Astronomisches Recheninstitut, Philosophenweg 12, 69120 Heidelberg, Germany}
\altaffiltext{6}{Jet Propulsion Laboratory,  California Institute of Technology, 4800 Oak Grove Drive, Pasadena, CA, 91109, USA}
\altaffiltext{7}{Laboratoire d'astrophysique (LASTRO), Ecole Polytechnique F{\'e}d{\'e}rale de Lausanne (EPFL), Observatoire de Sauverny, CH-1290 Versoix, Switzerland}

\begin{abstract}
The alignments between galaxies, their underlying matter structures, and the cosmic web constitute vital ingredients for a comprehensive understanding of gravity, the nature of matter, and structure formation in the Universe. We provide an overview on the state of the art in the study of these alignment processes and their observational signatures, aimed at a non-specialist audience. The development of the field over the past one hundred years is briefly reviewed. We also discuss the impact of galaxy alignments on measurements of weak gravitational lensing, and discuss avenues for making theoretical and observational progress over the coming decade. 
\end{abstract}


\tableofcontents


\section{Introduction}
\label{sec:introduction}

Galaxies are the most abundant and readily observed objects in the Universe
beyond our own Milky Way. Deep exposures in otherwise empty patches of sky
contain large numbers of faint and small galaxy images at increasingly larger distances from Earth and
thus at an earlier time in the history of the Universe \citep{FDW00}.
 This makes galaxies prime
candidates for studying the properties and the evolution of the large-scale
structure of the Universe. Analysing their spatial and spectral light
distributions in combination with their dynamical properties, one can infer the physical processes that govern galaxy formation
and attempt to explain the plethora of galaxy types and structures one observes. In addition, galaxies physically trace the density peaks of the overall
matter distribution, and their light samples the properties of spacetime along
the line of sight, which allows for direct constraints on cosmological models
and the non-luminous, more exotic ingredients of the Universe \citep[e.g.][]{Zwicky37,R66}.

The fundamental properties of galaxy images that can still be extracted from the
faintest, most distant objects include the position on the sky, the total flux,
the apparent size, as well as the lowest-order deviation from a circular
morphology given by a measure of ellipticity (e.g.\ the ratio of minor to major
axis) and orientation (e.g.\ the angle of the major axis with respect to a
reference direction). These quantities have been exploited to varying degrees in
studies of cosmology and galaxy evolution, for instance in galaxy clustering
(using the distribution of galaxy positions; e.g. \citealp{ZZW+11,SMK+14}) and gravitational magnification as well as measurements of the inherent size distribution (using size and flux; e.g. \citealp{SLM+12,MWF+12}).

Of particular interest are the orientations and ellipticities of galaxy images, because one expects
them to be random for a sufficiently large sample of galaxies in a homogeneous and
isotropic universe. Physical processes that locally violate isotropy may be indicated by either (a) any net preferred orientation, or \emph{alignment}, with respect to some reference direction in an ensemble of galaxy images; or, (b) any non-vanishing correlation between galaxy orientations.
Such processes have been linked e.g.\ to tidal gravitational
forces acting on galaxies during formation and at later evolutionary
stages. Moreover, the gravitational lensing effect by the large-scale structure
induces a coherent apparent distortion in galaxy images. The resulting galaxy
alignments can thus be used to constrain models of lensing effects, but only in
the absence of (or with a well-defined model for) other sources of
coherent alignment, which constitute a nuisance signal in this case.

Interest in galaxy alignments dates back to the early twentieth century (see
\Cref{subsec:first}), when the extragalactic nature of \lq nebulae\rq\ was not
even established. Contradictory results obtained from the slowly increasing
galaxy samples indicated that galaxy alignments are challenging to measure
reliably. Substantial stochasticity in most signals suppresses their signal-to-noise ratio, as
does the fact that three-dimensional alignments are diluted due to projection on
the sky. The low signal-to-noise ratio in samples of a few thousand galaxies, which
were typical for most of the last century, was paired with large spurious
ellipticities and alignments induced e.g.\ by telescope movement, optics, or
photographic plate artefacts \citep[e.g.][]{HWS+06}. Furthermore, it is a relatively recent insight
that the non-linear propagation of noise from the pixels to the shape of the image alone causes biases in
the measurement of galaxy orientation and shape \citep{MKJ}.

However, the past decade has seen a dramatic acceleration of progress in this
field, which can largely be attributed to the following developments:
\begin{itemize}
\item With the cosmological concordance model firmly established \citep{Planck15b}, there is now a
  robust framework upon which the more intricate models of galaxy alignments can
  build. This includes the cold dark matter paradigm, in which
  dark matter with negligible kinetic energy governs structure formation, as
  well as the bottom-up scenario of structure formation starting with small dark
  matter haloes that coalesce into ever-larger objects which eventually host
  galaxies.
\item Astrophysics has entered a golden era of large imaging and spectroscopic
  surveys, the first and foremost of which was the Sloan Digital Sky Survey
  (SDSS; \citealp{YAA+00}). These surveys are finally able to provide the galaxy sample sizes and
  the quality of shape measurements for significant and robust detections of
  galaxy alignments.
\item Computational power is constantly increasing such that one can nowadays run $N$-body
  simulations in cosmological volumes \citep{EDW+85}, with sufficient mass and spatial resolutions
  to obtain precise measurements of galaxy or halo shape and orientation, which implies that 
  alignment signals can be predicted robustly and at high statistical significance. This
  numerical effort is critical to better understand and model the highly
  non-linear physics expected in alignment processes. Some recent hydrodynamic simulations have also incorporated the physical processes of gas and stars, thus enabling a more direct link between
  theory and observations \citep[e.g.][]{VGS+14}.
\item A number of on-going and planned cosmological galaxy surveys (for details see \Cref{sec:outlook}) will use weak
  gravitational lensing by large-scale structure as a key probe of our
  cosmological model. The small but coherent galaxy shape distortions due to
  gravitational lensing are
  partly degenerate with local, physically-induced (and hence dubbed
  \emph{intrinsic}) galaxy alignments, which could thus constitute a limiting
  systematic effect. This has further boosted the interest in a better understanding of
  galaxy alignments.
\end{itemize}

The current research into galaxy alignments can roughly be split into two
branches according to the main drivers of this field: the study of galaxy
alignments with the elements of the cosmic web, such as clusters of galaxies,
filaments, and voids, with the purpose of directly testing models of galaxy
formation and evolution, and the measurement of pairwise alignments in large,
broadly-defined galaxy samples with the goal of quantifying and mitigating bias
in cosmological surveys, using similar datasets, statistics, and analysis
methodology as in the corresponding measurements of gravitational lensing.
One goal of this review is to take a synoptic viewpoint on these branches
and treat them as two approaches to the same science goal -- a deeper
understanding of the physics of galaxy alignments and its implications for
galaxy evolution and cosmology.

This work provides an overview on the subject and attempts to limit
the previous knowledge required to follow its contents to the basics of
extragalactic astrophysics and cosmology. There is a plethora
of ways in which galaxies can align with the multitude of structures that populate the
Universe, and we attempt to categorise and structure the vast amount of research
done over the past century. We will focus on correlations
which involve galaxy ellipticities or position angles, as well as the ellipticities of the underlying dark matter distribution, including the
latter\rq s observational proxies such as the distribution of satellite
galaxies.

After brief summaries of the basics of galaxy formation and evolution, the relevant gravitational lensing effects, and fundamentals of alignment processes in \Cref{sec:fundamentals}, this overview highlights the developments of the twentieth century (\Cref{sec:history}). We then review
recent work, proceeding from small-scale alignments inside an overdense region such as a galaxy cluster
(\Cref{sec:clusters}), to alignments between clusters and with the cosmic web
(\Cref{sec:lss}), to alignments of broadly defined galaxy
populations (\Cref{sec:galaxyalign}). \Cref{sec:impact} summarises the impact of
alignments on cosmological signals and corresponding mitigation strategies,
followed by an outlook on future developments of the field in
\Cref{sec:outlook}. This work is part of a topical volume on galaxy alignments consisting of three
papers in total. The two companion papers take a more
detailed and technical approach, covering theory, modelling, and simulations \citep{Paper2} as well
as observational results and the impact on cosmology \citep{Paper3}. \citet{TI15} also provided a recent review on galaxy alignments, with a focus on aspects related to weak gravitational lensing.

\section{Fundamentals}
\label{sec:fundamentals}

In this section we provide background information intended to help the reader follow and interpret the developments in the field of galaxy alignments. This includes a brief and qualitative overview of key concepts in galaxy formation and evolution processes, an introduction to weak gravitational lensing and associated measurements, as well as a summary of the theory of tidally induced galaxy alignments.

\subsection{A primer on galaxy formation and evolution}
\label{sec:intro_galaxyformation}

We give a brief introduction to the current picture of galaxy formation and
evolution, with a focus on processes that are thought to be relevant for a
galaxy's morphology and alignment with surrounding structures. Detailed accounts
of this topic can be found in \citet{B10} as well as \citet{L08} and \citet{MBW10}. Technical introductions to the theory of structure formation are for instance given in \citet{P99} and \citet{D03}.

\begin{figure}[t]
\centering
\includegraphics[scale=.18]{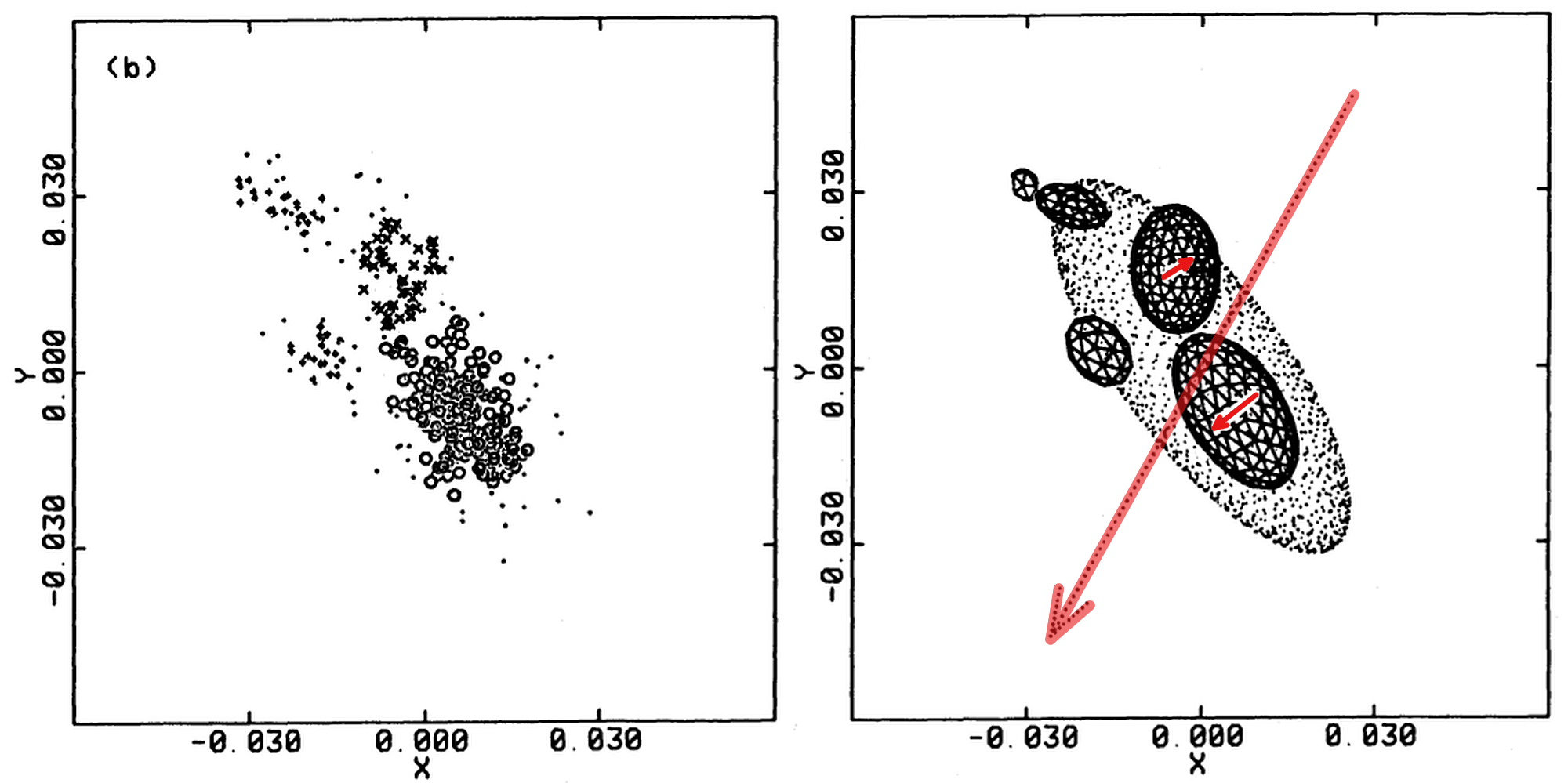}
\caption{\textit{Left}: Sample of simulation particles subsumed into a common halo in an $N$-body simulation. The halo was identified by a variant of the friends-of-friends algorithm, identifying arbitrarily shaped regions with a density above a certain threshold. Increasing this threshold, the halo is decomposed into a number of sub-haloes indicated by the different symbols. \textit{Right}: Representation of the halo and its substructure by ellipsoids which are determined by the eigenvalues and directions of the inertia tensor. The directions of the angular momenta of the larger sub-haloes as well as of the parent halo are given by the red arrows. Halo shapes and spins are key ingredients for the study of halo and galaxy alignments. \permapj{BE87}}
\label{fig:barnes87_halodef}
\end{figure}

Structure formation takes place on the background of an expanding, homogeneous and isotropic spacetime described by a Friedman-Lema\^itre-Robertson-Walker (FLRW) metric. The mean total matter density at a given time is determined by the curvature of spacetime as well as
the abundances of radiation (matter with relativistic velocities and thus
exerting radiation pressure), non-relativistic matter (including ordinary matter of which stars, planets, etc. are composed, and dark matter which makes up a large fraction of cosmological structures and dominates their dynamics), and dark
energy. The latter is usually assumed to permeate space smoothly and therefore
only affects structure formation via its impact on the background density and
the expansion rate. Radiation can easily escape gravitational wells (free-streaming)
and thereby suppresses the build-up of structures.

\begin{figure}[t]
\centering
\includegraphics[scale=.22]{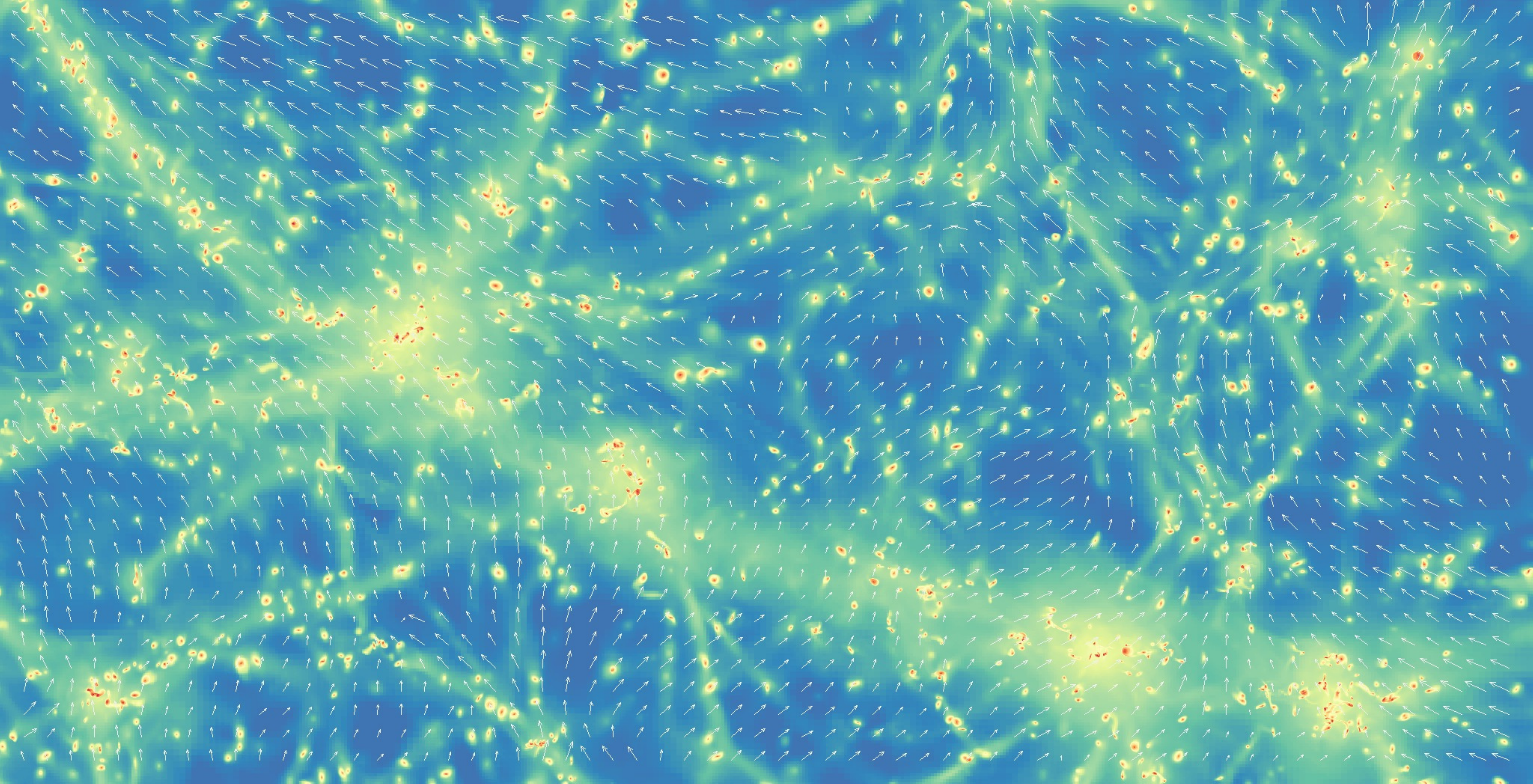}
\caption{Gas density tracing the cosmic web in a subvolume ($12.5\,{\rm Mpc}/h$ comoving horizontally, stacked over $25\,{\rm Mpc}/h$ comoving along the line of sight) of the HORIZON-AGN simulation. The arrows indicate the direction of the smallest eigenvector of the gravitational tidal tensor (given by the Hessian of the gravitational potential at that point), which is expected to align with filamentary structures on average (see e.g. the top left corner). See \Cref{sec:lss} for more details about the classification of filaments. \permmn{CGD+14}} 
\label{fig:codis14-tidalfield}
\end{figure}

About four to five times more abundant than ordinary matter, it is dark matter that governs the formation of structures via gravitational interaction. The absence of elastic collisions between dark matter particles implies a pressureless and non-viscous fluid. The subdominant ordinary matter, comprising mostly
neutrons, protons, and electrons (summarised henceforth under the common but
somewhat inaccurate term \emph{baryons}), closely follows the dark matter after
decoupling at recombination on scales above the Jeans length where gravity dominates over pressure in the baryon fluid. The distribution of initially tiny fluctuations around the mean matter density
is thought to be provided by the process of inflation in the early
Universe, which enlarges quantum fluctuations to macroscopic
scales. Inflationary models also predict that these density fluctuations are
well described by a Gaussian random field. The fluctuation strength is scale-invariant with a power-law exponent close to one, meaning that there are identical amounts of fluctuation in the density field in each logarithmic wavelength interval.

Once pressureless matter dominates the expansion history, density fluctuations within
the horizon (i.e. in regions in causal contact, which can physically interact) grow via gravitational interaction, initially by processes that are
well understood in linear perturbation theory. Eventually, over-densities (of generally triaxial shape; \citealp{BBK+86}) begin
to evolve non-linearly and collapse into a \emph{halo}, an approximately stable
state in which the random motions of the constituent particles or objects
balance gravity. Structures continue to grow in a bottom-up scenario, i.e.\ small
haloes form first and then coalesce into ever larger haloes. The abundance of dark matter haloes for a given mass can be estimated analytically \citep{PS74,EPS91}, while it was found empirically from simulations \citep{NFW97,GNC+08} that they have a near-universal radial density distribution. Interpreted as quasi-stable bodies with only little exchange of matter with their surrounding, haloes can be assigned an angular momentum vector and a shape, often approximated by an ellipsoid which is determined by the eigenvalues and eigenvectors of the inertia tensor (see \Cref{fig:barnes87_halodef}). These quantities, which are believed to be central to alignment processes, depend (sometimes strongly) on practical implementation choices such as the definition of a halo and the weights assigned to particles in the computation of the inertia tensor \citep{BEF+07}.

$N$-body simulations illustrate in a striking manner how the initially Gaussian
density fluctuations evolve under gravity into the cosmic web, a network of
overdense \emph{filaments} which intersect at massive haloes. The filaments are
in turn embedded into medium-density walls or sheets which surround large,
underdense regions of space called \emph{voids}. The general direction of
gravitational acceleration will cause matter to flow away from the centres of
voids onto sheets, in the plane of sheets towards filaments, and along filaments
into the massive haloes at the nodes (possibly having collapsed into smaller
haloes well before). \Cref{fig:codis14-tidalfield} displays part of the cosmic web obtained in a simulation that contained collision-less dark matter particles as well as gas and star particles \citep{CGD+14}.

Baryons follow the dark matter and thus accumulate in the centres of haloes to
eventually form galaxies, via processes that are still poorly
understood. When baryons fall into haloes, pressure becomes important in the
form of accretion shocks, which convert the ordered infall motion into the
random velocities of a virialised gas. Further collapse only takes place when
the gas can cool, i.e.\ lose kinetic energy via radiation. If the cooling is
efficient, as is expected for low-mass haloes up to Milky Way size \citep{RO77}, the baryonic gas can
essentially free-fall into the centre of the halo. Otherwise, the hot gas fills
the halo from where it gradually descends towards the centre, with the more
dense central regions cooling faster.

In any case, the gas reaching the central part of the halo will build up angular
momentum, for instance by free-falling with a certain impact parameter from a preferred
direction given by the adjacent filaments, or by being subjected to the tidal
gravitational field of surrounding structures which exerts a torque on
in-falling material. This leads to the generation of a rotationally-supported
disc of gas and, eventually, of stars. Note that the vast majority of alignment studies have been conducted in optical passbands in which the brightness distribution of galaxies is determined by the light of the stars. In these scenarios, the orientation of the angular
momentum vector and hence the disc is expected to be linked to the configuration
of the surrounding large-scale structure \citep[see e.g.][]{PJH+14}, which could
act as a seed for galaxy alignments. However, since only a fraction of the gas in a galaxy is converted into stars, there must exist feedback processes that limit (or even reverse) the gravitational collapse of gas, which in turn will also impact on the angular momentum amplitude and orientation of the stellar distribution.

The complex history of mergers of a dark matter halo with other haloes has a
strong impact on the evolution and appearance of the galaxy that it may
host. Accreted smaller haloes can survive for a long time orbiting inside the
large halo. The satellite galaxies inside these subhaloes may be tidally
stripped, may have their orientation tidally locked with respect to the centre
of the host halo, and could eventually be tidally disrupted \citep[see e.g.][]{PBG08}. A major merger,
i.e.\ the coalescence of two haloes or galaxies with comparable mass, may disrupt
the progenitors completely, erasing any memory of alignments generated during
galaxy formation, and lead to the formation of a dynamically hot, spheroidal
system, such as elliptical galaxies or the central bulges of disc galaxies\footnote{For an alternative formation hypothesis of spheroids involving cold gas streams, which is also closely linked to alignments with the surrounding  dark matter and gas distribution, see e.g. \citet{D+09}.}.

New alignments may be formed in major mergers by the re-arrangement of
stellar orbits, and thus the light distribution. Elliptical galaxies seem to
have similar shapes and orientations as their underlying dark matter haloes,
which in turn are well-described by triaxial ellipsoids (though reality can be
more complex, with different effective ellipticity or orientation as a function
of radius; see, e.g., \citealt{SFC12}, and substantial misalignment between galaxy and halo; see \Cref{sec:galaxyhaloalign}). The characteristics of
these ellipsoids are determined in a complex way by the surrounding matter
distribution and details of the merger history, such as the provenance of
progenitor haloes.  These, and possibly other effects, lead to alignments of
halo shapes out to separations of tens of megaparsecs \citep[e.g.][]{Hopkins05}, which
are stronger the more massive the haloes.

The more secular processes of galaxy evolution, such as star formation and
subsequent chemical enrichment of the interstellar medium \citep{MO07}, or the balance of
continued accretion of intergalactic gas versus feedback processes by supernovae
\citep{GE00} and active galactic nuclei \citep{SR98}, could also play some role in the evolution of galaxy
alignments, for instance via the re-distribution of angular momentum and the
modification of the galaxy's spectral energy distribution, which implies that we
may study the alignments of different components of a galaxy at different times
for a given passband.

All in all, the processes of galaxy formation and evolution are intimately linked at various stages to the creation or destruction of galaxy alignments. We can
therefore expect alignment signals to depend on the
galaxy's large-scale environment, its morphological type (\lq late\rq\ for disc
galaxies, \lq early\rq\ for elliptical galaxies), its colour
(blue for star-forming disc galaxies, red for ellipticals dominated by old
stellar populations), its mass or luminosity, redshift or age, and more. Both galaxy
evolution and alignments depend on highly non-linear physics acting over a wide
range of spatial and mass scales, and involve dark and baryonic matter, which
makes it a challenge to model them accurately.

\subsection{A primer on weak gravitational lensing}
\label{sec:intro_WL}

Correlations induced by the distortions of the images of distant galaxies
due to gravitational lensing are a sensitive probe of the large-scale matter
distribution as well as the geometry of spacetime. Intrinsic galaxy alignments
partly mimic this correlation signal and can thus bias cosmological constraints
inferred from galaxy shape correlations if those correlations are presumed to be
entirely due to lensing. To facilitate the insight into this link, we
sketch the basic methodology and formalism of weak gravitational lensing in this
section. Note that galaxy alignment observations share many
challenges with weak lensing, in particular the measurement of the ellipticities and
orientations of faint galaxy images. A standard technical introduction to weak lensing is given in \citet{BS01}; see also \citet{S06}. \citet{Ba10} provided an overview on the whole theme of gravitational lensing, while more specialised reviews are presented in \citet{MVW+08} and
\citet{HJ08} on cosmological applications of weak lensing, in \cite{MKR10} on the study of dark matter particularly via weak lensing, and in \citet{K14} on recent progress in weak lensing by the large-scale structure.

The gravitational deflection of light is accurately described in the framework
of general relativity and served as the first successful observational test of
the validity of Einstein's theory \citep{DED20}. According to Fermat's principle, light follows paths, called geodesics, for which the light travel time is stationary, i.e. the derivative of the light travel time with respect to position is zero. In spatially flat
FLRW cosmologies without structures, this results in a straight path for the light ray, while in a spacetime curved by a large mass such as a galaxy cluster, light will generally travel
along curved geodesics. For background light sources close in angular position to this massive object (the \lq lens\rq ), several stationary points in light travel time may exist, corresponding to multiple images of the same object according to Fermat\rq s principle. The differential deflection of light from extended sources distorts the images into
arcs tangentially around the centre of the lens, and can also magnify them. These sometimes spectacular effects visible on individual objects define the regime of \emph{strong gravitational lensing}.

\begin{figure}[t]
\centering
\includegraphics[scale=.25]{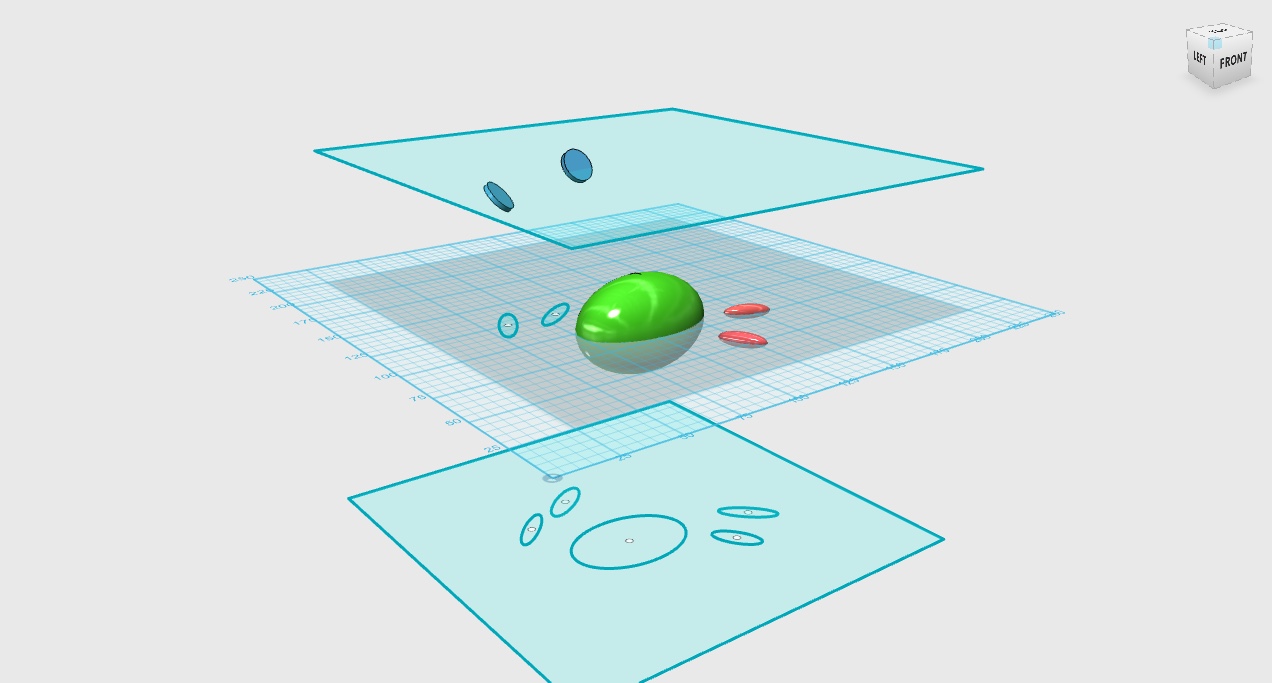}
\caption{Sketch of the gravitational lensing signal and its intrinsic alignment contamination. Light travels from the top of the sketch downwards, from the source plane via the lens plane to the plane at the bottom containing the images as seen by an observer. The matter structure (green ellipsoid) deflects the light from the background source galaxies (blue discs) and distorts their images tangentially with respect to the apparent centre of the lens (as seen in the bottom plane). As a consequence, the galaxy images become aligned (GG signal). Galaxies which are physically close to the lens structure (red ellipsoids) may be subjected to forces that cause them to point towards the structure, which results in the alignment of their images (II signal). Images of galaxies close to the lens are then preferentially anti-aligned with the gravitationally sheared images of background galaxies (GI signal).}
\label{fig:sketch_lensing}
\end{figure}

At larger angular distances from the lens no multiple images occur, and distortions and magnification only cause small
modifications to the original light profile of the source, usually also a galaxy (see \Cref{fig:sketch_lensing}). Yet,
by averaging in an annulus around the lens over the shapes of source galaxies,
one may still be able to recover the net effect induced by
gravitational lensing. If the changes to galaxy images are so small that statistical tools need to be employed to detect a
signal, one refers to \emph{weak gravitational lensing} effects. The changes to an image are captured to first order in the Jacobian matrix $\vek{A}$ of the mapping between the source and the image,
\eq{
\label{eq:jacobian}
\vek{A}=\left( \begin{array}{cc} 1- \kappa- \gamma_1 & - \gamma_2\\ - \gamma_2 & 1- \kappa+ \gamma_1 \end{array} \
\right) = \br{ 1-\kappa} \br{\begin{array}{cc} 1 & 0 \\ 0 & 1 \end{array}}  - |\gamma| \br{\begin{array}{cc} \cos 2\varphi & \sin 2\varphi\\ \sin 2\varphi & -\cos 2\varphi \end{array}}\;,
}
where $\kappa$ is the convergence and $\gamma= \gamma_1 + \ic \gamma_2$ the \emph{gravitational shear}. The second equality provides an illustrative understanding for this mapping. The convergence $\kappa$ yields isotropic focusing, whereas $\gamma$ quantifies distortions of the image (and anisotropic focusing). Sources with circular isophotes are mapped into elliptical images, where a combination of $\kappa$ and $|\gamma|$ determines the length of the major and minor axes, while the polar angle of $\gamma$, denoted by $\varphi$, describes the orientation of the ellipse. The factor of 2 in the phase takes into account that the shear is a polar (i.e. spin-2) quantity which maps onto itself after a rotation by $180\,$degrees. The magnification of the image is given by $\mu=1/|\det J|$, and both the flux and the size are modified by factors of $\mu$ since lensing does not change the surface brightness. To date, magnification effects have not been used as extensively as gravitational shear in studies of galaxies or cosmology because one generally expects lower signal-to-noise than for equivalent shear statistics (for applications see e.g. \citealp{SMR+05} on magnification bias, \citealp{FHW+14} on flux magnification, and \citealp{HG11} on size magnification). They come in principle with their own intrinsic correlations of galaxy observables, which we will not discuss further here.

Galaxies as light sources are intrinsically non-circular in general, and the deviation from a circular image can to first order be described by an intrinsic ellipticity $\epsilon^{\rm s}$. This ellipticity is intrinsic in the sense that it is a property of the galaxy itself rather than induced by gravitational deflection as the light travels to the observer, after leaving the galaxy. The observed ellipticity under the gravitational lens mapping is then given by \citep{SS97}
\eq{
\label{eq:epsgamma}
\epsilon = \frac{\epsilon^{\rm s}+g}{1+\epsilon^{\rm s}g^*} \approx \epsilon^{\rm s}+ \gamma ~\
~\mbox{with}~~ g \equiv \frac{\gamma}{1-\kappa}\;,
}
where $g$ is called the reduced shear. Both ellipticities and shear are understood
as complex numbers in this equation (with the complex conjugate denoted by a star), encoding the shape in the absolute value and
the orientation with respect to some reference axis in the phase, e.g.\ $\epsilon
= |\epsilon| \expo{2\ic \varphi}$. The simple summation of shear and ellipticity in the second equality of \Cref{eq:epsgamma} only holds in the limit of very weak lensing effects\footnote{There is a subtlety involved in this approximation: for an individual galaxy, as \Cref{eq:epsgamma} has been written, the expansion produces another term that is first order in the shear and proportional to $g^* (\epsilon^{\rm s})^2$. However, since the relation is only considered in practice when averaging over large numbers of galaxies, this term (as well as all higher-order terms) becomes negligible if the intrinsic galaxy shapes are uncorrelated, or only weakly correlated, with the shear acting on them.}, i.e.\ $|\gamma|, \kappa \ll 1$. It is
important to note that the term \lq ellipticity\rq\ is not uniquely defined in
general and, even if galaxy images were simple solid ellipses with
semi-minor to semi-major axis ratio
$b/a$, could correspond to several quantities which are functions of $b/a$. The
formalism presented in this section applies to the definition $|\epsilon| =
(a-b)/(a+b)$. 

Under the assumption of randomly oriented galaxies, $\ba{\epsilon^{\rm s}}=0$
(angular brackets denote ensemble averages), so that in the weak limit of \Cref{eq:epsgamma} the observed ellipticity is
an unbiased estimator of gravitational shear, $\ba{\epsilon}=\gamma$. Averaged
over large areas of sky, the shear is expected to vanish as well due to the
isotropy of the Universe. Hence, to lowest order, one generally considers
two-point statistics to detect weak gravitational lensing effects,
i.e.\ correlations between pairs of galaxy shapes or alignments of galaxy shapes
with reference positions. Typically, gravitational shear modifies the ellipticity of a galaxy only at the percent level, so that large samples of source galaxies are required to obtain sufficient signal-to-noise. Correlations of gravitational shears measured over
large patches of sky yield a signal referred to as \emph{cosmic shear} which measures the net lensing effect by the intervening large-scale structure.

Specifically, weak lensing shear provides a measurement of the projected tidal gravitational field through its distorting effect on a galaxy shape and combines geometrical information (due to the mapping of spatial derivatives to angular derivatives) with information about structure growth and gravity (relating tidal shear to the density field). In the most basic form, the average weak lensing shear is given by a line-of-sight integration,
\begin{equation}
\label{eq:wl_los}
\bb{\gamma_1+\mathrm{i}\gamma_2}(\theta_x,\theta_y) = 
\int_0^{\chi_{\rm H}}\mathrm{d}\chi\:p(\chi)
\int_{\chi}^{\chi_{\rm H}}\dd\chi^\prime\:
\frac{f_{\rm K}(\chi^\prime-\chi)f_{\rm K}(\chi^\prime)}{f_{\rm K}(\chi)}\left[(\partial^2_{y}-\partial^2_{x}+2\mathrm{i}\partial_x \partial_y)\right]\Phi(x,y,\chi)\;,
\end{equation}
collecting second derivatives of the gravitational potential $\Phi$ at the spatial position $(x,y,\chi)$ which are, in the small angle limit, related to the angular position through $\theta_x=x/\chi$ and $\theta_y=y/\chi$. Here, $\chi$ is the comoving distance, $\chi_{\rm H} \sim c/H_0$ is the comoving horizon distance ($H_0$ is the Hubble constant and $c$ is the speed of light), and $f_{\rm K}(\chi)$ is the comoving angular diameter distance, given by 
\eq{
\label{eq:fk2}
f_{\rm K}(\chi)=\left\{ \begin{array}{ll}
  1/\sqrt{K} \sin\left(\sqrt{K}\chi\right)    &K>0 ~(\mbox{open})\\
  \chi                             &K=0 ~(\mbox{flat})\\
  1/\sqrt{-K} \sinh\left(\sqrt{-K}\chi\right) &K<0 ~(\mbox{closed})\;,
\end{array} \right.
}
where $1/\sqrt{|K|}$ is interpreted as the curvature radius of the spatial part of spacetime. The line-of-sight average is taken over the probability distribution of comoving distances for the source galaxies, $p(\chi)$. 

Analytically, the power spectrum, i.e.\ the Fourier transform of the correlation function, is the most convenient two-point statistic to work with. The angular power spectrum, $C_{\gamma\gamma}(\ell)$, of weak gravitational shear is derived from the line of sight expression in \Cref{eq:wl_los} by Limber-projection \citep{K92},
\eq{
\label{eq:limberlensing}
C_{\gamma \gamma}^{(ij)}(\ell) = \int^{\chi_{\rm H}}_0 \dd \chi\; \frac{q^{(i)}(\chi)\, q^{(j)}(\chi)}{f^2_{\rm K}(\chi)}\; P_{\delta \delta} \br{\frac{\ell}{f_{\rm K}(\chi)},\chi}\;.
}
With increasing $\chi$, ever smaller wavenumbers $k=\ell/f_{\rm K}(\chi)$ contribute to the fluctuation on the multipole $\ell$, from which one can obtain the angular scale $\theta=\pi/\ell$. The shear correlations are generated by the continuous deflection of light by the matter distribution between the source galaxies and Earth, hence the line-of-sight integration over the power spectrum $P_{\delta \delta}$ of the matter density contrast $\delta=\rho/\bar{\rho}-1$, where $\bar{\rho}$ denotes the mean matter density. The projected and three-dimensional power spectra are formally defined as
\eqa{
\ba{ \widetilde{\gamma}^{(i)}(\vek{\ell})\, \widetilde{\gamma}^{(j)}(\vek{\ell}') } &= (2 \pi)^2\, \delta_{\rm D}^{(2)}(\vek{\ell} + \vek{\ell'})\; C_{\gamma \gamma}^{(ij)}(\ell) \;; \\ \nn
\ba{ \widetilde{\delta}(\vek{k})\, \widetilde{\delta}(\vek{k}') } &= (2 \pi)^3\, \delta_{\rm D}^{(3)}(\vek{k} + \vek{k'})\;  P_{\delta \delta}(k)\;,
}
where the tilde denotes Fourier transforms, and $\delta_{\rm D}^{(n)}$ is the $n$-dimensional Dirac delta distribution. Angular frequencies and wavenumber in bold denote vectors in two and three dimensions, respectively. The integral in \Cref{eq:limberlensing} is weighted by the lensing efficiency
\eq{
\label{eq:weightlensing}
q^{(i)}(\chi) = \frac{3 H_0^2\, \Omega_{\rm m}}{2\, c^2} \frac{f_{\rm K}(\chi)}{a(\chi)}
\int_{\chi}^{\chi_{\rm H}} \dd \chi'\; p^{(i)}(\chi')\; \frac{f_{\rm K}(\chi' -
  \chi)}{f_{\rm K}(\chi')}\;,
}
which is proportional to the ratio of the distance between source and lens over the distance between source and observer, weighted by the line-of-sight distribution of source galaxies, $p^{(i)}(\chi)$. Different galaxy samples can be \mbox{(cross-)}correlated, and these are indexed by the superscripts in parentheses. Note that $a=1/(1+z)$ in the equation above refers to the cosmic scale factor, $z$ denotes redshift, and $ \Omega_{\rm m}$ is the matter density parameter. In practice, lensing is most efficient when the lens is approximately midway between us and the source.

While the shear power spectrum can
be obtained from a catalogue of shear estimates directly, most analyses to date
are based on its Fourier transforms, the shear correlation functions
\eq{
\label{eq:xipm}
\xi_\pm(\theta) = \langle \epsilon_+ \epsilon_+ \rangle (\theta) \pm \langle
\epsilon_\times \epsilon_\times \rangle (\theta)\;,
}
as they are insensitive to the generally very complex angular selection function of weak-lensing quality photometric survey data. Since the shear is a complex quantity, one can obtain three real-valued
correlation functions of which only the two given above contain cosmological
information (the third vanishes if parity is conserved). The correlation
functions are given in terms of the tangential ellipticity component
$\epsilon_+=- {\rm Re} (\epsilon\, {\rm e}^{2 \ic \varphi})$ and the cross
component $\epsilon_\times=- {\rm Im} (\epsilon\, {\rm e}^{2 \ic \varphi})$,
where the polar angle $\varphi$ is measured against the line connecting the pair
of galaxies\footnote{The minus sign in these definitions ensures that the tangential alignment of shear around an object yields a positive signal. As a caveat, measurements of galaxy alignments tend to omit the minus sign in related statistics because in this situation the generally expected radial alignment is desired to yield a positive signal.}. The averages in \Cref{eq:xipm} are calculated by summing the
corresponding products of ellipticity components over all galaxy pairs in a
given angular separation bin centred on $\theta$.

Other statistical measures with desirable properties can be derived from the
shear correlation functions, for instance the aperture mass dispersion \citep{SWJ+98},
which to a good approximation separates the field of gravitational shears into a
curl-free and a divergence-free part, called $E$- and $B$-modes respectively, in
analogy to decompositions of polarisation. Gravitational lensing effects only
generate a negligible level of $B$-modes through higher-order effects, so estimates of B-mode shear correlations can therefore be employed as a
test for a range of systematic effects, e.g. in the shape measurement process. Moreover, the source galaxies are often split into
redshift slices, which improves cosmological constraints \citep{H99},
particularly on those parameters that encapsulate evolutionary effects (e.g.\ the
dark energy equation of state parameters). To perform this \emph{tomography}, a
large number of redshifts for faint galaxies are required, which is too costly
to obtain via spectroscopy\footnote{Note that even for a non-tomographic cosmic shear analysis the overall redshift distribution of source galaxies is needed, although the requirements on accuracy and precision are less stringent in this case.}. Instead, multi-band photometry, usually in the
optical and supplemented by near-ultraviolet and near-infrared passbands if
available, is used to obtain very low-resolution information on the spectral
energy distribution of a galaxy. The precision of these \emph{photometric redshifts} is
limited to a scatter typically of order $0.05(1+z)$. Catastrophic failures can occur
e.g.\ due to the confusion of spectral features like the Balmer and Lyman breaks,
leading to potentially large systematic offsets in redshift and hence to groups of outliers in the line-of-sight distribution of source galaxies that enters \Cref{eq:weightlensing}. The estimation of photometric redshifts and the
characterisation of their quality via calibration samples or clustering
measurements is an active field of research \citep[e.g.][]{HAC10}.

Cosmic shear was first detected at the turn of the millennium
\citep{BRE00,KWL00,WME00,WTK00} and is developing into an increasingly mature
cosmological probe \citep[e.g.][]{SHJ+10,HGH+13,SHP+13,KAH+14}. The large scatter of intrinsic galaxy
ellipticities limits the signal-to-noise of these measurements, introducing a
shot noise-like term in the statistical errors, so that the efforts to measure cosmic shear are driven
towards faint galaxy samples in deep surveys with high number densities. This in
turn renders the estimation of gravitational shear from noisy, small and
pixelated galaxy images a challenge, which has spawned large community effort to
develop more powerful algorithms \citep{HVB+06,MHB+07,BBB+10,KBB+12,MRA+14}. In
addition to shear
estimation biases and accurate photometric redshift determination, a further key
issue for the forthcoming generation of cosmic shear measurement campaigns are
intrinsic galaxy alignments\footnote{An aside on nomenclature: galaxy alignments
  often receive the attribute \lq intrinsic\rq, especially if the physical
  alignments inherent to the galaxy population need to be distinguished from the
  apparent alignments on galaxy images induced by gravitational lensing
  (occasionally denoted as \lq extrinsic\rq; see \citealp{CKB01}). The term is
  also applied in a slightly different context to distinguish between the
  physical three-dimensional shape of a galaxy and its projected shape we
  observe on the sky (see \citealp{SFS70} for the earliest occurence that we
  could trace).} which can mimic the correlations expected from cosmic
shear. Using \Cref{eq:epsgamma} in its weak limit, a generic correlator of two galaxy
ellipticities, as is for instance found in \Cref{eq:xipm}, reads
\eqa{
\label{eq:iadef}
\underbrace{\ba{{\epsilon_i} {\epsilon_j}}} &= \underbrace{\ba{{\gamma_i}
      {\gamma_j}}} + \underbrace{\bigl\langle {\epsilon_i^{\rm s}}
    {\epsilon_j^{\rm s}}} \bigr\rangle + \underbrace{\bigl\langle {\gamma_i}
    {\epsilon_j^{\rm s}} \bigr\rangle+\ba{{\epsilon_i^{\rm s}}
      {\gamma_j}}}{}\;.\\ \nn \mbox{observed} & \hspace*{0.5cm} \mbox{GG} \hspace*{1cm}
  \mbox{II} \hspace*{1.5cm} \mbox{GI}
  }
In the following we will adopt a common shorthand notation for the resulting
terms: GG for the shear correlation, which is the desired quantity for
cosmological analysis, II for correlations between the intrinsic ellipticities
of two galaxies, and GI for correlations between the gravitational shear acting
on one galaxy and the intrinsic shape of another galaxy.
Note that one of the GI terms in \Cref{eq:iadef} is expected to vanish because the shear acting on a galaxy in the foreground cannot be affected by a galaxy behind the source galaxy, unless their positions along the line of sight are confused because of errors in the redshift measurement.
If galaxy shapes are
intrinsically randomly oriented, only GG is non-zero. However, since galaxies are known to
align with other galaxies (generating II) and with the large-scale structure
that in turn contributes to the gravitational deflection of light from
background galaxies (generating GI), cosmic shear measurements may be severely
biased if these alignment effects are not accurately modelled or removed from
the signal. An illustration of the generation of II and GI correlations is provided in \Cref{fig:sketch_lensing}. Their two-point correlations for the tomographic case can be expressed analogously to \Cref{eq:limberlensing} as \citep[e.g.][]{JB10}
\eqa{
\label{eq:limberia}
C_{{\rm GI}}^{(ij)}(\ell) &= \int^{\chi_{\rm H}}_0 \dd \chi\; \frac{p^{(i)}(\chi)\, q^{(j)}(\chi)}{f^2_{\rm K}(\chi)}\; P_{\delta {\rm I}} \br{\frac{\ell}{f_{\rm K}(\chi)},\chi}\;; \\ \nn
C_{{\rm II}}^{(ij)}(\ell) &= \int^{\chi_{\rm H}}_0 \dd \chi\; \frac{p^{(i)}(\chi)\, p^{(j)}(\chi)}{f^2_{\rm K}(\chi)}\; P_{\rm II} \br{\frac{\ell}{f_{\rm K}(\chi)},\chi}\;,
}
where $p^{(i)}(\chi)$ is the distribution of galaxies in the $i$th tomographic bin and the resulting lensing efficiency function is $q^{(i)}(\chi)$; see \Cref{eq:weightlensing}. The power spectra, $P_{\delta {\rm I}}$ and $P_{\rm II}$, quantify the correlation between the matter distribution and the intrinsic shear, $\gamma^I$, and among the intrinsic shears of different galaxies, respectively. The intrinsic shear can be understood as the correlated part of the intrinsic ellipticity of a galaxy, which is not an observable quantity per se. Yet, when considering ensembles of galaxy shapes, it is conceptually useful to split the intrinsic ellipticity into $\gamma^I$, which determines the alignments, and a purely random part, which only leads to a noise contribution in correlations.

The gravitational lensing effect is correlated across tomographic bins because two bins share the common light path through the large-scale structure in front of the less distant bin. In contrast, intrinsic alignments are locally generated processes. This means that intrinsic ellipticities are only correlated with each other (II) within the same tomographic bin, unless the galaxy distributions of different bins overlap due to scatter in the photometric redshift bins. However, the cross-correlation with gravitational shear (GI) extends beyond adjacent redshift bins and becomes stronger with increasing distance between bins due to the contribution by the lensing efficiency,  $q^{(i)}(\chi)$. 
The challenge is to obtain a good model for the underlying power spectra, $P_{\delta {\rm I}}$ and $P_{\rm II}$. This has prompted an interest in galaxy alignments from the cosmology community in recent years. The new large and high-quality datasets for cosmic shear surveys and the advancements in data analysis techniques, especially the accurate measurement of galaxy shapes in the presence of noise, complex models of the telescope\rq s point-spread function (PSF), and image artefacts, have also greatly improved the power and fidelity of galaxy alignment observations \citep[see][]{Paper3}.

Finally, we briefly mention the rich field of weak galaxy lensing (often referred to as 
galaxy-galaxy lensing), probing the (dark) matter environment around individual
galaxies \citep[first detected by][]{BBS96}. Since the lensing signal from a single
galaxy is too weak to detect, lens galaxies selected with certain properties
(e.g.\ colour, luminosity, redshift) are \lq stacked\rq. Alternatively, one can
carry out this process with individual galaxy clusters or ensembles of galaxy
clusters and groups.  Stacking is
performed statistically by measuring correlation functions of the form
$\xi_{{\rm g}+}(\theta) = \ba{\delta_{\rm g} \epsilon_+}(\theta)$, where
$\delta_{\rm g}=N_{\rm g}/\ba{N}_{\rm g}-1$, denotes the number density contrast of \lq lens\rq\
galaxies, with $N_{\rm g}$ the galaxy number count. The tangential ellipticity is measured with respect to the line
connecting the background galaxy to the lens. For large separations $\theta$
this correlation probes the large-scale matter distribution and how it is traced
by the lens galaxies, while on megaparsec scales and below it measures the average tangential matter profile
of the lenses. If the same type of statistic is applied to a source galaxy sample that resides at
the same redshifts as the lens galaxies, one does not expect any gravitational
lensing effects but instead obtains a measure of the alignment of galaxy shapes
towards the positions of physically close neighbours. Measurements of this kind
will be further discussed in \Cref{sec:galaxyalign}.

\subsection{Tidally induced alignment processes}
\label{sec:tidaltheory}

We briefly review the basic concepts of how tidal gravitational fields are thought to generate alignments between the shapes of galaxies or larger bound structures such as galaxy clusters, and the large-scale matter distribution or the shapes of other galaxies. We will distinguish between the tidal processes thought to apply to the two major types of galaxies: angular momentum generation for disc galaxies (linked to \emph{tidal torque theory}), and the coherent modification of stellar orbits for elliptical galaxies (as described by the \emph{tidal alignment} paradigm). For the plethora of observational signatures that will be discussed in this work, most -- possibly all -- analytic models are based on the assumption of tidally generated alignments. This theory has been quite successful at predicting the general form of correlations between tidal fields and observables, but is bound to fail at making quantitative statements about the amplitude of signals as these depend strongly on highly non-linear and stochastic processes (such as the later stages of gravitational collapse that lead to galaxy formation). For this reason we shall for simplicity limit ourselves to establish proportionalities in most equations. Note that other alignment theories will be discussed in \Cref{sec:galaxyalign}.

Tidal interaction of galaxies with the surrounding gravitational field can in many cases be understood as a perturbative process, in particular at early stages of galaxy formation. Commonly, the interaction is described by Lagrangian perturbation theory, where the basic quantities are trajectories of dark matter constituent particles. These trajectories are determined by the strength and direction of gravitational fields, and the perturbation series is constructed by considering higher-order derivatives of the gravitational potential that lead successively to more detailed curved trajectories. 

At lowest order, however, the ZelÕdovich approximation \citep{Z70} tells us that all dark matter particles follow straight lines parallel to the gravitational field at their initial positions, or
\eq{
\label{eq:zeldovich}
x_\alpha ( \vek{\zeta},a) = \zeta_\alpha - D(a)\; \partial_\alpha\Psi(\vek{\zeta})\;,
}
in Cartesian coordinates with $\alpha=1,2,3$, where $\vek{x}$ is the particle's comoving coordinate and $\vek{\zeta}$ its coordinate in the Lagrangian frame. The linear growth factor of structure is given by $D(a)$ and $\Psi$ denotes a displacement potential, which is proportional to the gravitational potential $\Phi$.

The dynamics of objects such as protogalaxies are obtained by integrating over the Lagrangian trajectories of all particles that make up the object. The notion of an actual object allows the definition of a centre of gravity $\bar{\vek{\zeta}}$, relative to which the motion of a test particle can be expanded in a Taylor-series,
\eq{
\label{eq:zeldovich:taylor}
x_\alpha ( \vek{\zeta},a) \approx \zeta_\alpha - D(a) \bb{ \partial_\alpha \Psi(\bar{\vek{\zeta}}) + \sum_\beta (\zeta_\beta -\bar{\zeta}_\beta)\; \partial_\alpha \partial_\beta \Psi(\bar{\vek{\zeta}}) }\;,
}
such that one obtains the peculiar motion $\propto\partial_\alpha \Psi(\bar{\vek{\zeta}})$ of the object as a whole and the differential motion of the particles relative to the centre of gravity $\propto\partial_\alpha \partial_\beta \Psi(\bar{\vek{\zeta}})$. The latter term, proportional to the Hessian of the gravitational potential, plays therefore an important role in determining the distribution of the relative positions and velocities of the constituent particles.

Gradients of the gravitational force across the object, as described by the second derivatives of the potential, can lead to a change in the protogalaxy's shape (\lq tidal stretching\rq) as well as generate angular momentum (\lq tidal torquing\rq). The strength of both effects depends on the orientation of the protohalo's \emph{inertia tensor}, $I_{\alpha\beta}$, relative to the quadrupole of the gravitational potential, given by the \emph{tidal shear tensor} $T_{\alpha\beta} \propto \partial_\alpha \partial_\beta\; S\{\Phi\}$, where $S\{\}$ is a smoothing operator that removes structures below a certain scale, typically that of the galaxy, and keeps only the large-scale contributions. For a more rigorous argument on the foundation of tidal theories of galaxy alignment see the review by \citet{Schaefer08}.

The protogalaxy is expected to contract fastest in the direction of the strongest positive curvature of the gravitational potential, so that an initially spherical mass distribution will evolve into an ellipsoid whose principal axes are collinear with those of the tidal shear tensor. Using a similar argument, but applying it directly in projection onto the sky, \citet{CKB01} proposed the following model for the intrinsic shear, $\gamma^{\rm I}$,
\eqa{
\label{eq:ia_ansatz}
\gamma^{\rm I}_+ &\propto \br{\partial_x^2 - \partial_y^2}\; S\{\Phi\}\;; \\ \nn
\gamma^{\rm I}_\times &\propto 2\, \partial_x \partial_y\; S\{\Phi\}\;,
}
where $x$ and $y$ are Cartesian coordinates in the plane of the sky, and the shear tangential and cross components are measured with respect to the $x$-axis. As argued by \citet{HS04}, the form in \Cref{eq:ia_ansatz} is also the lowest-order (and hence simplest) combination of derivatives of the potential that yield the same symmetry as gravitational/intrinsic shear, i.e. the symmetry of a polar or spin-2 quantity. The constant of proportionality absorbs the response of the shape of the visible galaxy to the tidal field, as well as any stochastic misalignments once an ensemble of galaxies is considered. \citet{HS04} adopted this model, related the gravitational potential to the matter density contrast $\delta$, and derived a cross-power spectrum between matter density contrast and the intrinsic shear,
\eq{
\label{eq:gipowerspec}
P_{\delta {\rm I}}(k,z) \propto - \rho_{\rm crit}\; \frac{\Omega_{\rm m}}{D(z)}\; P^{\rm lin}_{\delta\delta}(k,z)\;.
}
They assumed that the tidal field at the time of galaxy formation determines the alignment, so that the correlation with the matter field is frozen in since then, whence the growth factor $D(z)$ is divided out\footnote{As we will discuss later, this model describes the observed galaxy alignments of bright early-type galaxies rather well, including its redshift evolution, albeit within large error bars \citep[see also][]{Paper3}. This is somewhat puzzling because these galaxies are thought to have been created only recently (typically at redshifts below two) by major mergers, disruptive events that one would naively expect to erase all memory of alignment processes during galaxy formation. Hence, the assumptions underlying \Cref{eq:gipowerspec} may not be fully valid, but it could be that any modifications would primarily affect the amplitude of the predicted correlations, which is unconstrained by the model anyway.}. The constant $\rho_{\rm crit}$ corresponds to the total matter density today in a spatially flat universe. This result can be used to predict a GI signal via \Cref{eq:limberia}, and \citet{HS04} also obtained a similar expression for the power spectrum of pairs of intrinsic shears that leads to an II signal,
\eq{
P_{\rm II}(k,z) \propto \br{ \rho_{\rm crit}\; \frac{\Omega_{\rm m}}{D(z)} }^2\; P^{\rm lin}_{\delta\delta}(k,z) + {\cal O}\bc{\br{P^{\rm lin}_{\delta\delta}}^2}\;.
}
Recently, \citet{BVS15} extended these calculations to higher-order perturbation theory, including non-linear evolution of the matter density, non-linear galaxy bias, and the density weighting due to the fact that alignments are only observed at the position of galaxies. \citet{HS04} also related \Cref{eq:gipowerspec} to a correlation function between matter and tangential intrinsic shear, projected along the line of sight,
\eq{
\label{eq:wgplus}
w_{\delta +}(r_p,z) = - \int_0^\infty \frac{\dd k\, k}{2 \pi}\;  J_2(k r_p)\; P_{\delta\, {\rm I}}(k,z)\;,
}
where $J_2$ is a cylindrical Bessel function of the first kind. Using galaxies as tracers for the distribution of mass, one can in practice obtain the correlation function $w_{{\rm g}+}$ from data. This correlation function is measured in bins of transverse separation $r_p$ and line-of-sight separation between pairs of galaxies and then summed over the line of sight to boost signal-to-noise and wash out effects of redshift space distortions. This statistic can be related to \Cref{eq:wgplus} if one knows how galaxies trace the matter distribution, e.g. in the case of linear and deterministic galaxy bias, $b_{\rm g}$, this relation is simply $w_{{\rm g}+} = b_{\rm g}\, w_{\delta +}$. Correlation functions of this form are more readily determined than the power spectrum from data with complex spatial geometry and selection functions, and are therefore the most widely used observables of large-scale galaxy alignment measurements; see \Cref{sec:galaxyalign} and \citet{Paper3}.

A large body of work has considered alignment of orientation angles rather than full ellipticities. Assuming that the galaxy ellipticity components and the galaxy distribution follow a multivariate Gaussian, \citet{BMS11} derived the relation between $w_{{\rm g} +}$ and the average of $\cos (2 \theta)$ over all galaxy pairs where $\theta$ is the angle between the major axis of one galaxy and the line connecting the pair. It reads
\eq{
\label{eq:angle_corr}
w_{{\rm g} +} (r_p,z) = \frac{4 \bar{\epsilon}}{\pi}\; \ba{\cos (2 \theta)}(r_p,z) \bb{w_{\rm gg}(r_p,z) + 2\, \Pi_{\rm max} } = \frac{4 \bar{\epsilon}}{\pi}\; \bc{2 \ba{\cos^2 \theta}(r_p,z) - 1} \bb{w_{\rm gg}(r_p,z) + 2\, \Pi_{\rm max} } \;,
}
where $\bar{\epsilon}$ is the mean absolute value of the ellipticity of the sample, and $\Pi_{\rm max}$ is the maximum (positive and negative) line-of-sight separation between galaxy pairs included in the measurement. The term in square brackets accounts for the fact that the average of $\cos (2 \theta)$ is taken over all galaxy pairs whose numbers are modified due to galaxy clustering, as quantified by the clustering correlation function $w_{\rm gg}$. Note that the definition of mean alignment introduced by \citet{FLW+09} is subtly different in that it averages over the clustering correlation function and thus over the number of \emph{excess} galaxy pairs.
We added the second equality to \Cref{eq:angle_corr}, which is a useful expression in a slightly different context, using trigonometric relations. The average of $\cos^2 \theta$ over all pairs is a popular statistic in galaxy cluster alignment studies, where in this context $\theta$ corresponds to the angle between the major axis of the satellite distribution and the line to the other cluster in the pair \citep[e.g.][]{SMB+12}. All these predictions originate from the ansatz in \Cref{eq:ia_ansatz}, which leads to a linear scaling of the intrinsic shear with matter density contrast and is thus termed the \emph{linear alignment model}. It is widely considered appropriate for elliptical galaxies, as well as the distribution of galaxies within clusters, as these are believed to trace the shape of the underlying dark matter haloes.

For disc galaxies it seems physically reasonable to consider a model that relates galaxy angular momentum to the tidal field. Using \Cref{eq:zeldovich}, \citet{White84} showed that the angular momentum of a proto-galaxy is given by the expression
\eq{
J_\alpha \propto \sum_{\beta\gamma\sigma} \varepsilon_{\alpha\beta\gamma}\; I_{\beta\sigma}\; T_{\sigma\gamma}\;,
}
where $\varepsilon_{\alpha\beta\gamma}$ is the Levi-Civita symbol. If the inertia and tidal tensors are perfectly aligned, i.e. if they are diagonal in the same coordinate system, no angular momentum is generated. However, since $I_{\beta\sigma}$ is determined solely by the proto-galaxy's shape, whereas $T_{\sigma\gamma}$ is dominated by large-scale distribution of matter (especially when the potential has been smoothed), this is generally not expected. Simulations indicate that significant correlation between the inertia and shear tensors is present \citep{PDH02b}, which will suppress the magnitude of the resulting angular momenta and their correlations. To account for this, \citet{LP00} proposed the following effective one-parameter model for the Gaussian angular momentum distribution $p(J_\alpha|T_{\alpha\sigma})$ for a given tidal shear, by making an ansatz for the covariance matrix $\ba{J_\alpha J_{\alpha^\prime}}$ between the angular momentum components,
\eq{
\label{eq:lee_spin_modeldef}
\ba{ J_\alpha J_{\alpha^\prime} } = \frac{1}{3}\, \delta_{\alpha\alpha^\prime} + C\, \bb{\frac{1}{3}\, \delta_{\alpha\alpha^\prime} - \sum_\sigma \hat{T}_{\alpha\sigma} \hat{T}_{\sigma\alpha^\prime}} \;,
}
where $\hat{T}_{\alpha\sigma}$ is the normalised, trace-less tidal shear tensor and $\delta_{\alpha\alpha^\prime}$ is the Kronecker-$\delta$. It can be derived from the gravitational quadrupole by subtraction of the trace, $\tilde{T}_{\alpha\alpha^\prime}=T_{\alpha\alpha^\prime}-\delta_{\alpha\alpha^\prime} {\rm Tr}(\vek{T})/3$ and subsequent normalisation, $\hat{T}_{\alpha\sigma} = \tilde{T}_{\alpha\sigma}/\sqrt{\tilde{T}_{\alpha\rho}\tilde{T}_{\rho\sigma}}$, where the Einstein summation convention over repeated indices is implied. For $C=1$, \Cref{eq:lee_spin_modeldef} reproduces the result for angular momentum correlations obtained when assuming completely uncorrelated inertia and shear tensors \citep{LP01}, which yields the tightest coupling of the angular momentum to the tidal field. A value $C<1$ is expected due to non-linear evolution because of the partial alignment of the inertia and shear tensors, while in the limit $C=0$ the tensors are perfectly aligned and the angular momentum directions are randomised. Assuming a Gaussian distribution of the components of the angular momentum vector, \citet{LP01} then deduced the correlation of the directions of angular momenta (as opposed to the correlation of the full angular momentum components in the foregoing equation),
\eq{
\label{eq:lee_spin_align}
\ba{ \hat{J}_\alpha \hat{J}_{\alpha^\prime} } = \frac{1+a_{\rm T}}{3}\, \delta_{\alpha\alpha^\prime} - a_{\rm T}\, \sum_\sigma \hat{T}_{\alpha\sigma} \hat{T}_{\sigma\alpha^\prime} \;,
}
where $\hat{J}$ is the normalised spin vector. This equation looks deceptively similar to \Cref{eq:lee_spin_modeldef}, but note that $a_{\rm T}$ is now a parameter that runs between 0 (random angular momenta) and $3/5$ (maximum alignments).

Assuming that the ellipticity of a galaxy is given by the projection of a circular disc which is orthogonal to the angular momentum direction of the galaxy, the intrinsic shear can be computed as \citep{CKB01,CNP+01}
\eqa{
\gamma^{\rm I}_+ &\propto \br{\hat{J}_x^2 - \hat{J}_y^2} \;; \\ \nn
\gamma^{\rm I}_\times &\propto 2\, \hat{J}_x \hat{J}_y \;,
}
where the coordinate system is defined as above: the $z$-direction is parallel to the line-of-sight and the indices run over the $x$- and $y$-directions. \citet{CNP+01} proceeded to show that
\eq{
\xi^{\rm II}_+(r) \propto a_{\rm T}^2\; \xi_{\delta\delta}^2(r)\;
}
holds for a correlation function of intrinsic shears, measured as a function of three-dimensional pair separation $r$. Here, $\xi_{\delta\delta}$ is the correlation function of the matter density contrast (the Fourier transform of $P_{\delta\delta}$). This proves that the parameter $a_{\rm T}$ only modifies the amplitude of correlations, and is thus degenerate with the impact of finite disc thickness and a stochastic misalignment between the spins of the stellar and dark matter components of a galaxy. This \emph{quadratic alignment model} predicts correlations that scale with the square of the matter correlation function or power spectrum to lowest order, which suppresses the alignment signals. Besides, in linear theory there are no correlations between matter and quadratically aligned intrinsic shear if both fields are Gaussian, as these would be third-order in the matter density contrast \citep{HS04}. Note however that non-linear evolution may introduce a linear scaling on large scales, as qualitatively argued by \citet{HZ02} using higher-order perturbation theory, so that disc galaxy alignments could in principle feature signals of the form predicted by the linear alignment model. \Cref{eq:lee_spin_align} has also been used to derive distributions of alignment angles between angular momentum directions and the principal axes of the tidal tensor; e.g. in the case of void surfaces one obtains for the angle $\theta$ between the angular momentum direction and the surface normal (\citealp{LE07}; in the corrected form by \citealp{SW09})
\eq{
\label{eq:voidalign}
p(\theta) = \sin \theta\; \frac{2 (1-a_{\rm T}) \sqrt{2+a_{\rm T}}}{(2+3a_{\rm T} \cos^2 \theta - 2a_{\rm T})^{3/2}}\;.
}

\section{Developments of the twentieth century}
\label{sec:history}

In this section we provide a rough guide to the study of galaxy shapes, orientations, and alignments
throughout the twentieth century, deferring the discussion of more recent research since about 2000 to the main part
of this paper and its companion works. We do not claim to be comprehensive but rather highlight
influential works and key developments. 

\subsection{The very first works: 1914-1926}
\label{subsec:first}

The first statement regarding the orientations of galaxies that we have been able to trace was made
by \cite{Fath14}, who published a list of 1,031 galaxies observed with the 60-inch reflector at Mt.\
Wilson Solar Observatory. It contains position angles for approximately 60\% of the objects and,
while there was no detailed study in the text, it was stated in the summary that ``[o]n the assumption
that most of the nebulae are approximately disc-shaped the planes of the discs appear to be oriented
at random in space.''

A more explicit analysis of the orientations of galaxies was performed by \cite{Reynolds20}, using 
seven galaxies in the southern and 16 in the northern Galactic hemispheres whose
diameters\footnote{These sizes are difficult to relate to modern measurements such as the half-light 
radius, but it is enough to consider that these are 23 of the largest galaxies, in apparent size, 
that there are.} exceeded 
$10'$. \citet{Reynolds20} noted that 16 of the 23 galaxies (70\%) had inclination 
angles with respect to the line-of-sight of 30$^\circ$ or less, consistently for both Galactic
hemispheres, and briefly suggested 
that this could imply some connection between our own Galaxy and those he observed in the sense 
that the planes tended to coincide. Soon after, \cite{Reynolds22} confirmed his previous result 
using 263 spirals, including galaxies up to $\sim$3 times smaller than before. Of these 263 
spirals, 172 (65\%) had inclination angles $<30^\circ$, consistent with his previous measurement.  
He found a relation between size (which he regarded as a measure of distance) and 
average angle, with galaxies with diameters \{$3'-5'$, $5'-10'$, $>10'$\} having average inclination 
angles \{$26.7^\circ$, $24.2^\circ$, $21.9^\circ$\}. He used his findings to construct an argument that supported Shapley\rq s view that these \lq spirals\rq\ were \emph{not} objects similar to the Milky Way.

Subsequently, a lively debate between Reynolds and \"Opik over the validity of these results ensued \citep{Oepik23a,Reynolds23a,Oepik23b,Reynolds23b}. While the discussion had no definite conclusion, it marked the first occurrence of a persistent theme in galaxy alignment observations: the importance of selection effects and systematic trends. Remarkably, at least two points were raised in this regard which are still relevant today. \citet{Oepik23a} argued that Reynolds\rq\ results stem from a selection effect rather than a real alignment, in that edge-on spirals with fixed apparent magnitude have a higher surface brightness than face-on spirals, thus favouring the detection of the former. \citet{Reynolds23a} replied that the suggested larger surface brightness would be compensated by an ``absorbing medium along the periphery of the spiral arms,'' such that spiral galaxies should appear brightest at 20$^\circ$--30$^\circ$. Indeed, selection biases due to orientation and dust can be important in modern datasets as well and may have impacted on some of the results discussed later on in this section.

For elliptical galaxies their flattening generally was the only clearly discernible morphological
feature in the images of the time. Therefore ellipticity determined the subdivisions on Hubble's
early-type branch, E$n$, where $n$ ran from 0 to 7 and was given by the integer part of $10(1-
b/a)$, where $b/a$ again denotes the axis ratio of the galaxy image. \citet{Hubble26} also
presented a de-projection scheme for elliptical galaxies, noting that the distribution of projected
galaxy ellipticities is sensitive to whether galaxies have preferred orientations or are randomly
distributed. Thus Hubble initiated another recurrent theme in galaxy alignment studies, the relation between three-dimensional correlations and their projection on the sky.

\subsection{A second wave: 1938-1958}

\citet{Brown38a} revisited Reynolds's claim of preferred inclinations of disc galaxies using a
sample of 600 galaxies observed at Heidelberg observatory, covering the whole northern sky down to
declination $-20^\circ$. Among disc galaxies exceeding $2\arcmin$ in diameter he found a $14\,\%$
excess of galaxies with axis ratios below $1/3$ and an even more pronounced depletion of nearly
circular objects. He presented evidence for the completeness of his sample, thus arguing in favour
of the physical nature of this effect. This claim was disputed by \citet{Knoxshaw38} who did not
find deviations from random orientations of disc galaxies larger than $2\arcmin$ in a sample of
similar size in the Shapley-Ames catalogue. \citet{Brown38b} in turn challenged those results,
finding an excess of nearly edge-on objects in the same catalogue after correcting for
incompleteness of strongly elliptical objects due to the limiting magnitude. He presented additional
data that suggested the excess is seen down to galaxy diameters of $30$\arcsec.

Brown was also among the first to investigate the systematic alignment of galaxies in the plane of
the sky. In the Horologium supercluster he found a strong preference in a sample of 355 galaxies for
a narrow range of position angles (e.g.\ more than twice the number of galaxies than expected for a
random orientation in a $10^\circ$ range of position angle; \citealp{Brown39}). \citet{WB55} later
made a very similar observation for a region in Cetus, claiming a detection in excess of
$4\sigma$. However, \citet{Reaves58} noted that a new, more complete dataset showed no preferred
galaxy orientations in the same Horologium area of sky. It is instructive to read Brown's review of
these works and others of that period in the introduction of \citet{Brown64}.

As the examples above showcase, many observations of galaxy alignments were -- and arguably are to
the present day -- tentative, depending strongly on the approach and dataset used. While it was
clear from the earliest efforts that selection effects played a pivotal role (see
e.g.\ \citealp{BHF91} for a detailed discussion), it is interesting to note that neither distortions
of shape and spurious alignments caused by telescope tracking, optics, or the photographic plates
nor biases due to the human-led object detection and morphology measurement were mentioned in these works.

\subsection{A case study: alignments in the Palomar Sky Survey}

\begin{figure*}[t]
\begin{minipage}[c]{.5\textwidth}
\centering
\includegraphics[scale=.36]{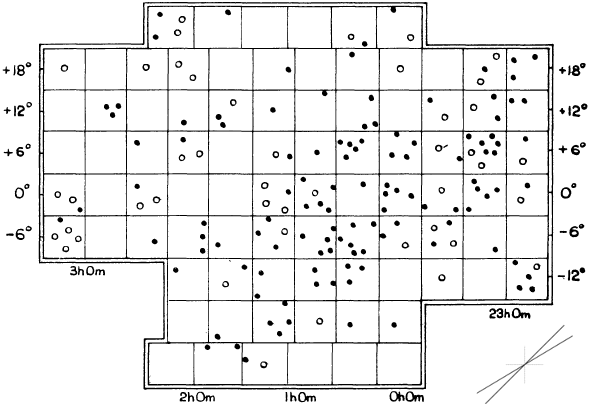}%
\end{minipage}%
\begin{minipage}[c]{.5\textwidth}
\includegraphics[scale=.25]{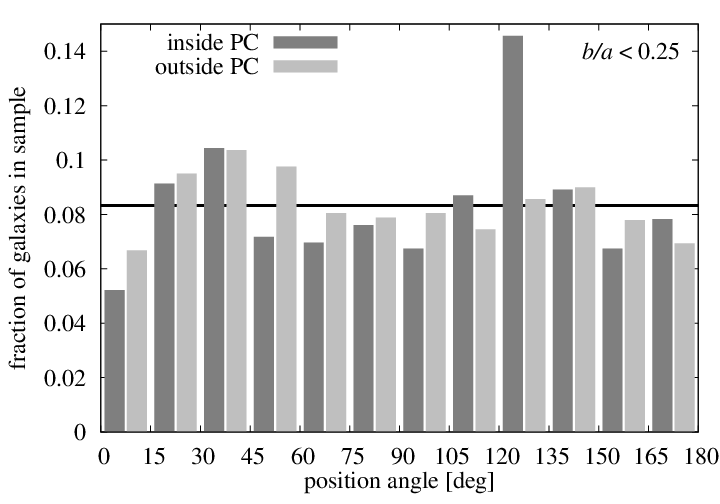}
\includegraphics[scale=.25]{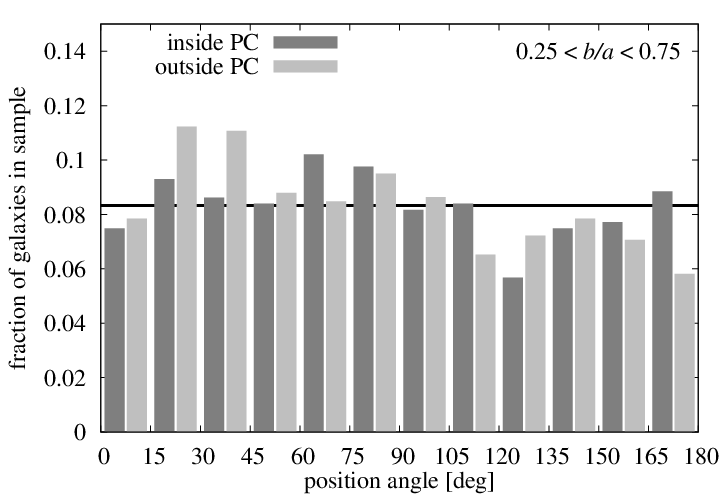}
\end{minipage}
\caption{\textit{Left}: Area of the Palomar Sky Survey analysed by \citet{Brown64}. Dots (circles)
  mark the positions of galaxies with diameters in excess of 40$\arcsec$ (60$\arcsec$),
   axis ratios of less than 0.25, and position angles in the range 121 to 135 degrees
  (East of North, indicated by the lines at the bottom right). Rectangular lines correspond to plate
  boundaries. \textit{Right}: Normalised histograms of position angle distributions compiled from
  table VII of \citet{Brown64}. The top (bottom) panel shows galaxies with axis ratios $b/a < 0.25$
  ($0.25 < b/a < 0.75$). Light grey bars correspond to galaxies inside the \lq Pisces
  Concentration\rq\ (PC, roughly comprises the overdensity seen in the left panel), dark grey bars
  to those outside the concentration. The black horizontal line indicates the expected fraction for a
  random distribution of galaxy orientations. \permmn{Brown64}}
\label{fig:brown64}
\end{figure*} 

It is worth taking a closer look at the work of \citet{Brown64}, which was based on a galaxy sample size
(close to 5000) approaching the order of magnitude that is 
routinely used nowadays, and claimed a significant detection of
alignments, both among galaxies and between galaxies and large-scale structure. The analysis was
performed on several thousands of square degrees situated around the vernal equinox, based on
observations with the Palomar Observatory Sky Survey (POSS). The actual measurements were done on
prints of the photographic plates which \lq\lq were clipped to a plywood base of the same size as
the print and covered by a sheet of highly transparent material sold commercially as
Polyglaze\rq\rq, which in turn was ruled with a grid pattern. Galaxy sizes and position angles
were then read off with a scale and protractor, while the type classification and rejection of
dubious images or artefacts was done by eye, helped by the colour information from the red and blue
POSS plates.

Limiting himself to galaxies with major axis diameters larger than 40$\arcsec$, Brown found
a strong excess of position angles in a bin $15^\circ$ wide, concentrated in an area he referred to
as the Pisces Concentration that covers part of the Pisces-Cetus supercluster and the Pegasus I and
II clusters. As can be seen in \Cref{fig:brown64}, the elongation direction of the concentration
matches closely the preferred direction of galaxies within. \citet{Brown68} extended the analysis to
other areas covered by POSS and again found several samples with marked excesses in position
angle. \citet{Reinhardt71} confirmed the significance of the detections via simple statistical
checks and identified another nearby cluster region in which galaxies align with themselves and with
the main axis of the cluster.

\citet{Brown68} tested for systematic errors in the position angles by repeating the measurements
after rotating the prints by $90^\circ$, and by comparing measurements of the same galaxy on
overlapping plates, in both cases finding biases well below the bin width of the
distributions. \citet{Reinhardt72} discussed in detail possible systematics, including physiological
ones related to the visual analysis, and conducted further significance tests on Brown's measurements, concluding that the effect
must have an astrophysical origin. In contrast to this, \citet{Oepik70} once again argued that the
preferential orientations of position angles could be explained by selection effects, his key argument
being that any preferred orientation of disc galaxies has to coincide with a preferred axis ratio,
i.e.\ disc inclination. Since this was not observed, \"Opik concluded that the effect cannot be
physical.

To our knowledge this debate remains without definite conclusion to the present day. It is
challenging to translate the selection criteria and measurements to modern CCD-based observations
and automated galaxy detection and morphology measurements, particularly for the key parameters, size
and axis ratio \citep[see][]{FP85}. Current databases have their own limitations, e.g.\ a
reproduction of Brown's results with standard products of the Sloan Digital Sky Survey pipeline is
hindered by a spurious pattern in the distribution of position angles caused by limitations in the
fitting procedure of galaxy light profiles \citep{VCD13}. Similarly, in an attempt to repeat the
analysis using a scanned catalogue of the Palomar Survey\footnote{\texttt{http://aps.umn.edu/}} we
found patterns of preferred position angles of $0^\circ$ and $90^\circ$ for large, edge-on galaxies
in the area covered by \citet{Brown64}, which may hint at a spurious alignment with the edges of the
photographic plates. This could be caused by systematic effects in the shape measurement (as image distortions are likely to be largest far from the optical axis of the telescope) or again due to selection effects (galaxies with orientations perpendicular to the plate boundary are more likely to have their light profile cut off by the edge, and thus be discarded, than those which are parallel).

\subsection{More recent observational works}
\label{sec:hist_recent_obs}

For a further discussion of alignments of nearby galaxies with the local large-scale structure we recommend the review by \citet{HWS+06} who not only covered recent work but also the historical development
of the field. The authors provided a detailed account of the limitations and systematic effects in
the majority of twentieth century datasets, arguing that complex selection effects, incompleteness
and contamination of galaxy samples, as well as an under-estimation of statistical and systematic
errors explains many discrepancies found in early papers. This applies in particular to the
Uppsala General Catalogue, which was derived from the Palomar Sky Survey and constituted a standard
dataset from the mid 1970s into the 1990s. It is likely that the arguments of \citet{HWS+06} also
apply to the alignment studies in the more distant Universe of that period.

In light of this context, the work of \citet{HP75} is remarkable in its rigorous error analysis and more
conservative conclusions. The authors did not claim detections of any significant physical galaxy
alignments in a large Palomar Sky Survey sample in excess of 5500 galaxies, except a tentative
alignment of galaxies in the Coma cluster towards the cluster centre. Using blind analysis
techniques, they did however identify various sources of significant systematic signals including: a
decrease in the measured size of galaxies over the duration of the project, a tendency for galaxies
to be aligned vertically on the prints (due to observer bias or distortions in the print), a
preferential selection of diagonally oriented galaxies if these have small angular size, and a
potential bias due to the analyst's assignment of measured position angles at histogram bin boundaries.

\citet{Helou84} detected spin correlations among about 30 close galaxy pairs, using a combination of
spectral line measurements, dust obscuration in the disc, and spiral patterns to assign a sense of
rotation in addition to the spin direction. He found evidence for a preferential anti-alignment of
galaxy pairs, and thus against the primeval turbulence
model, which predicted parallel alignment. \citet{LGP88} did not observe a departure from random orientations for spiral and lenticular
galaxies with respect to the surrounding galaxy distribution on scales of a few megaparsecs, but
found the major axes of ellipticals to align with the large-scale structures. On similar scales,
\citet{ML92} claimed a $2.8\sigma$ detection of major axis alignment with the position of the
nearest neighbour for spirals, while reporting significant alignments beyond the nearest neighbour
for elliptical galaxies.

Regarding the orientations of galaxies within individual clusters, early 
results were contradictory, sometimes even on a cluster-by-cluster basis: in the Coma 
cluster, \cite{RB67} found no evidence of a preferred orientation but \cite{Djorgovski83} found 
significant alignments both of satellite galaxies towards the central galaxy and between clusters \citep[see also][]{T76}. 
On a statistical level, however, most studies pointed to random orientations of these galaxies.
The first such statistical analysis was done by \cite{Dressler78}, who found no evidence for 
satellite galaxy alignments in 12 clusters. 

\begin{figure}[t]
\begin{minipage}[c]{.4\textwidth}
\centering
\includegraphics[scale=.24]{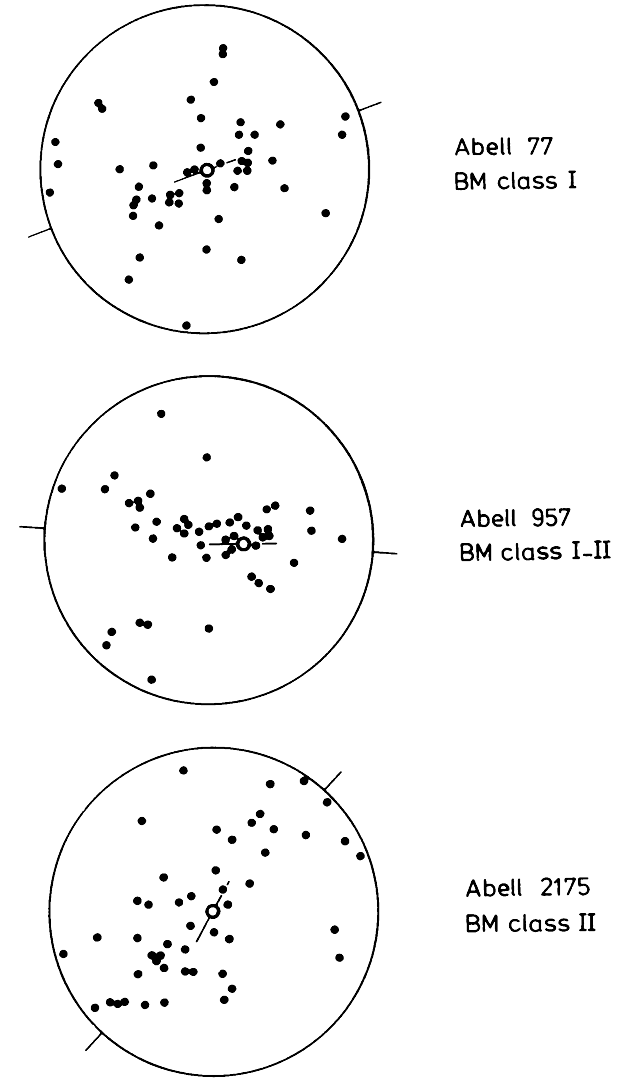}
\end{minipage}%
\begin{minipage}[c]{.6\textwidth}
  \caption{Three strongly elongated Abell clusters analysed by \citet{Binggeli82} . The 50 brightest
    galaxies in a radius of 2 Mpc are plotted as black dots in each case. The brightest galaxy (BCG)
    is indicated by the circle. Position angles of the BCG and cluster are given by the thin black
    lines. BM class stands for the morphological Bautz-Morgan classification \citep{BM70} of galaxy clusters, where classes I and II are dominated by elliptical BCGs. \permaa{Binggeli82}}
\label{fig:binggeli82}
\end{minipage}
\end{figure} 

Given the intricacies of measuring the shapes and orientations of individual galaxies, astronomers
hoped to obtain more robust results from clusters of galaxies, using the distribution of member
galaxies on the sky as a tracer of the projected cluster shape. In an early effort, \citet{Sastry68}
measured the alignment of the central bright (cD) galaxy with the shape of its host cluster for 
nine Abell clusters, each with around 30 securely identified galaxy members in the Palomar Sky
Survey. He found strong alignment within 10$^\circ$ for five clusters (later confirmed by \citealt{CM80}) and no alignment for the Coma
cluster; the remaining clusters were too close to circular to define an orientation. Soon after, \citet{Holmberg69} found the opposite result for spiral galaxies similar to the Milky Way: their haloes, as traced by the satellites, tend to be aligned with the galaxies\rq\ \emph{minor} axis. \citet{LyndenBell76} pointed out that this effect is also present in the Milky Way whose satellites and distant globular clusters lie in a tight plane within about $10^\circ$ from the 
Galactic poles.

In a pioneering work, \citet{Binggeli82} obtained the position angles and ellipticities of 44 regular (i.e. with shapes which are well approximated by an ellipse) Abell clusters and their brightest cluster
galaxies (BCG) in the same survey. The strong alignment 
of clusters and BCGs was confirmed; see \Cref{fig:binggeli82}. Additionally, the author
detected alignments between neighbouring clusters separated by less than about 30 Mpc, and between
cluster shape and the distribution of surrounding clusters out to 100 Mpc, arguing that tidal
interactions must play a critical role in the dynamics of these systems. \citet{SP85} repeated the
analysis with 237 clusters located in superclusters, finding only weak, if any,
detections. However, later works supported Binggeli's findings with increasingly large samples and varying degrees of significance \citep{AGP+86,LGP88b,LNM+90,Plionis94,FWB99}.

\citet{West89} extended this type of analysis to the alignment of galaxy group shapes with neighbouring
group positions and detected alignments out to tens of megaparsecs. Interestingly, he concluded that
this result would favour a top-down structure formation scenario and hence provides evidence against the Cold Dark Matter model. \citet{ML89} did not see alignments of spiral and lenticular galaxies with
the surrounding galaxy distribution, suggesting a galaxy type and/or environment dependence of galaxy
alignments when contrasted with the results for elliptical galaxies in the centres of clusters. Dropping the assumption that cluster galaxies trace its overall shape, \citet{WJF95} studied the correlation of the X-ray cluster ellipticity or its substructure distribution, as measured by the Einstein
satellite, with the large-scale structure, using 93 clusters with $z<0.2$, and claiming a marked
detection of alignments. \citet{CMM00,CMM02} confirmed a strong alignment with the position of nearby clusters using 103 different clusters and data from the Einstein and ROSAT satellites.

\subsection{Early Tidal Torque Theory}

\citet{FH51} was the first to suggest tidal torques as the origin of galactic
rotation.\footnote{\citet{GE03} provides an excellent description of the conference proceedings in
  which Hoyle first proposed his theory. In fact, as highlighted by \citet{GE03}, Hoyle makes a
  remarkably prescient statement in a later work \citep{H66} that \lq\lq the properties of the
  individual stars that make up the galaxies form the classical study of astrophysics, while the
  phenomena of galaxy formation touches on cosmology. In fact, the study of galaxies forms a bridge
  between conventional astronomy and astrophysics on the one hand, and cosmology on the
  other.\rq\rq} In this paradigm the matter falling into a halo to form a galaxy acquires angular
momentum through the interaction of the quadrupole of the matter distribution with the tidal
gravitational forces of the surrounding matter (see \Cref{sec:tidaltheory} for a more detailed account). As
\citet{Peebles69} noted in his seminal paper, tidal torque theory explains the formation of rotating
discs as a natural consequence of the gravitational instability picture of structure formation, as
opposed to earlier suggestions by \citet{W51} and \citet{G52} who postulated a \lq primeval
turbulence\rq\ which would have transferred an ab initio value of angular momentum to the galaxies.

As argued e.g.\ by \citet{SFS70}, the amount of angular momentum of the gas cloud collapsing at the
time of galaxy formation was considered to be critical to determine the Hubble type of the galaxy,
with little impact of the subsequent evolution. Low angular momentum was linked to rapid
fragmentation and thus early star formation which quickly depletes gas and leads the final stages
of the collapse to be governed by stellar rather than gas dynamics, creating a spheroidal system. Conversely, the systems with
high angular momentum collapse into the gas-rich discs associated with spiral galaxies \citep[see
also][]{Jones76,ES83}. \citet{Jones76} also proposed spin correlations between nearby galaxies as an
observational test to distinguish tidal torque theory from the primeval turbulence model.

\citet{SLW79} proceeded to demonstrate that disc galaxies in close pairs can have a non-random
distribution of the angle between their major axes, as a result of spin angular momentum alignment
due to tidal torques. However, they found that the observed distribution of this angle, using $79$
pairs of galaxies, was consistent with the spin of the galaxies being completely uncorrelated; this
was still the case even when a more stringent selection criterion was used to isolate galaxy
pairs. Their attempt at reducing observer bias in manually measuring the position angles was to 
employ at least two analysts to measure the angle of each galaxy pair member, whilst the other
galaxy was masked out. \citet{SLW79} suggested that since the data seemed inconsistent with the
assumption that the galaxy spin
was frozen in at formation, later interactions might further affect spin.

\citet{White84} then established the concept and formalism of tidal torque theory as it is currently used
in a succinct work that built on an earlier argument by \citet{Doroshkevich70}. While
\citet{Peebles69} calculated the build-up of angular momentum in a spherical region, which takes
place at second order in perturbation theory, White showed that galaxies as generally non-spherical
objects acquire angular momentum via the torques induced by the tidal quadrupole which is misaligned
with the moment of inertia, a first-order effect. Moreover, he demonstrated that in Peebles\rq\
picture angular momentum is generated solely via the transport of matter across the boundary of the
sphere. White confirmed the results for the first-order effect for groups of particles in $N$-body
simulations that were linked with a friends-of-friends algorithm, so long as the matter density is
low enough that first-order perturbation theory is applicable \citep[this was later confirmed
by][]{BE87}.

\subsection{More recent theoretical and numerical progress}

Galaxy clusters were also at the focus of theoretical research on the shapes and alignments of
matter structures in the last two decades of the twentieth century, in part to match the
observational progress but also because the resolution of $N$-body 
simulations allowed a detailed study of only the largest bound structures. For instance,
\citet{DWA84} measured alignments of cluster-size dark matter haloes with each other and the
position of neighbouring haloes in $N$-body simulations, with the aim of verifying the observations by
\citet{Binggeli82}. Significant detections that matched observations were indeed made but only in simulations that featured top-down
structure formation scenarios, while no agreement was found for the signals generated by tidal alignments in Cold Dark Matter models. It should be noted that the
simulation suites used in that work contained a total of $10^3$ to $10^4$ particles, with massive clusters
typically consisting of $10 - 30$ particles, which is at least an order of magnitude below current
limits for reliable measurements of halo shapes and angular momenta \citep{BEF+07,JSB+13}.

Cluster alignments were thought to be key probes in providing answers to the most important
cosmological questions of the time, such as discriminating between hot and cold dark matter models,
or determining the total matter density parameter $\Omega_{\rm m}$. Indeed, in their simulation-based
study of alignments of cluster-size haloes with the surrounding matter distribution, \citet{WDO89}
argued that alignments (but not shapes) can help discriminate between different cosmogony scenarios
\citep[see also][]{WWD90}. Subsequent works used these alignments to test structure formation scenarios, and probed their sensitivity to cosmological parameters, sometimes with contradictory conclusions \citep[e.g.][]{WVD91,HW93,SML+97}.

Regarding analytic work, a productive route was the analysis of the peaks in Gaussian random fields
which are expected to be a good representation of the primordial spatial distribution of matter
density fluctuations. Depending on the peak threshold or smoothing scale, these peaks are assumed to
describe the matter distribution around galaxies or clusters. \citet{PH85} calculated the
distribution of shapes of peaks in random fields, finding that typical peaks (corresponding to
massive galaxies) are prolate, nearly independent of the form of the matter power spectrum, while
the highest peaks (corresponding to clusters) are closer to spherical (see also the seminal paper by
\citealp{BBK+86} who also consider the shapes of peaks in the matter distribution). \citet{HP88}
extended this work to computing tidal torques and spin parameters of the peaks, concluding that the
low angular momentum of elliptical galaxies compared to spirals cannot be explained solely by the
larger host halo mass of ellipticals because the calculated anti-correlation between peak mass and
spin parameter was weak. Using a similar approach, \citet{CT96} later obtained realistic spin
parameter values for spiral galaxies, reasoning that highly non-linear processes must therefore
happen during the evolution of ellipticals, such as would occur during violent galaxy
mergers. \citet{SS93} analytically estimated cluster properties based on tidal alignments and could
explain the observed distribution of cluster axis ratios and (qualitatively) the alignment signals
of clusters. The authors concluded that tidal effects are important ingredients in the dynamics of
large-scale structure.

Generally, tidal forces emerged as the central driver for all forms of observed alignments in the
large-scale matter distribution. In addition to the well-known link to the angular momenta of bound
structures, \citet{D92} proposed that the tidal shear directly affects the shape of a
halo. However, in constrained $N$-body simulations with external tidal shear, he did not find a clear
correlation between the shapes of initial density peaks and those of the resulting
haloes. \citet{CKB01} built on this work to devise their \lq tidal stretching\rq\ model for intrinsic alignments (see \Cref{sec:tidaltheory}), predicting that the correlations of projected galaxy shapes should be stronger for elliptical
than for spiral galaxies, which current results suggest is indeed the case (see 
\Cref{sec:galaxyalign}). Note that, much earlier, \citet{BE87} had seen from simulations that the
alignments of the major axis of a halo with nearby objects was stronger than the corresponding
alignment of the angular momenta. Hence, while the repeated usage of tidal effects to explain the
origin of the dichotomy between spiral and elliptical galaxies may not have proven conclusive, they
do seem to imply a marked galaxy type dependence of alignment signals.

The paper by \citet{CKB01} was part of a suite of works which were contemporaneous with the first
detections of cosmic shear \citep{BRE00,KWL00,WME00,WTK00} and which established a link between the old field of galaxy alignments
and the novel, promising cosmological probe that equally relies on measurements of correlations
between galaxy shapes. \citet{HRH2000} used $N$-body simulations to quantify the expected
contamination of weak lensing signals by intrinsic galaxy alignments. They considered both an
elliptical and spiral model analogous to \citet{CKB01} for the translation of halo shape to galaxy
shape and concluded that deep surveys with a broad redshift distribution should not be affected
significantly. Assuming that halo shapes trace the galaxy shapes, \citet{CM00} found significant
intrinsic shape correlations in $N$-body simulations out to 20 Mpc/h, which could account for
$10-20\%$ of the signal seen in first cosmic shear detections. They anticipated the concern that
this contamination would become more significant once weak lensing source samples were defined in
narrow redshift slices.

\section{Intra-halo alignments}
\label{sec:clusters}

A dark matter halo above a certain minimum mass tends to host one or more galaxies, where in the latter case one distinguishes between a central galaxy and satellites. Frequently, the most massive or brightest of the galaxies in a halo is observationally defined as the central galaxy. It is generally expected to be situated close to the centre of the halo which attracts most of the baryonic matter, especially if the central is much more massive than any satellite. In this section we discuss the alignments of the central galaxy with its own halo, in most cases traced by the spatial distribution of the satellite galaxies\footnote{Constraints on potential misalignments could in principle also be obtained from halo shape measurements via galaxy-galaxy lensing, stacking galaxies and orienting them with respect to their major axes \citep{HYG04,MHB+06,UHS+12,ACD15}. }, as well as the alignments of the shapes of satellites measured relative to the centre of the halo (or the observable central galaxy). Since the satellites generally have to be identifiable in surveys beyond the local Universe, the haloes studied are mostly those of galaxy groups (with typical masses $10^{13}\lesssim M/M_\odot\lesssim10^{14}$) and clusters (with masses $M\gtrsim10^{14}M_\odot$).

While the alignment of central galaxies with their host haloes (measured through the distribution 
of satellite galaxies) was established early on, the twentieth century saw a persistent lack of 
consensus regarding the level -- or presence -- of alignments of satellite galaxies within haloes, 
ranging from field galaxies to massive galaxy clusters.
The history of such results can be understood in terms of the technical challenge in measuring the 
shapes of each of the objects: the shapes of central galaxies are relatively easy to measure, 
especially in massive galaxy clusters, since these are the largest galaxies in the Universe. In 
contrast, satellite galaxies are smaller, and their light is confused by neighbouring galaxies (an 
effect which is enhanced as we go from field galaxies to galaxy clusters).

\subsection{Alignment of the central galaxy with its halo}
\label{sec:galaxyhaloalign}

As detailed in \Cref{sec:hist_recent_obs}, \citet{Sastry68} first detected an alignment between the major axes of the central galaxy and the satellite distribution in clusters, while \citet{Holmberg69} established an anti-alignment in the case of field galaxies and their satellites. This \textit{Holmberg} effect is now well established locally in the Milky Way \citep{PPK12} and M31 \citep{CLI+13}. However, the picture is different for objects outside the Local Group. The \emph{inverse} 
Holmberg effect\footnote{In hindsight this nomenclature is unfortunate, since Sastry\rq s observation 
took place before that of Holmberg.}, i.e.\ the tendency of the central galaxy to align with its parent halo, has been observed from the scales of individual galaxies \citep[e.g.][]{Brainerd05} to galaxy clusters, the latter in support of the pioneering works by \citet{Sastry68} and \citet{Binggeli82}; see also \Cref{fig:binggeli82}. This is also strongly supported by simulations \citep[e.g.][and references therein]{WLK+14}. This correlation appears to extend even beyond the host halo and may reflect the preferred directions of accretion of satellites, as evidenced by the alignment of satellite position vectors and local filament directions seen in SDSS as well as simulations \citep{TGK+15}. Note that a direct comparison with the results from the Local Universe is difficult because the work on the satellite distribution of more distant galaxies is limited to larger separations from the central galaxy and suffers from low numbers, so that alignments can only be detected statistically rather than on a per-galaxy basis.

Using SDSS galaxy groups, \cite{YBM+06} showed that the Sastry-type major-axis alignment is stronger for red galaxies and haloes defined by red satellites only, and is indeed not 
present in groups with blue central galaxies \citep[see also][and references therein]{WYM+08}. Additionally, this effect is more pronounced at low redshifts 
and for more massive central galaxies but is independent of cluster richness \citep{HKF+11}. \citet[note the reversal of alignment tendency published in the erratum]{SL04,SL09} showed that there also exists an inverse Holmberg effect for field galaxies (i.e.\ galaxies that do not have other galaxies above a certain brightness threshold in their neighbourhood), which is more prominent for red, passive central galaxies. Later works confirmed that the dichotomy between blue galaxies with no significant alignments and red galaxies with clear major-axis alignment extends down to galaxy-sized haloes, where the latter signals exists out to radii of $0.5\,{\rm Mpc}/h$ \citep{APP+07} and persists for stricter criteria of isolation of the host \citep{BPN+08}. \citet{AB10} confirmed these trends and were also able to qualitatively reproduce them with a simulation-based model, where blue central galaxies are assumed to have their spin aligned with the dark matter halo angular momentum and red galaxies are homologous to their halo. They explained the lack of blue satellite alignment with the more recent accretion of blue satellites compared to red ones, and a substantial number of interlopers in the sample, while the signal for blue central galaxies is diluted by averaging a major-axis alignment at small separations from the centre of the halo with a preferential minor-axis alignment (i.e. a Holmberg effect) at larger radii ($r_p \gtrsim 300\,{\rm kpc}$). 

A comparison of the strength of large-scale alignments in observations and simulations led \citet{HBH+04} to suggest that galaxies and their dark matter haloes have typical misalignments of $\sim30^\circ$, later constrained by \citet{OJ09} and \citet{OJL09} to $35\pm2^\circ$. \citet{WYM+08} found similar results through a detailed study of SDSS galaxy groups, inferring an average projected misalignment of $23^\circ$, which was stronger for blue galaxies and more massive haloes, reaching $\sim65^\circ$ for blue centrals in haloes of $10^{13}<M/M_\odot<10^{14}$. This can be compared to the recent analysis of a hydrodynamic simulation by \citet{TMD+14,TMD+14b}, who determined that the (three-dimensional) mean misalignment between the dark matter and the stellar distribution decreases from order $30^\circ$ to around $10^\circ$ for group and cluster haloes, and that the degree of alignment does not depend on redshift but is slightly higher for red galaxies compared to blue ones of similar mass. Interestingly, the misalignment of the minor axes of blue central galaxies and their haloes of $\sim40^\circ$ coincides roughly with the typical misalignment between a halo\rq s angular momentum and its minor axis \citep[e.g.][]{KBY+07}.

This suggests that the level of alignment between galaxy and halo angular momenta is the key quantity applicable to disc galaxies. The typical misalignment between the angular momentum directions of the gaseous and dark matter components in simulations is $\sim30^\circ$ \citep[e.g.][]{BAC+02,BEF+10}, in broad agreement with the aforementioned observations, where the degree of misalignment grows as a function of halo radius \citep{BKG+05,DMF+11}. Gas appears to be the main agent driving the misalignment of star-forming stellar discs, while the stronger alignment of gas-poor discs with their host halo is helped by minor merger events, which perturb the spin direction towards the most stable configuration parallel to the halo minor axis \citep{DBR+15}.

We caution at this point that, while hydrodynamic simulations are crucial to gain insight into the highly non-linear processes of small-scale alignment generation or destruction, they can currently only serve as exploratory tools as they still fail to reproduce fundamental galaxy properties such as colour, size, or stellar mass function, owing to the uncertainty in modelling small-scale physics below the resolution of the simulations \citep[see e.g. the discussion in][]{schaye15,crain15}. 
For instance, \citet{VCS+15}  found  typical variations of about $10^\circ$ (and up to $20^\circ$) in the alignment angles of galaxies with their host haloes for scenarios based on different feedback processes occurring in galaxy formation in the EAGLE simulations \citep{schaye15}.

\subsection{Alignment of satellite galaxy shapes}

The alignment of \textit{satellite shapes}, on the other hand, has a more complicated history (see \Cref{sec:hist_recent_obs}), 
primarily because of the difficulty of measuring the shapes of small satellite galaxies which are 
additionally more significantly contaminated by light from neighbouring galaxies, particularly in the dense environments of groups and clusters. Early SDSS studies (which had orders of magnitude more 
galaxies, and much better data quality, than previous measurements) suggested that satellites in 
clusters aligned radially towards their central galaxies \citep{PK05,FLM+07}, but \cite{HKF+11} 
demonstrated that this was due to systematic effects arising from the contamination of shape 
measurements by light from neighbouring galaxies. Note that shape measurement methods differ in their sensitivity to the outskirts of the light profile of a galaxy, and therefore in their susceptibility to this contamination \citep{SCF+13}, but note that there could also be physical differences in alignment strength between the inner and outer parts of a galaxy (see also the discussion in \citealp{Paper3}).

Subsequently, the majority of studies have 
found that satellite galaxies in groups and clusters are consistent with being randomly oriented 
\citep{HE12,SCF+13,CMS+14,SHC+15}. An exception is \citet{SMM14} who detected a non-zero density-shape correlation function, as given in \Cref{eq:angle_corr}, for small $r_p$ when restricting the sample with shape measurement to galaxies classified as satellites. Note, however, that their underlying sample consisted of bright early-type galaxies, so that these satellites are likely to be similarly bright as the central galaxy and only very sparsely populate haloes. This signal could indicate a galaxy type dependence of satellite alignments or an increase of alignment strength with the luminosity of the satellite.

\begin{figure}[t]
\centering
\includegraphics[scale=.25]{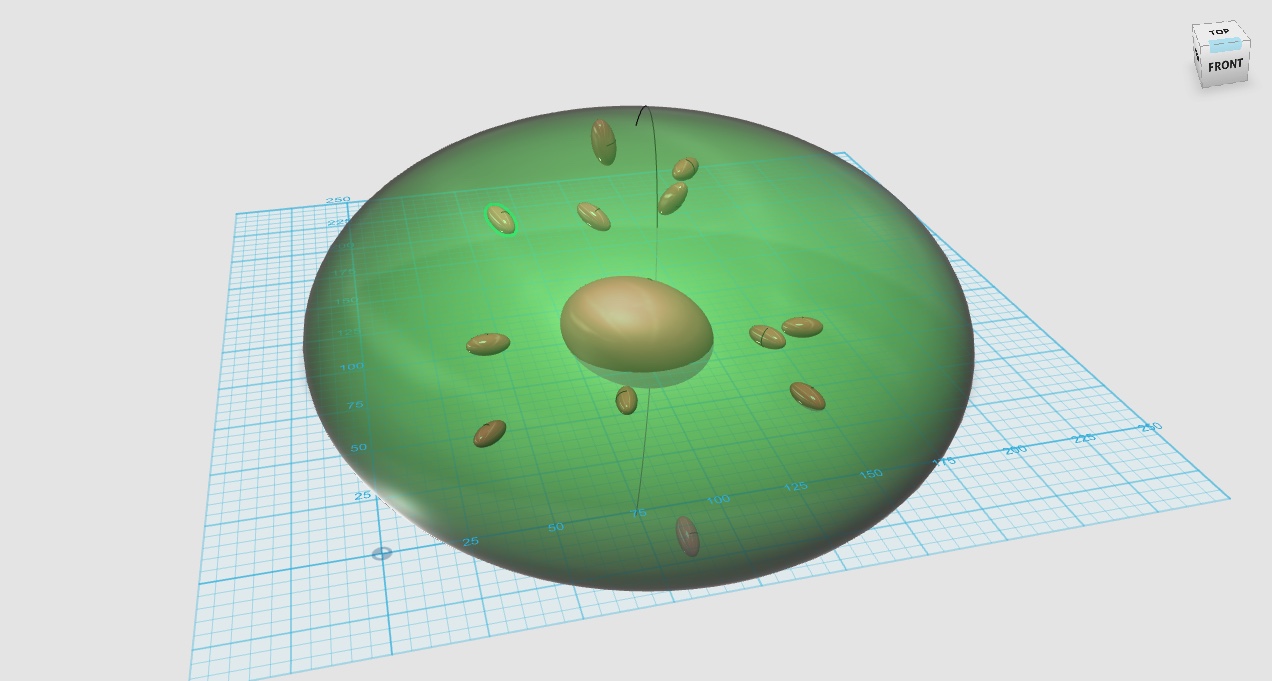}
\caption{Sketch of galaxy alignments in galaxy groups and clusters. Satellite galaxies (small red ellipsoids) are depicted as preferentially pointing their major axes towards the centre of the cluster whose shape and orientation is indicated using a green ellipsoid. The large red ellipsoid represents the central galaxy of the cluster, which generally does not have the same ellipticity and orientation as its host halo.}
\label{fig:sketch_cluster}
\end{figure} 

$N$-body simulations typically use sub-haloes as a proxy for satellite galaxies. There is agreement in these simulations that there is a strong preferential radial alignment of satellites towards the centre of mass of the host halo \citep[e.g.][]{KDM07,FJL+08,KYK+08,KDP+08,PBG08}.  \citet{FJL+08} investigated this trend over a wide range of scales and found that this alignment is strongest within the virial radius of the host halo and drops off rapidly with increasing radius. Intriguingly, the strength of this alignment is inconsistent with the null signals seen in observations, and a possible explanation is the tidal torque origin of this alignment \citep[see e.g.][]{KDP+08}: since the sub-haloes are subject to strong tidal torquing, the loosely bound outer particles will be more highly distorted than the inner particles. Consequently, the shape of the outer sub-halo would be more distorted than the inner region and the centre of the sub-halo (where the luminous satellite would reside) may have a weaker alignment with the centre of mass of the host \citep[see also][]{PB10}. Therefore it is desirable to use hydrodynamic simulations \citep[see e.g.][]{KLK+10} to account for such an effect, as well as the impact of baryons, and thus make predictions for quantities that are as close as possible to the observed ones. \citet{TMD+14b} employed such a hydrodynamic simulation in a cosmological volume to measure the correlation functions of the position angles and projected shapes of the star particles in simulated galaxies. As these statistics are sensitive to satellite alignments on small scales, the authors could confirm a preferentially radial alignment whose strength increases with redshift and the mass of the host halo. Similar trends were previously seen for the same type of measurement in pure $N$-body simulations \citep{LSP+08}.

A tool that has proven successful in describing the clustering of galaxies and that has recently been extended to also model galaxy alignment statistics \citep{SB10} is the halo model. A sketch of the halo model view of the intra-halo galaxy alignment is shown in \Cref{fig:sketch_cluster}. Small red ellipsoids represent satellite galaxies whereas the large green ellipsoid represents the shape of the dark matter halo whose extension encompasses the whole figure. Note that satellites are depicted as being preferentially radially oriented, where the strength of this orientation can potentially depend on halo mass and galaxy position within the halo. The halo model postulates that each galaxy resides in a dark matter halo (usually assumed to be spherical, at least effectively after averaging over a large number of haloes) whose mass is the main physical driver behind observable galaxy properties. Exploiting the fact that properties of haloes can be read directly from numerical simulations and that they can be presented in compact form via analytical fitting functions, the halo model allows for the 
computation of alignment statistics by assuming simple analytical parametric functions for the number of galaxies residing in a halo and for the dependence of their position and shape on the host halo mass.

\citet{SB10} proposed a single new parameter for the intra-halo alignment term which governs the strength of an assumed radial alignment of satellite galaxies, encoding a combination of average satellite
ellipticity and stochastic misalignment. The simplicity of this model has led various observational studies to phrase their findings in terms of the same parameters \citep{SCF+13,SHC+15,SMM14}. Arguably, the greatest advantage of using the halo model is the simplicity of the computation together with the fact that, if needed, most of its assumptions can be refined in view of new results, both observational and theoretical, which make it the current framework of choice for modelling the small-scale, highly non-linear parts of alignment signals. However, a halo model of alignments is likely to fail on mildly non-linear scales outside the largest haloes \citep[see][]{Paper2} where alignment processes unrelated to haloes, e.g. with the filamentary structure of the cosmic web, are relevant. These effects will be discussed in the following section.

\section{Inter-halo and large-scale structure alignments}
\label{sec:lss}

\begin{table}[t]
\begin{center}
\begin{tabular}{rccc}
\hline\hline
structure & $\lambda_1$ & $\lambda_2$ & $\lambda_3$ \\
\hline
clusters/knots & + & + & + \\
filaments         & - & + & + \\ 
sheets/walls    &  - & - & + \\
voids                 & - & - & - \\
\hline
\end{tabular}
\caption{Large-scale structure categories defined by the gravitational tidal field tensor, using the eigenvalues $\lambda_1<\lambda_2<\lambda_3$. Plus (minus) signs correspond to positive (negative) eigenvalues.}
\end{center}
\label{tab:lsscategories}
\end{table}

\begin{figure}[t]
\centering
\includegraphics[scale=.55]{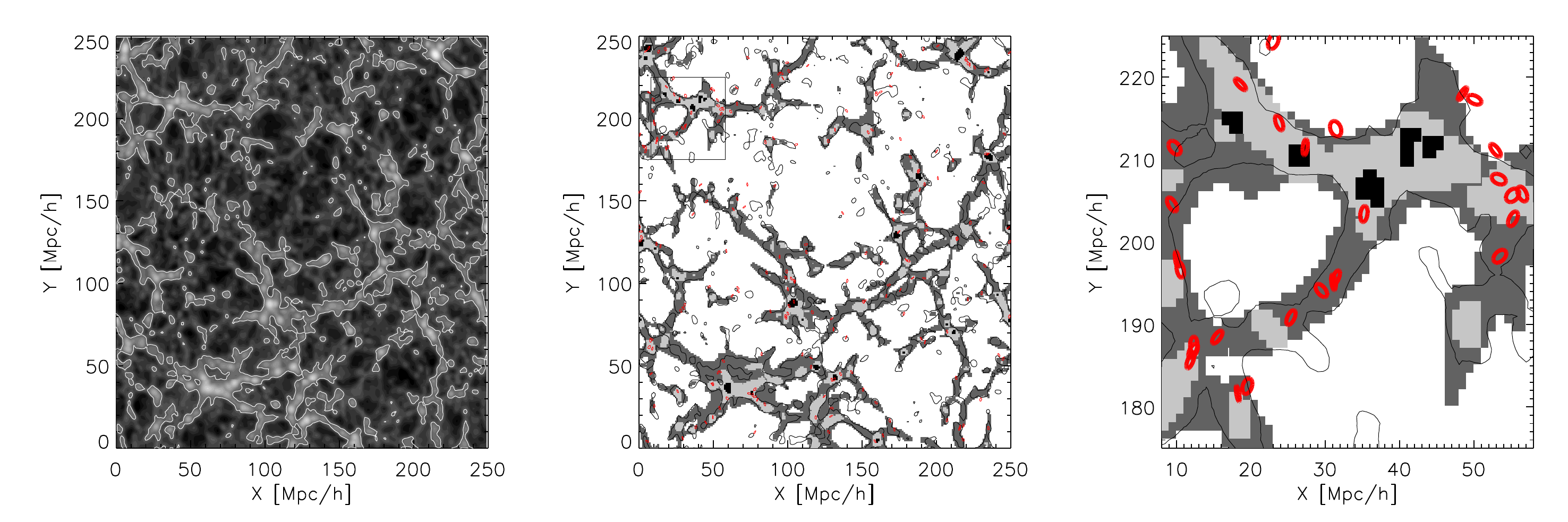}
\caption{Slice of thickness $\sim 1\,{\rm Mpc}/h$ from the Bolshoi simulation. \textit{Left}: Dark matter density, with brightness increasing from low to high density. Contours separate regions above and below the mean density. \textit{Centre}: Cosmic web classification including clusters (black), filaments (dark grey), walls (light grey), and voids (white). Massive haloes are shown in red. Contour lines are the same as in the left panel. \textit{Right}: Zoom into the region marked in the centre panel. Clusters are marked as red ellipses which indicate the eigendirection and -values of the inertia tensor. Note that the sizes of the ellipses are \emph{not} representative of the halo sizes. \permmn{LHF+13}}
\label{fig:libeskind13_cosmicweb}
\end{figure}

While the previous section dealt with alignments of the objects within haloes hosting galaxies, groups or clusters, we now proceed to discuss alignments between clusters as a whole, as well as alignments of galaxies with the defining elements of the cosmic web, such as filaments and the surfaces of voids. The gravitational tidal field tensor is widely used to classify the large-scale matter distribution according to its environmental characteristics. It determines the local deformation of a group of test particles in the gravitational field generated by the matter distribution as can be shown by Taylor-expanding the equation of motion of the test particles; see \Cref{eq:zeldovich:taylor}. The form of the tidal tensor at any given point in space describes in which directions gravitational forces contract (positive eigenvalues) or expand (negative eigenvalues) a distribution of objects. For eigenvalues $\lambda_1<\lambda_2<\lambda_3$, one can thus define the four categories \citep[e.g.][]{HCP+07} given in \Cref{tab:lsscategories}. For filaments, the eigenvector corresponding to $\lambda_1$ specifies the direction of expansion, i.e. it points along the direction of the filament (see also \Cref{fig:codis14-tidalfield}). For sheet-like structures, the eigenvector corresponding to $\lambda_3$ is the local normal vector to the plane of the sheet. Clusters, or knots of the cosmic web, can be assigned an ellipsoid via the eigenvalues and eigendirections of the tidal tensors, and these ellipsoids tend to be prolate \citep[e.g.][]{BEF+07}, while voids are expected to be closer to spherical (see the argument in \citealp{SW04}). \Cref{fig:libeskind13_cosmicweb} shows an example classification of the cosmic web in an $N$-body simulation.

Observationally, these structures are difficult to identify. Candidates for galaxy clusters can be obtained from apparent overdensities in the galaxy distribution and then be confirmed if redshift information is available to place the cluster member galaxies at the same distance, or if the candidates cluster in a region of a colour-magnitude diagram known as the red sequence, where the typically early-type cluster member galaxies are expected to lie \citep[e.g.][]{RRB+14}. \citet{EB09} provided observational evidence, using the stacked weak lensing signal of galaxy clusters compiled from SDSS data, that the distribution of cluster members is indeed a good tracer of both ellipticity and orientation of the underlying dark matter halo. Besides, the hot intra-cluster gas leaves clear signatures in X-ray maps and in observations of the cosmic microwave background whose photons are inverse Compton-scattered by the gas (the Sunyaev-Zeldovich effect; see \citealp{Planck15}). While such observations are valuable for cluster detection, they have not been used widely for studies of cluster shapes and alignments \citep[see e.g.]{CMM00,CMM02}. Filaments and, to a lesser degree, sheets are traced by the galaxy distribution and can thus be identified from spectroscopic galaxy surveys, which have a dense sample of galaxies with accurate redshifts and cover a large volume \citep[e.g.][]{ARD+14}. Using the same datasets, voids have often been approximated as the largest spherical regions devoid of galaxies above a certain brightness threshold \citep[e.g.][]{TCP06}. More sophisticated methods for void identification and characterisation now exist \citep[e.g.][]{SLW+14}, but have not been used in the context of galaxy alignment studies yet.

\subsection{Alignments between galaxy clusters}

\begin{figure}[t]
\centering
\includegraphics[scale=.35,angle=270]{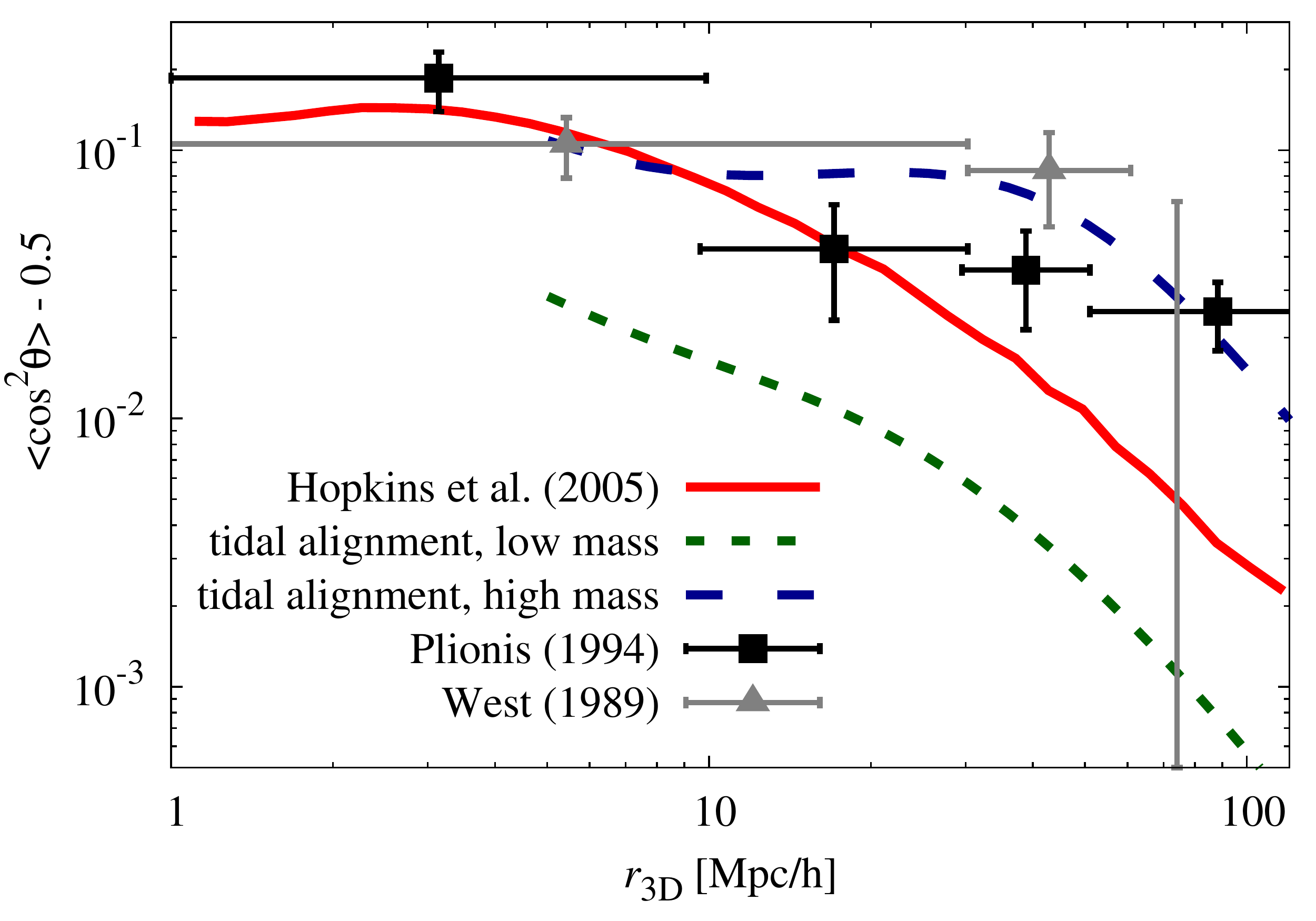}
\caption{Mean of $\cos^2 \theta$, where $\theta$ is the angle between the major axis of a galaxy cluster in projection onto the sky and the line connecting it to another cluster, as a function of the three-dimensional galaxy cluster pair separation $r_{\rm 3D}$. We show the excess signal beyond the expectation of 0.5 for random orientations. Data from \citet{West89} (groups/poor clusters) and \citet{Plionis94} (rich clusters) are shown as grey triangles and black squares, respectively (note the rightmost grey point is negative), while the simulation results from \citet{Hopkins05} are given by the red solid line. The blue dashed and green dotted curves correspond to tidal alignment predictions at $z \sim 0.25$ (mean redshift of the simulation signal) based on \Cref{eq:angle_corr}, assuming a mean ellipticity of 0.22 (after conversion to the definition of ellipticity used in this work) read off from Fig. 3 of \citet{Hopkins05} and a linear cluster bias from the halo model prediction \citep{ST99,MGW+13}. We have extrapolated the luminosity scaling of the alignment amplitude using the best-fit model of \citet{JMA+11} and the relation between cluster masses and integrated luminosities from \citet{WHL12}. The blue dashed line results for a cluster of $10^{15}\,M_\odot/h$, adequate for the Plionis sample, while the green dotted line corresponds to a group/cluster of $5 \times 10^{13}\,M_\odot/h$, the typical mass of the haloes considered in the simulation. The analytical predictions cannot be trusted below $\sim 5\,{\rm Mpc}/h$ as the assumption of linear biasing breaks down. \textit{Adapted from} \citet{Hopkins05}.}
\label{fig:cluster_align}
\end{figure}

The relative ease with which massive galaxy clusters could be identified has led to a comparatively early advent of observational data on cluster alignments; see \Cref{sec:hist_recent_obs}. Since then there has been general consensus that the shapes of galaxy clusters are aligned with each other, and even more strongly with the large-scale matter distribution as e.g. traced by neighbouring clusters. \citet{Hopkins05} used $N$-body simulations with a box size of $1.5\,{\rm Gpc}/h$ to study the alignments between the position angles of pairs of haloes, and between the position angle of one halo and the line connecting to another halo, with haloes of mass $2 \times 10^{13}M_\odot/h$ and above. They established that these alignments are stronger and extend to larger pair separations if the haloes have higher mass, and if they are at higher redshift. By also considering the alignment signals shortly after the formation of haloes, \citet{Hopkins05} deduced the following picture: haloes are endowed with alignments at the time of their formation, with little dependence on formation time and initial mass. They grow via hierarchical structure formation, where major merger events lead to a period of rapid decrease in coherent long-range alignment over a redshift range suggested to be 0.5 to 1. Outside these periods the alignment strengths decrease only slightly. Since low-mass haloes tend to undergo this process earlier, they reached a level of weak alignments in the past, whereas rich galaxy clusters have acquired their mass only recently and thus still have stronger alignments. \citet{KE05} obtained very similar results with another large $N$-body simulation and using the same statistics, while \citet{LSP+08} confirmed the scalings with redshift and mass also for correlations that involve the full halo ellipticity (as opposed to just orientation), using weak lensing-like statistics.

\citet{RP07} additionally considered in their simulations the
dependence of the alignment signal on the amount of substructure
within haloes, which they interpreted as a sign of recent merging
activity. Haloes without substructure displayed significantly stronger
alignments than those with substructure, which may support the
scenario drawn up by \citet{Hopkins05} that structure growth tends to
destroy alignments. However, since subhaloes preferentially fall into
the cluster along the filaments, they are expected to introduce a
memory of the surrounding large-scale matter distribution into the
halo, so the loss of alignment could be a transient state shortly
after a merger \citep{RP07}. However, this picture seems to contradict observational results from large low-redshift samples of clusters, which show stronger alignments with neighbouring cluster positions and orientations if they have got more substructure \citep{PB02}. An important caveat is that the mass resolution of the simulation, the exact definition of a halo and the particles that are assigned to it, as well as the radius out to which particles are included (as the inner and outer part of haloes can be misaligned) all have a strong impact on the strength of alignment correlations \citep{BEF+07,SFC12}.

Recent observations, all using SDSS data but different compilations of galaxy clusters and groups, also found significant detections of alignments, in good qualitative agreement with the theory, as did the earlier works by \citet{West89} and \citet{Plionis94} albeit with large error bars; see \Cref{fig:cluster_align}. Alignments between clusters and surrounding cluster and galaxy positions were seen out to $20\,{\rm Mpc}/h$ by \citet{PSM+11} and out to $100\,{\rm Mpc}/h$ by \citet{SMB+12}, while the latter work also made a tentative detection of pairwise correlations between cluster orientations. \citet{WPY+09} measured various alignment statistics down to group scales with masses as low as $10^{11.5}M_\odot/h$, including an alignment of the shape of the distribution of cluster members with nearest neighbour direction. They, and \citet{PSM+11}, confirmed that the alignment signal increases with object mass, while none of the studies had a sufficient redshift baseline to test for a redshift dependence.

\citet{PSM+11} even obtained excellent quantitative agreement with their own simulations that contained the angular and radial selection functions of the data and were processed in the same way as the observations. \citet{SMB+12}, in contrast, found that the simulations by \citet{Hopkins05} over-predict their signals, even when incorporating observational complications, for instance due to photometric redshift errors and scatter due to the fact that they could typically use only 5-6 galaxies per cluster for estimating the position angle. \Cref{fig:cluster_align} may also point in that direction since the typical halo mass in the simulation sample was about an order of magnitude lower than the clusters in the Plionis sample, and yet a similar alignment amplitude was found. The figure also indicates that the prediction by the linear alignment model for this statistic can in principle reproduce the correct order of magnitude of observations and simulations, although the model prefers a strong scaling with mass/luminosity. It will be interesting to reconsider cluster alignments and analytic models thereof in order to test whether the haloes of clusters are subject to the same alignment mechanism as elliptical galaxies and their lower-mass haloes for which we know the models to work well.

\subsection{Alignments with filaments}
\label{sec:filaments}

\begin{figure}[t]
\centering
\includegraphics[scale=.25]{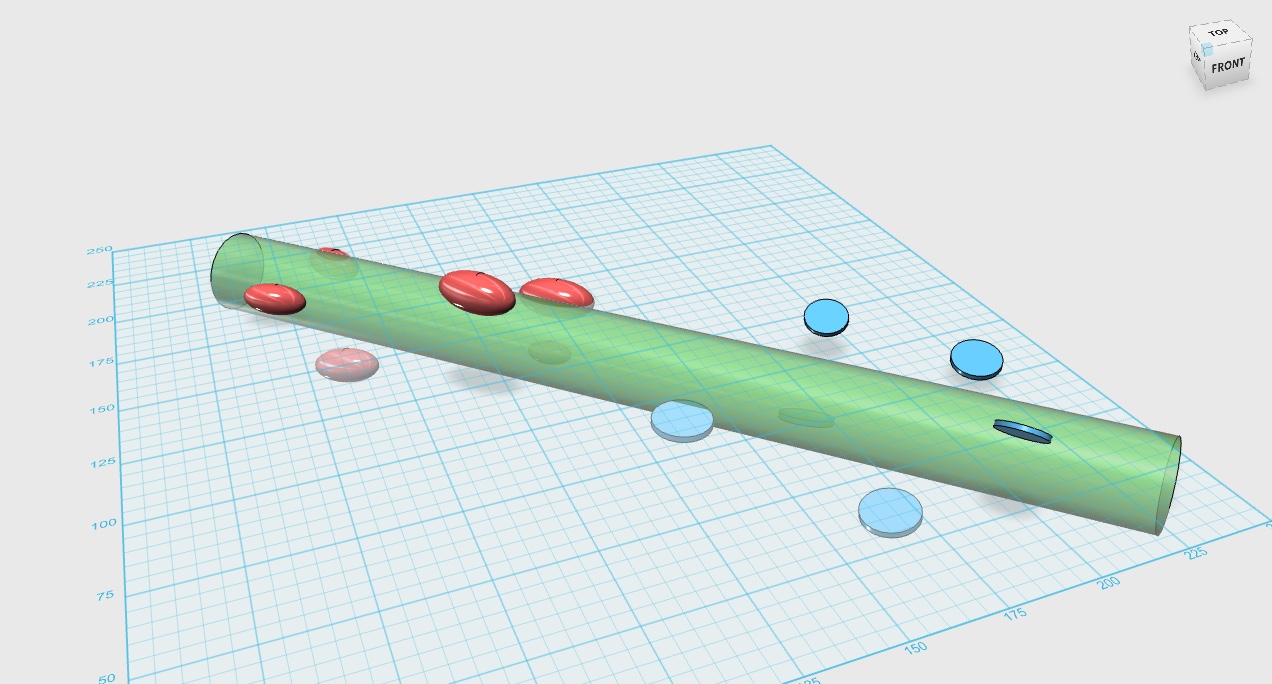}
\caption{Sketch of galaxy alignments with a filament (shown in green). Elliptical galaxies (red ellipsoids) tend to align their major axes with the filament direction, while disc galaxies (blue discs) tend to align their spin perpendicular to the filament direction. Note that the latter alignment trend only holds for massive objects; see \Cref{sec:filaments} for details.}
\label{fig:sketch_filament}
\end{figure}

The growing body of literature using $N$-body simulations to study the alignments of dark matter halo shapes and spins with the filamentary structures of the cosmic web points towards an at least qualitatively consistent picture, which is sketched in \Cref{fig:sketch_filament}, assuming that disc galaxy alignments are driven by spin, and that elliptical galaxy alignments are well described in terms of the major axis orientation. Galaxy-size haloes tend to have their angular momenta aligned parallel to filaments \citep[e.g.][]{HPC+07, ZYF+09, LHF+13}. As we move to larger halo masses, an inversion occurs in the alignment signal with respect to filaments, whereby the halo angular momentum preferentially aligns perpendicular to the filament direction; see e.g. \citet{LHF+13}. These authors also noted that the mass at which the angular momentum alignment flip happens with respect to filaments is dependent on the halo environment, suggesting that the origin of this transition is linked to the interaction of haloes with the cosmic web \citep[see also][]{AY14,CPP15}, while earlier works claimed a connection to the qualitatively different merger histories of the haloes above and below the transition mass \citep{CPD+12}.

There appears to be less consensus in observations, however. \citet{LP02} studied alignments in the IRAS Point Source Catalogue Redshift survey (PSCz), and found that the galaxy spin typically aligned perpendicular to filaments. On the other hand, \citet{TSS13} observed a parallel alignment between the spins of bright spiral galaxies and filaments, while early-type (mostly lenticular) galaxies had their spins aligned perpendicular to the filament direction. \citet{TL13} supported the findings of \cite{TSS13} in relation to spiral galaxies, whereas \citet{ZYW+14} found a perpendicular alignment between the spin axes of spiral galaxies and the filament direction. All recent works used SDSS data but differed in the selection of the galaxy samples and in the definition and reconstruction of filaments. While this is a likely explanation for the apparently discrepant results, physical effects related to the differences in the alignments of low- and high-mass haloes and the dependence of the mass scale of the spin alignment transition on environment may also have an impact. In particular, the findings of \cite{TSS13} of a different alignment signal in spiral galaxies compared to early-type galaxies could reflect the fact that the latter tend to be more massive than their spiral counterparts. And while \cite{ZYW+14} find a perpendicular alignment signal with spiral galaxies, they note that this signal is stronger in cluster environments, perhaps pointing to a lower mass threshold for the spin alignment flip in higher density environments. However, such a scenario would be in conflict with the results of \cite{LHF+13}, who found that the spin flip mass threshold was higher in clusters compared with voids. 

Considering the shapes of dark matter haloes in $N$-body simulations, the halo major axis preferentially lies along the direction of filaments on all mass scales \citep{BS05,ACC06, LSP+08, ZYF+09, LHF+13}. Using observations, \citet{Zhang13} confirmed the prediction from simulations, finding a strong alignment signal for red galaxies and a weaker signal for blue galaxies. This could reflect the mass dependence of the strength of the alignment signal seen in $N$-body simulations, as red galaxies tend to be older and more massive than blue galaxies.

\subsection{Alignments with sheets and void surfaces}

As voids by definition lack sufficiently bright galaxies suitable for shape measurement, observational analyses deal with the alignments of galaxies in loosely defined regions at the surfaces of voids. These environments should in principle be equivalent to the \lq sheet\rq\ category defined above, but the practical algorithms to identify them are substantially different. There have been three recent works \citep{TCP06,SW09,VBT+11} in this area which all employed the same void finder \citep[by][]{PBP+06}, searching for the largest non-overlapping spheres within the survey volume that is devoid of galaxies above a certain brightness threshold. They all worked with SDSS data (from data releases 3, 6, and 7, respectively; \citealp{TCP06} additionally considered data from the 2dF Galaxy Redshift Survey) and defined similar rest-frame magnitude thresholds just below $M_r = -20$. Substantial differences prevailed in the selection of the galaxy samples used for the estimation of the spin direction, and the measurement process for the latter, with \citet{TCP06} limiting themselves to edge-on and face-on disc galaxies only, while \citet{VBT+11} fitted a thick-disc model to all galaxies classified as spirals by GalaxyZoo \citep{LSS+08}.

\begin{figure}[t]
\centering
\includegraphics[width=2.5in,angle=270]{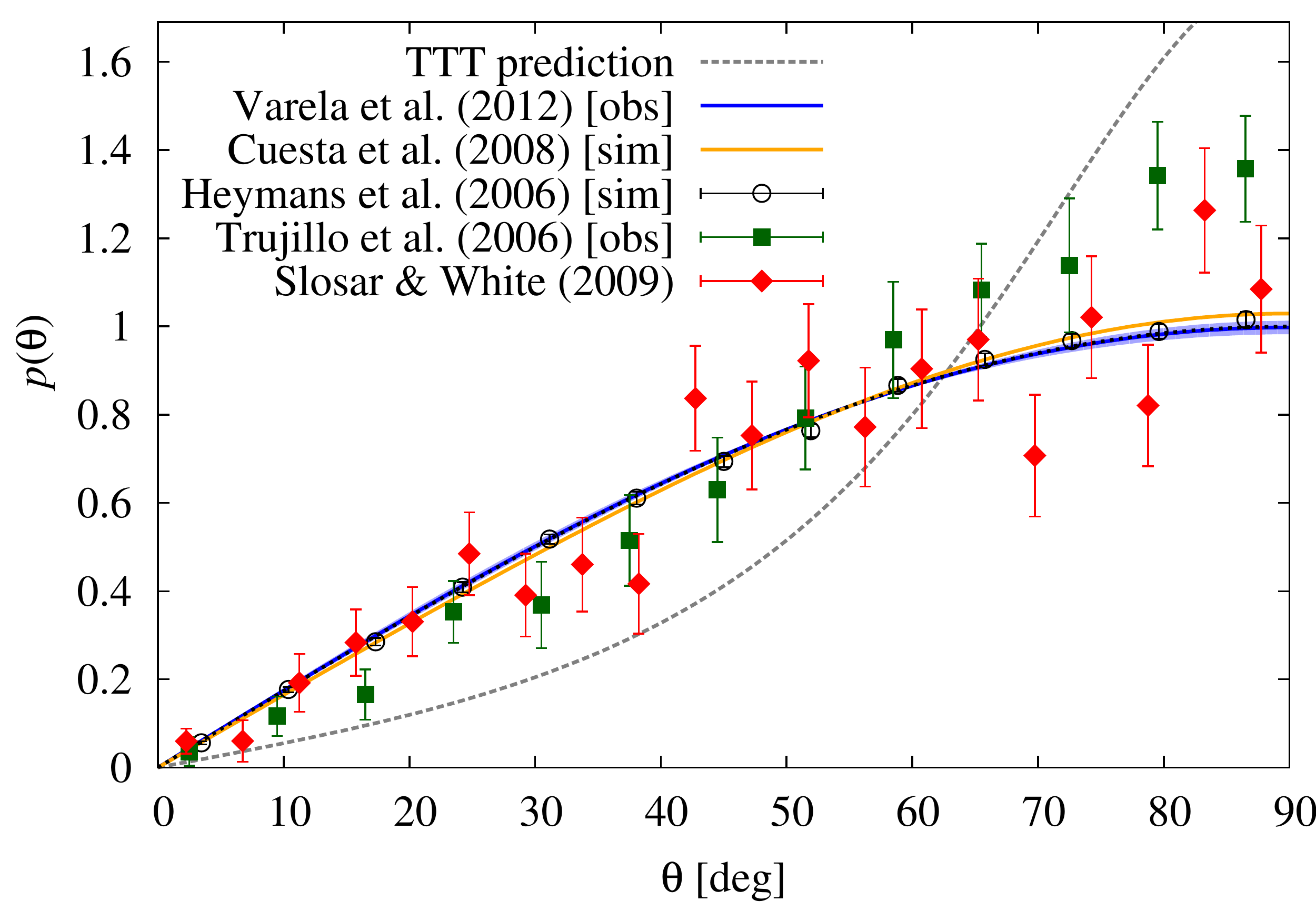}
\caption{Probability density of the angle, $\theta$, between the galaxy spin vector and the vector 
connecting the void centre with the position of the galaxy. For random galaxy orientations one 
expects a $\sin(\theta)$ dependence, as marked by the black dotted line (which almost completely overlaps with the blue line). Note that all distributions integrate to unity over the angular range shown in the figure. Signals and $1\sigma$ constraints are shown for the case of void radii larger than $10\,{\rm Mpc}/h$, and a shell width at the surface of $4\,{\rm Mpc}/h$ thickness, from the observations by \citet{TCP06} (green squares), \citet{SW09} (red diamonds), and \citet{VBT+11} (blue line and error band), and the simulations by \citet{HWH+06} (black circles) and \citet{CBG+08} (yellow line and error band). The grey dashed line corresponds to the prediction by tidal torque theory (TTT) with $a_{\rm T}=0.6$.}
\label{fig:void_alignment}
\end{figure}

Results for the distribution of the angle between the galaxy spin vector and the vector 
connecting the void centre with the position of the galaxy from these three papers are compared in \Cref{fig:void_alignment} for the choice of minimum void radius of $10\,{\rm Mpc}/h$ and a shell $4\,{\rm Mpc}/h$ thick at the void boundary in which the alignments are measured. The data were fitted to \Cref{eq:voidalign} with the free parameter $a_{\rm T}$ which encodes a combination of the effects of the primordial coupling between tidal shear tensor and inertia tensor of the forming galaxy, as well as the degree of non-linear and stochastic effects acting on the galaxy spin after galaxy formation. A value of $a_{\rm T}=0.6$ is expected for the tidal torque theory prediction with independent tidal and inertia tensors, while $a_{\rm T}=0$ corresponds to random spin vectors, and the distribution reduces to be proportional to $\sin \theta$. \citet{TCP06} claimed a strong detection conformal with tidal torque theory, but the significance was reduced to less than $2\sigma$ in a re-analysis by \citet{VBT+11} as the original measurements were selected a posteriori to yield a maximum signal-to-noise. With a more homogeneous dataset and a somewhat more sophisticated analysis, \citet{SW09} and \citet{VBT+11} agreed on a null detection. 

However, \citet{VBT+11} considered a range of void radii and found a significant signal for minimum void radii around $15\,{\rm Mpc}/h$ which corresponds to negative values of $a_{\rm T} \approx -0.5$. This result is in disagreement with the standard picture of tidal torque theory \citep{LP00,LP01,LE07} which posits a preferential alignment of the spin vector with the intermediate principal axis of the tidal shear tensor which lies tangentially to the void surface, whereas \citet{VBT+11} found an alignment of the spin with the radius vector. Simulations which explicitly attempted to reproduce the void alignment signals found either no signal or a very small signal (\Cref{fig:void_alignment}). While \citet{HWH+06} employed a thick-disc galaxy model oriented with some random misalignment around the halo angular momentum vector, \citet{BTP+07} and \citet{CBG+08} did not use any galaxy model but studied angular momentum alignments directly, so that the link to observations is less clear in these cases. In this respect it is interesting to note that the significant detection by \citet{CBG+08} was lost when limiting the angular momentum measurement to the central part of a halo.

\begin{figure}[t]
\centering
\includegraphics[scale=.25]{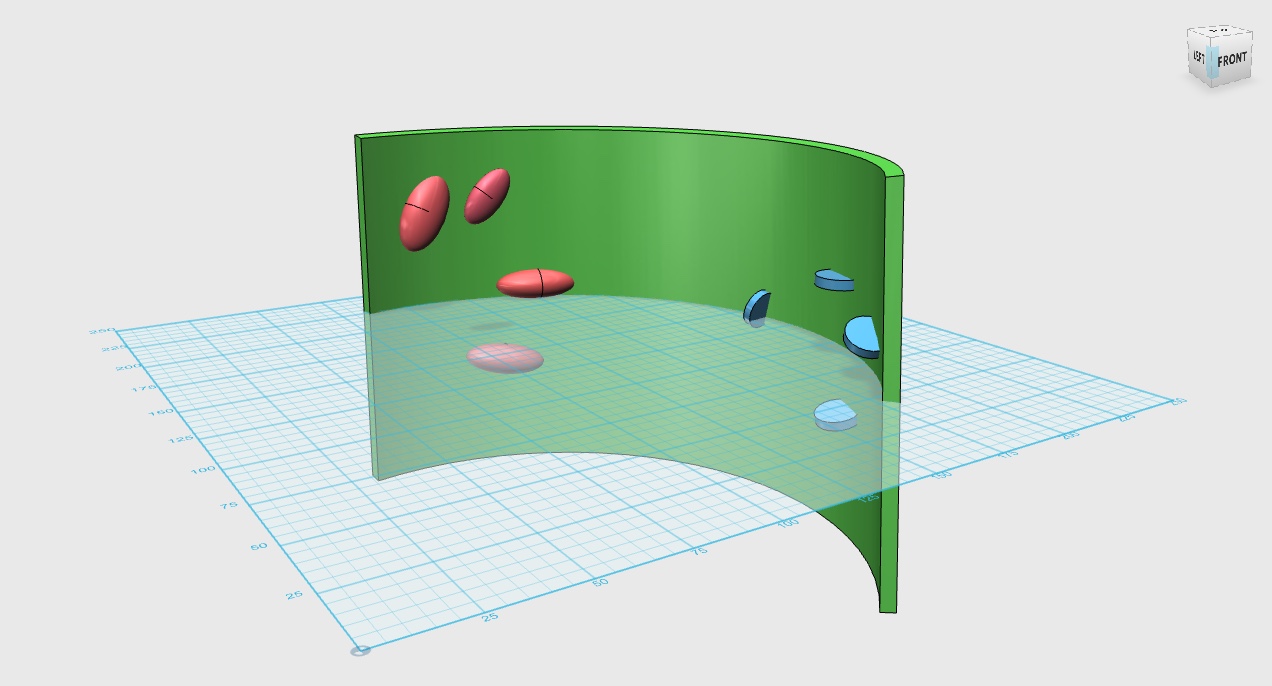}
\caption{Sketch of galaxy alignments at the surface of a void (shown in green). Elliptical galaxies (red ellipsoids) tend to align their major axes perpendicular to the radius vector from the centre of the void, while disc galaxies (blue discs) tend to align their spin along this direction.}
\label{fig:sketch_void}
\end{figure}

The observational picture for the case of alignments of galaxy spins with sheets as defined in \Cref{tab:lsscategories} remains unclear. \citet{LP02} and \citet{LE07} concluded from observations still based on photographic plate data that galaxy spin axes tend to lie within the sheets, while with SDSS, \citet{TL13} and \citet{ZYW+14} found galaxy angular momenta to point perpendicular to the plane of the sheet on average, although in both cases the alignment signal was weak. The latter measurement appears to be consistent with the void result of \citet{VBT+11}, whereas simulations generally agree that angular momenta are preferentially parallel to planar structures \citep[e.g.][]{LHF+13}, i.e.\ ($a_{\rm T}>0$), and this tendency becomes stronger for more massive haloes \citep[e.g.][]{FCP14}. 

We lack observations for the alignments of halo shapes (as opposed to spins) with void structures because the objects we consider as faithful tracers of halo shape, elliptical galaxies and galaxy groups or clusters, are rare in the under-dense regions close to voids. \citet{Zhang13} found that galaxy major axes tend to lie within the sheets. This is in agreement with $N$-body simulations 
for which there is clear consensus that the major axes of haloes lie parallel to the sheet or the surface of the void, with a clear trend towards stronger alignments for more massive haloes \citep[see also references in the latter work]{BTP+07,ZYF+09,CBG+08,FCP14}. All in all, while observational evidence is inconclusive and possibly still affected by selection effects and systematic measurement errors, at least numerical studies agree on a general picture for alignment tendencies of halo spins and shapes with respect to planar structures of the cosmic web that is illustrated in \Cref{fig:sketch_void}, neglecting any misalignments between the observable galaxies and their dark matter counterparts.

\section{Galaxy alignments with the matter density field}
\label{sec:galaxyalign}

Rather than studying galaxy alignments in particular environments, as done in the previous sections, one can alternatively analyse spatial correlations involving galaxy ellipticities or position angles in more generally selected, usually large, galaxy samples, and the matter density contrast as traced by (the same or different) galaxies. While the interpretation of the resulting signals in terms of galaxy evolution processes may be more involved in this case, these measurements are closely linked to the intrinsic alignment contamination of weak lensing statistics and benefit from the rigorous measurement and analysis techniques developed for galaxy clustering and cosmic shear.

In the standard picture of galaxy formation, the large-scale structure interacts with the forming
galaxy through accretion and by exerting tidal torques (\Cref{sec:tidaltheory}) and therefore determines the boundary conditions for the luminous components inside the dark matter host halo. Primarily, there are six models for explaining galaxy alignments, which differ substantially in their physical picture (see \Cref{tab:alignment_models} for a summary):
\begin{itemize}
 \item The simplest model of alignment is the alignment of the stellar component with the dark matter host halo
structure due to identical dynamical properties, for example identical velocity tensors of both
components due to equilibration. These \textbf{halo shape} models are thought likely to apply to elliptical galaxies, and the correlation in shape between neighbouring galaxies would be set by the correlation in host halo shapes in the initial conditions of structure formation.
 \item Tidal alignment models, dubbed \textbf{linear alignments}, assume that elliptical galaxies are
embedded in the gravitational tidal field generated by the ambient cosmic large-scale structure (at least over some period in the galaxy\rq s history) and that the galaxy halo ellipsoid is tidally distorted. The stellar distribution then follows this distortion (see \Cref{sec:tidaltheory}) and reflects the orientation and strength of the tidal field, which can in principle be computed from initial conditions.
 \item In angular momentum alignments, the host halo builds up angular momentum by the tidal shearing mechanism and one assumes that the galactic disc's symmetry axis is aligned with the angular momentum of the host halo. Due to the fact that angular momentum generation is a perturbative effect, these particular alignments are already present in the initial conditions of structure formation. Dubbed \textbf{quadratic alignment} models because of their dependence on the gravitational potential, these angular momentum alignments are thought to govern the shape alignment of spiral galaxies (see \Cref{sec:tidaltheory}).
 \item \textbf{Vorticity alignments} have recently attracted much interest, as strong alignments between
dark matter haloes and the large-scale vorticity field were found in simulations. In contrast to the models discussed above, vorticity generation (and possibly alignment as well) is a purely non-linear effect.
 \item \textbf{Accretion models} stipulate that the shape and orientation of the stellar component is
related to the pattern of accretion of matter onto the galaxy. Like vorticity alignments, accretion alignments
would require an understanding of non-linear structure formation for predicting the accretion pattern
e.g. in terms of its multipolarity.
 \item Closely related to the preceding case are \textbf{merging} models which consider the infall of galaxies rather than of \lq raw\rq\ gas and dark matter. Assuming that angular momentum defines the orientation of the system, as applicable to spiral galaxies, the angular momentum of the system results from the conversion of the orbital angular momenta of progenitor galaxies or haloes in merging processes. Both the accretion and merging model can effectively show a linear dependence on the the surrounding tidal field, but their description might require higher than second order derivatives of the gravitational potential.
\end{itemize}

While the alignments listed above illustrate the variety of physical process, much of their phenomenology may be captured by effective alignment models, which in the case of weak alignments, are generally linear. This argument would apply to the halo shape, accretion, merging and of course the tidal alignment model, because despite the difference in their physical picture, an effective linear and local relation between observed shape and surrounding tidal field could be established.

A common feature of these models, further discussed in \Cref{sec:otheralignment}, is the notion that galaxy alignment is a local process, i.e. alignments of the stellar component with any surrounding field are determined by the interaction of the galaxy with that field. Any neighbouring galaxy contributes to e.g. the gravitational tidal shear experienced by a given galaxy, but commonly one excludes any direct tidal interaction on short scales just involving the gravitational interaction between pairs of galaxies, leading to tidal streams. Instead, one considers coherent alignments of galaxies due to their interaction with the same large-scale gravitational field.

\begin{table}[t]
\begin{center}
\begin{tabular}{llll}
\hline\hline
model & field & mechanism & galaxy type\\
\hline
halo shape & halo inertia & identical velocity tensor of baryons and dark matter & ellipticals \\
linear alignment & tidal shear & tidal field acts on stellar component & ellipticals \\
quadratic alignment & angular momentum & stellar disc reflects angular momentum & spirals\\
vorticity & vorticity direction & halo alignment due to non-linear dynamics & unknown\\
accretion & accretion pattern & accretion pattern determines ellipticity & unknown\\
merging & angular momentum & conversion of orbital momentum to spin & spirals \\
\hline
\end{tabular}
\caption{Comparison of the six major alignment mechanisms, categorised in terms of the aligning field,
the alignment mechanism and galaxy type.}
\label{tab:alignment_models}
\end{center}
\label{table}
\end{table}

Our discussion below is divided into descriptions of generally large-scale alignments for different galaxy
types for which there are different physical mechanisms that are believed to cause the alignments.
Observations of such alignments are critical for
understanding the amplitude of, and mechanism behind, intrinsic
alignments on the scales that will be used for cosmological weak
lensing measurements.  At the same time, these are challenging
measurements to carry out because they require a large contiguous area
(to find many galaxy pairs with large separations) that has high
enough imaging quality that galaxy shapes or spins can be robustly
estimated, as well as either spectroscopic redshifts or high-quality
photometric redshifts. While a large contiguous area is necessary,
the sampling rate (number density of usable galaxies) also cannot be
too low, since that will also increase the noise.  Unlike the types of
observations described in \Cref{sec:history}, the necessary
datasets for this type of observation did not exist with sufficient quality until the
start of this century.

\subsection{General observational studies}

First we describe some general observational studies that sought to observe intrinsic alignments
without making a distinction between galaxy types that might have different underlying alignment
mechanics.

The first  study of intrinsic alignments that had many galaxy pairs on the
scales used for cosmological weak lensing analysis was that of
\cite{BTH+02}.  They  used digitized photographic data (which is subject to a number of technical
difficulties in shape measurement, see \Cref{sec:introduction}) from the
SuperCOSMOS Sky Survey \citep{HMR+01}.  This study took advantage of
the fact that at low redshift, cosmic shear is extremely small, and
hence non-zero galaxy shape correlations should be due to intrinsic
alignments.  Unlike later approaches to measuring large-scale
intrinsic alignments, they used all galaxy shapes without any redshift
information, except the general knowledge that the overall galaxy
redshift distribution was at quite low redshift around $z=0.1$.  Using $2\times
10^6$ galaxies, they found galaxy shape correlations (corresponding to II correlations; see \Cref{sec:tidaltheory}) out to
separations of 100~arcmin, the maximum scale for which measurements
were made. This measurement is still commonly used to normalise the amplitude of alignment signals \citep{BK07}.

 After this finding, \cite{HBH+04} carried out a
re-analysis of the COMBO-17 survey \citep{WDK+01} weak lensing results
\citep{BTB+03}, allowing for the possibility of some II-type intrinsic
alignments contaminating the results for galaxy pairs that are nearby
in redshift, taking advantage of the good photometric redshifts from
COMBO-17 to reliably create samples of galaxy pairs that are close and
distant along the line-of-sight.  The intrinsic alignments were found to
lead to at most a few percent-level effects. Some later
papers used the cross-correlation between spectroscopic samples and
those with potentially much poorer photometric redshifts to constrain
large-scale intrinsic alignments of fainter galaxy samples \citep{BMS+12,CMS+14}, thus far producing only upper limits due to their focus on faint samples. For Chisari et al. the galaxy samples were mixed in type, whereas Blazek et al. considered both a mixed sample, red, and blue separately, finding only upper limits in each case.

\subsection{Early-type galaxies}

Early-type galaxies are thought to align themselves with the large-scale structure through the
linear alignment model \citep{HS04}; see \Cref{sec:tidaltheory}. If there is dynamical equilibration between the dark matter and the stellar components, one can expect that the shape of the luminous component reflects the shape of the dark matter halo. If the halo is now embedded into the large-scale structure, gravitational tidal fields will distort the potential of the halo and the luminous component rearranges itself while maintaining virial equilibrium as long as the gravitational shear is not strong enough to disrupt the halo \citep{CL15}. For sufficiently small tidal shears one can expect a linear relationship between the observed (projected) shape and the strength of the tidal fields. In addition, this linear relationship is the simplest one allowed by symmetry as the tensor of second moments of the stellar brightness distribution is a symmetric tensor of rank two, just like the tidal shear tensor. Typically, there is a single parameter involved which characterises the proportionality between ellipticity and tidal shear. Alignments of this type are correlated on scales on which the tidal shear is correlated, which is identical to the correlation length of the density field. This straightforward picture of alignments for elliptical galaxies would be challenged if the stellar component contains a (velocity) structure on its own or if this could not be directly related to the host halo properties, for instance due to a strong anisotropy of the stellar velocity dispersion. Likewise, incomplete virialisation or remnants from previous merging processes would add further complication.

Observational constraints on intrinsic alignments of red galaxies come from studies starting in
2006 or later.  These studies focused on
samples with both redshifts and shear estimates for each galaxy, and used the galaxy density-shape correlation function $w_{{\rm g}+}$, which is related to the matter density correlation function of \Cref{eq:wgplus} by assuming a linear galaxy bias, $w_{{\rm g}+}=b_{\rm g}\, w_{\delta +}$ (see \citealp{BVS15} for modelling attempts that go beyond the linear galaxy bias assumption). The density-shape correlation function can be measured at higher signal-to-noise for typical
intrinsic alignment models 
than shape-shape correlations (and are a more important contaminant to
cosmic shear via the GI term, as well; see \Cref{sec:impact}).  The SDSS has emerged as the leading survey for these studies,
with highly significant measurements of density-shape alignments out
as far as $100\,{\rm Mpc}/h$ for red galaxy samples, and upper limits for
blue galaxy types.  The analyses have been carried out by several
groups \citep{MHI+06,HMI+07,PSP08,OJ09,JMA+11,Lee11,LJF+13,SMM14}, including
in \cite{JMA+11} the development of a formalism to use high-quality
photometric redshifts, and model out the contamination from
galaxy-galaxy lensing due to incorrectly identified galaxy pairs \citep[see also][]{BMS+12}.  The highest redshift measurements
with SDSS for red galaxies extend to $z\sim 0.6$, by \cite{JMA+11} and \cite{LJF+13}.  These works employed two different statistical measures of the alignments, the former using shapes and
the latter using position angles alone, and they both had a clear detection to tens of megaparsec scales.

\begin{figure*}[t]
\begin{center}
\includegraphics[width=0.45\textwidth]{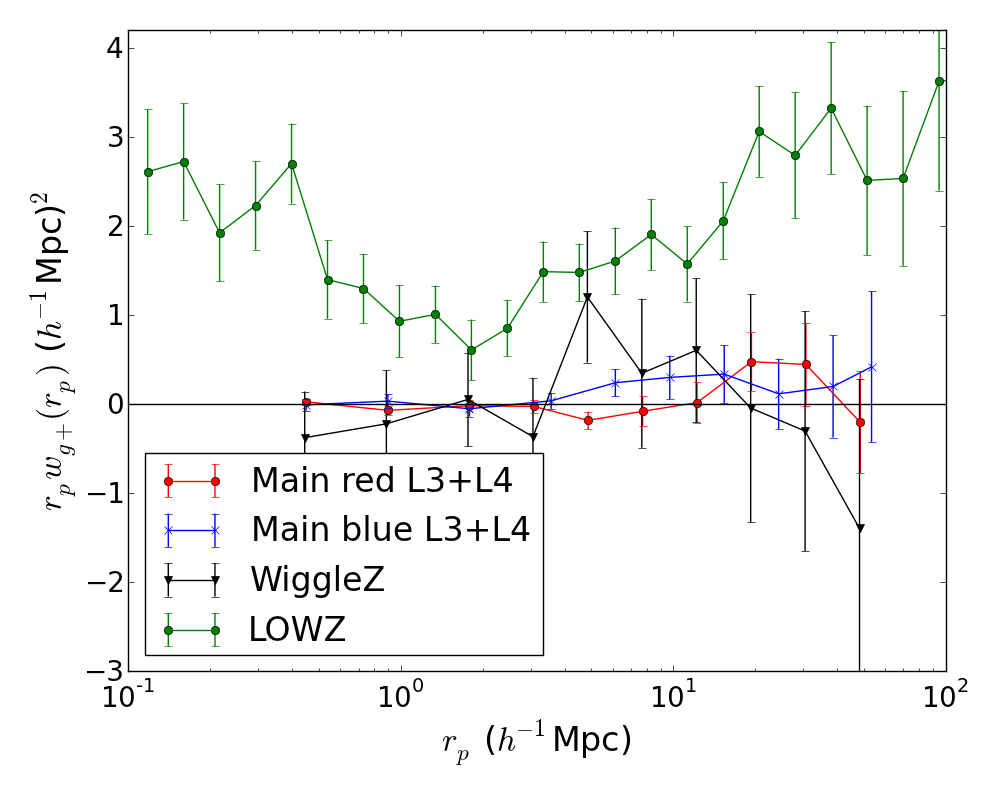}
\includegraphics[width=0.5\textwidth]{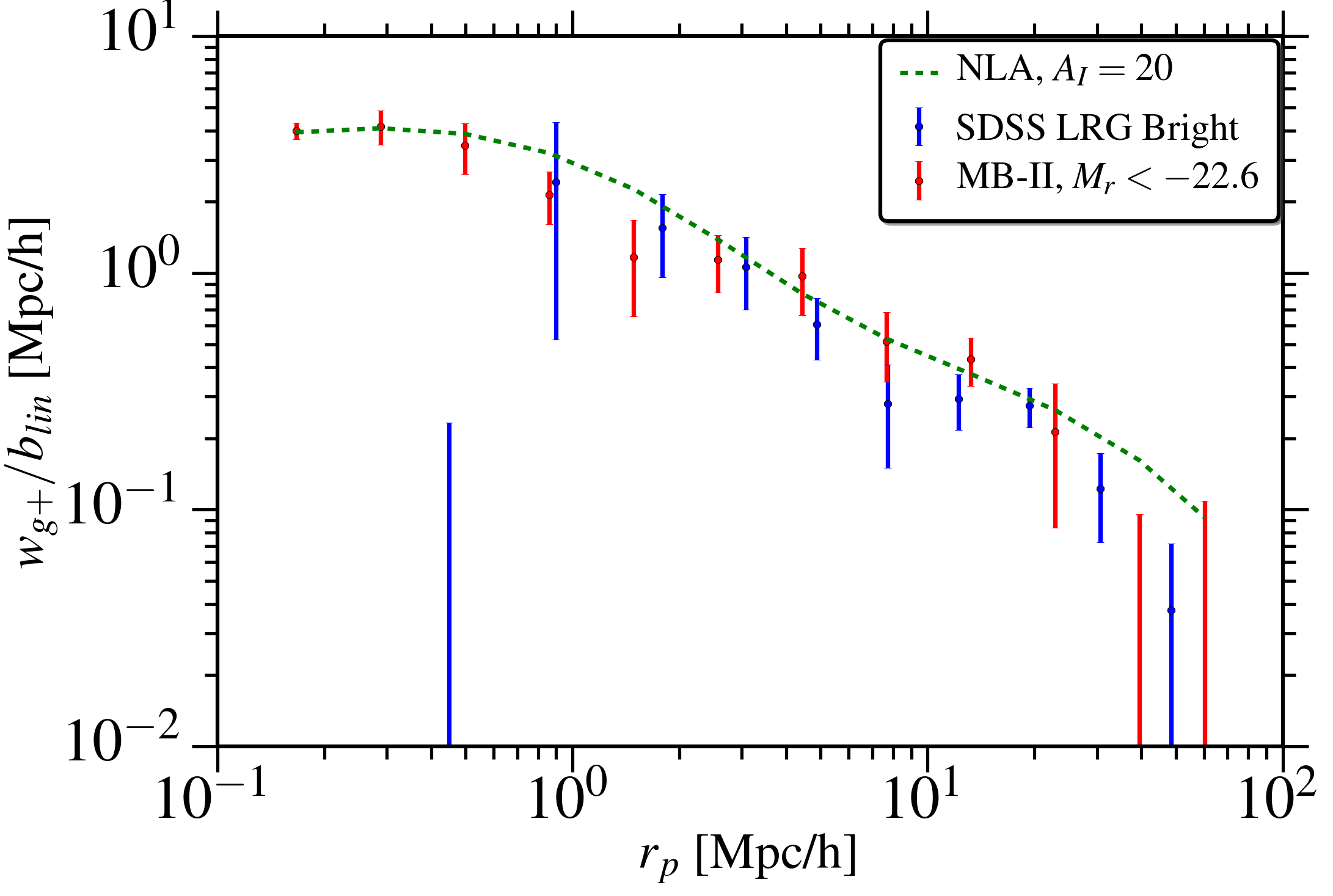}
\caption{{\em Left:} Examples of some large-scale intrinsic alignments
  measurements in the literature, employing a galaxy density-shape correlation function, $w_{{\rm g}+}$, as a function of comoving transverse separation between galaxies, $r_p$.  The samples called ``Main'' refer
  to the SDSS main (flux-limited) spectroscopic sample, divided into
  two subsamples, both at intermediate (Milky Way-type) luminosities.
  The red sample results use the sample from \cite{HMI+07}, but were re-measured
by \cite{JMA+11} using a different colour cut that is more consistent with
ones used by later works.  The WiggleZ results come from \cite{MBB+11}, and
the LOWZ (a low-redshift sample from the SDSS BOSS survey) results come from \cite{SMM14}. {\em Right:} A comparison of the observed density-shape
correlation for LRGs in SDSS, a prediction from the MassiveBlack-II (MB-II) hydrodynamic simulation,
and the non-linear alignment model.  As shown, both hydrodynamic simulations and this simple analytic
model are well able to reproduce the scaling of the observed density-shape correlations with
separation.  The data and predictions have been normalised by the linear galaxy bias, here referred to as $b_\text{lin}$, 
relating the galaxy and matter overdensities, $\delta_\text{g}=b_\text{lin}\delta$. The analytic model labelled \lq NLA\rq\ corresponds to a slightly modified version of \Cref{eq:wgplus}; see also \citet{BK07}. \textit{Right-hand figure
based on data from \citet{TMD+14b}, with credit to Sukhdeep Singh.}}
\label{fig:largescale_ia}
\end{center}
\end{figure*}

An illustration of the measurements that led to the aforementioned
conclusions about large-scale intrinsic alignments of early-type
galaxies is shown in the left panel of \Cref{fig:largescale_ia}.  For early-type galaxies, the amplitude of the intrinsic alignment signal has been found to scale roughly linearly with the galaxy luminosity
\citep{JMA+11,SMM14}, although \citet{BVS15} argued that this may largely be explained by the luminosity dependence of galaxy bias not accounted for in linear perturbation theory. The right panel of \Cref{fig:largescale_ia} shows that both hydrodynamic
simulations \citep{TMD+14b} and analytic models, based on \Cref{eq:wgplus}, are able to describe the scaling of red galaxy intrinsic alignments
with transverse separation.  The curve from hydrodynamic simulations is a prediction, while the
curve from an analytic non-linear extension of the linear alignment model \citep[for details see][]{Paper2} is a fit to a functional form with a free amplitude.

\subsection{Late-type galaxies}
\label{sec:latetype}

The alignment of late-type galaxies follows an equally persuasive physical picture, but there are two competing mechanisms as possible explanations. Late-type spiral galaxies have formed a galactic disc which, depending on the angle of inclination, is perceived to have a certain ellipticity. The inclination of the disc must be determined by its angular momentum, but how the angular momentum is ultimately linked to the surrounding large-scale structure is not as clear. It is worth emphasising that the ellipticity in these models, being an orientation effect, depends only on the angular momentum direction.

A widely employed picture of angular momentum generation is tidal torquing (\Cref{sec:tidaltheory}), where the gravitational field of the large-scale structure exerts a torquing moment on the protohalo prior to collapse. When, at a later stage, a galactic disc is formed through the cooling of the baryonic component followed by star formation, the galactic disc can be expected to follow the host halo angular momentum imprinted by tidal torquing \citep{CNP+01, CNP+2002}. Employing this picture for predicting intrinsic alignments requires a model for the dynamics of dark matter in tidal shearing and the possibly strong assumption that there are no misalignments between the disc symmetry axis and the host halo angular momentum. An environment dependence would be introduced by the typical tidal field strength and the typical orientation of the tidal shear tensor. Based on tidal torquing one can expect relatively short-ranged correlations of not much more than $1\,$Mpc \citep{SM11}. Although tidal torquing has been tested as a source of angular momentum of dark matter haloes \citep{PDH02a, PDH02b}, there are alternatives such as the conversion of orbital angular momentum to spin, as suggested by \citet{CPD+12, DPW+14}, which would yield similar correlation functions, but with a strong environmental dependence. Recently, \citet{CPP15} analytically investigated the alignment of tidal shear and inertia tensors in the vicinity of filaments and walls and calculated the resulting alignment of galaxy spins with these structures, finding good qualitative agreement with observational and simulation results.

Clearly, tidal torquing would be challenged if merging of subhaloes is important, as each merging process adds new angular momentum to the halo. In this model, the angular momentum direction and ultimately the disc orientation would fluctuate strongly between merging events in contrast to the steady orientation in the tidal torquing picture. There might be scenarios, however, where the merging takes place along preferred directions, which would stabilise the angular momentum direction. These effects can be important in sheets and filaments. Whereas in the first case there would be little correlation between haloes, the second case would suggest relatively long-ranged correlations, typically of the size of filaments. Tidal torquing must have a straightforward parameterisation describing the orientation of angular momentum as a function of tidal shear. As merging of subhaloes can be expected to take place along directions defined by the orientation of the tidal shear tensor, there is in fact a similar parameterisation applicable to these cases.

For rotationally-supported galaxies there are clearly two direction vectors of interest, one defined
by the angular momentum and the other defined by the projected galaxy shape.  Observational studies
have considered alignments of both of these vectors.  \cite{LP02}, \cite{LE07}, and \cite{Lee11}
approached this question observationally using the direction of the disc angular momentum for nearby
galaxies.  \cite{LP02} correlated the spin direction of disc galaxies (determined based on axis
ratios and position angles, assuming disc galaxies are infinitely thin with spin axis perpendicular
to the disc) with the local tidal shear field reconstructed from the
IRAS PSCz survey.  They detected a correlation at more than $3\sigma$ confidence using the model of \Cref{eq:lee_spin_align}.  \cite{LE07} later carried out a similar analysis, but using the 2MASS Redshift
Survey to reconstruct the tidal field and $\sim 12$k galaxies from the Tully Galaxy Catalog.  They
again found a highly significant ($6\sigma$) detection of correlation between the spin direction and
the intermediate principal axis of the tidal field, which was found to be stronger in high-density
regions than in low-density regions.  Finally, \cite{Lee11} used data from SDSS to measure the
spatial correlation function of the 3D spin directions of pairs of nearby $(z\le 0.02$) disc
galaxies, finding a $\sim 3\sigma$ detection around separations of $1\,{\rm Mpc}/h$ and no detection for
scales above $3\,{\rm Mpc}/h$.

On the other hand, attempts to measure large-scale intrinsic alignments of the projected shapes of
late-type galaxies in \cite{HMI+07} led only to upper limits on the effect for the SDSS Main galaxy
sample at $z\sim 0.1$.  The later emergence of additional spectroscopic surveys that targeted
somewhat deeper or differently-selected galaxy samples in the SDSS region also proved useful, as it
allowed for additional measurements beyond those enabled by SDSS spectroscopy alone. Examples of
such surveys include the WiggleZ survey \citep{DJB+10}, which was used by \cite{MBB+11} to constrain
the intrinsic alignments of a very blue starburst galaxy population at intermediate redshifts,
$z\sim 0.6$.  Again, only upper limits were placed in this case (see \Cref{fig:largescale_ia}). It
is worth commenting on the possible explanations for the detections of spin correlations for
$\lesssim 1\,{\rm Mpc}/h$ scales for nearby disc galaxies by \cite{Lee11}, versus non-detections of
shape correlations by \cite{HMI+07} and \cite{MBB+11}.  First, it is important to bear in mind that
the large-scale studies of shape correlations have little statistical power on $\lesssim 1\,{\rm Mpc}/h$
scales, making it hard to compare them with the results for spin alignments on those scales.
Second, projection along the line-of-sight separations out to tens of Mpc to make 2D correlation
functions can wash out 3D correlations that are present at scales of $\lesssim 1\,{\rm Mpc}/h$, making
them entirely undetectable.  Finally, it has been shown \citep[e.g.,][]{HWH+06} that putting in
models for disc angular momentum with the local density field can result in quite small projected
shape correlations.   Given these three caveats, it is not clear that there is any discrepancy at
all between the measurements using spin and shape alignments.

\subsection{Alternative alignment mechanisms}
\label{sec:otheralignment}

In contrast to the specific alignment models for early- and late-type galaxies discussed above, there are a number of alternative hypotheses on how dark matter haloes, irrespective of the type of the galaxy they contain, can be aligned with the large-scale structure. It is important to emphasise that these are conceptually new, and use a different aligning large-scale field other than the tidal shear as is the case for the linear and quadratic alignment models.

One such model posits that the stellar component of a galaxy would simply follow the halo shape, as would be expected from virial equilibrium, very much like in the case of elliptical galaxies. The individual galaxies would appear correlated through correlations in the dark matter halo shapes and orientations, which are already present in the initial conditions of structure formation and can in principle be determined as an extension to the random process that seeds haloes into the large-scale structure as peaks in the density field \citep{BBK+86, BKP96, SFC12, Rossi12, Angrick13}. In this picture, halo shapes are related to the curvature of the density field, which would imply short ranged correlations.

Secondly, alignments of dark matter haloes with the local vorticity field have been observed in numerical simulations \citep{LHK+12, LHF+13}. The actual mechanism is, in contrast to tidal torquing, difficult to grasp via perturbation theory or other analytic means because vorticity generation only occurs in non-linear structure formation. If the stellar component follows the orientation of the dark matter halo, one would observe a shape correlation which would be induced by correlations in the vorticity field, which is expected to show correlations only on small scales.

Thirdly, accretion of matter can determine shape and orientation of the stellar component. Numerical simulations of individual (spiral) galaxies have shown the presence of cold gas accretion streams \citep{D+09,KKF+09,SNT+12}, which have a significant impact on the orientation of the disc and are able to tilt the orientation away from the initial one \citep{KDS+11,DDH+12,DDH+14}, up to the point where the angular momentum direction shows a random walk behaviour at successive merging events \citep{DPW+14}. It is difficult to make a statement about shape correlations which would arise in such a model, but it seems reasonable to assume that such correlations would strongly depend on the topology of the region of large-scale structure the corresponding galaxies reside in \citep{PJH+14}. Strong correlations between the flow patterns of accreted matter and the strength and orientation of the local tidal field could serve to explain why tidal alignment and tidal torque models describe alignment signals well in certain regimes \citep{SWM15}.

\section{Impact on cosmology}
\label{sec:impact}

\begin{figure}[t]
\centering
\includegraphics[scale=0.32]{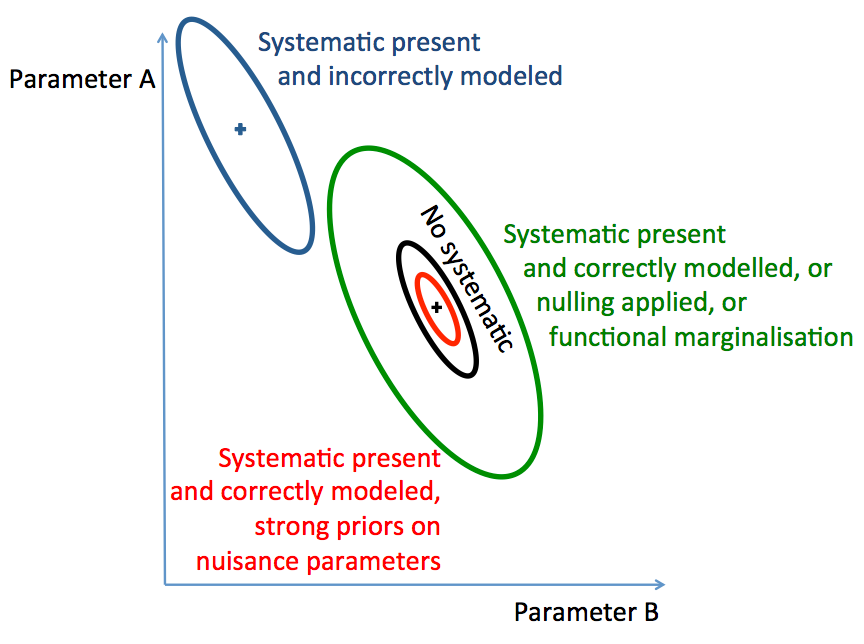}
\caption{Diagram illustrating the influence of systematic effects on cosmological parameter inference. The various ellipses represent 
the confidence regions for two cosmological parameters A and B, marginalised over some large multi-parameter space. In the case that no systematic effects were present, the parameters 
could be measured in an unbiased way (black ellipse). If there is a systematic effect and it is either ignored or incorrectly modelled, 
the constraints are biased by an amount and direction that depends on how incorrect the modelling is (blue ellipse). In the case that the modelling of the systematic effect 
is correct, the parameter errors are increased because of the larger parameter space but the result is unbiased (green ellipse); this would also be the case if nulling was applied, 
or a non-parametric marginalisation. If the extra systematic effects are also dependent on the parameters of interest, a strong prior 
on the additional parameters can even cause the original error bars to be reduced (red ellipse).}
\label{fig:bias_diagram}
\end{figure}

The most widely employed method that uses weak lensing data to infer information on cosmological parameters is to calculate 
the two-point statistics of the observed ellipticity field. As shown in \Cref{sec:intro_WL}, the intrinsic 
alignment of galaxies leads to additional terms being present in the observed correlation function, or power spectrum. In this section we discuss the impact of intrinsic galaxy alignments on cosmological inference and ways to mitigate it. When calculating the likelihood of cosmological parameters given a statistic, a model is required and a loss function is 
typically constructed that involves the observed data $\vek{D}$ and the model of that data $\vek{M}(\vek{\pi})$ which is dependent on some 
parameters of interest $\vek{\pi}$, such as a set of cosmological parameters. In the case that the data is Gaussian distributed this loss function is simply 
\begin{equation} 
\label{like}
-2\ln L (\vek{\pi}) = \sum \bb{\vek{D}-\vek{M}(\vek{\pi})}^\tau\; \vek{C}^{-1} \bb{\vek{D}-\vek{M}(\vek{\pi})}\;,
\end{equation} 
where the sum is over the data points, and $\vek{C}$ is the observed covariance matrix of the data. This is the case in which the 
parameter dependence is in the mean of the data vector. If the model that is used to describe the 
data is complete, in the sense that all physical effects that can occur in the data are captured in the model, then the 
maximum likelihood of the inferred parameters will be unbiased. We
do not have a complete model for galaxy alignments, and in this case the maximum likelihood values of the
inferred cosmological parameters can be biased. In addition, any extra set of parameters 
$\vek{\phi}$ that extend the set of inferred parameters to $\{\vek{\pi}$, $\vek{\phi\}}$ will increase the parameter-space volume and thereby increase the error bars on the parameters of interest, with respect to the case where no additional parameters were required. Such 
additional parameters are referred to as `nuisance parameters'. We show this diagrammatically in \Cref{fig:bias_diagram}. 

If a model is complete, then there is no need 
to re-label an effect; indeed if the extra effect also depends on the parameters of interest then the error bars on those 
parameters can even be reduced with correct modelling. However, because it has been established that the intrinsic alignment model is uncertain, the phenomenon is referred to 
as a systematic effect in weak lensing. The impact of intrinsic alignments on cosmology 
is therefore two-fold: the error bars on cosmological parameters are likely to be increased, and the maximum likelihood of the cosmological parameters may 
be biased. 

It was first shown, within the context of predicting cosmological parameter performance, 
that incorrect modelling can bias cosmological parameter inference in \citet{Kim04} who showed that 
cosmological inference using Type Ia supernovae can be biased when systematic effects are poorly modelled. This 
was re-derived and applied to the case of measurement of the matter power spectrum in \citet{HT05}, and applied to the 
case of weak lensing systematics in \citet{HTB+06} and \citet{Taylor06}; the approach was subsequently re-derived in 
\citet{amara08}. The fact that intrinsic alignment modelling can bias the maximum 
likelihood in cosmological parameter estimation was first measured and inferred empirically in 
\cite{HMI+07}, and first shown analytically and in a predictive sense in \citet{KTH08}, who used a simple two-parameter model from 
\citet{HH03} to show that, in the case that intrinsic alignments were not modelled (and that the \citealp{HH03} model was correct), parameters could be biased by several percent. However such a bias was implicit in the conclusions of several earlier papers 
such as \citet{HRH2000} and \citet{HS04}, who found that intrinsic alignments could suppress the power spectrum by large amounts. 
There have been several elaborations that have used increasingly sophisticated and realistic models, for example \citet{JB10}, \citet{KBS10}, and \citet{KRH+12}. 

\begin{figure}[t]
\centering
\includegraphics[scale=.6]{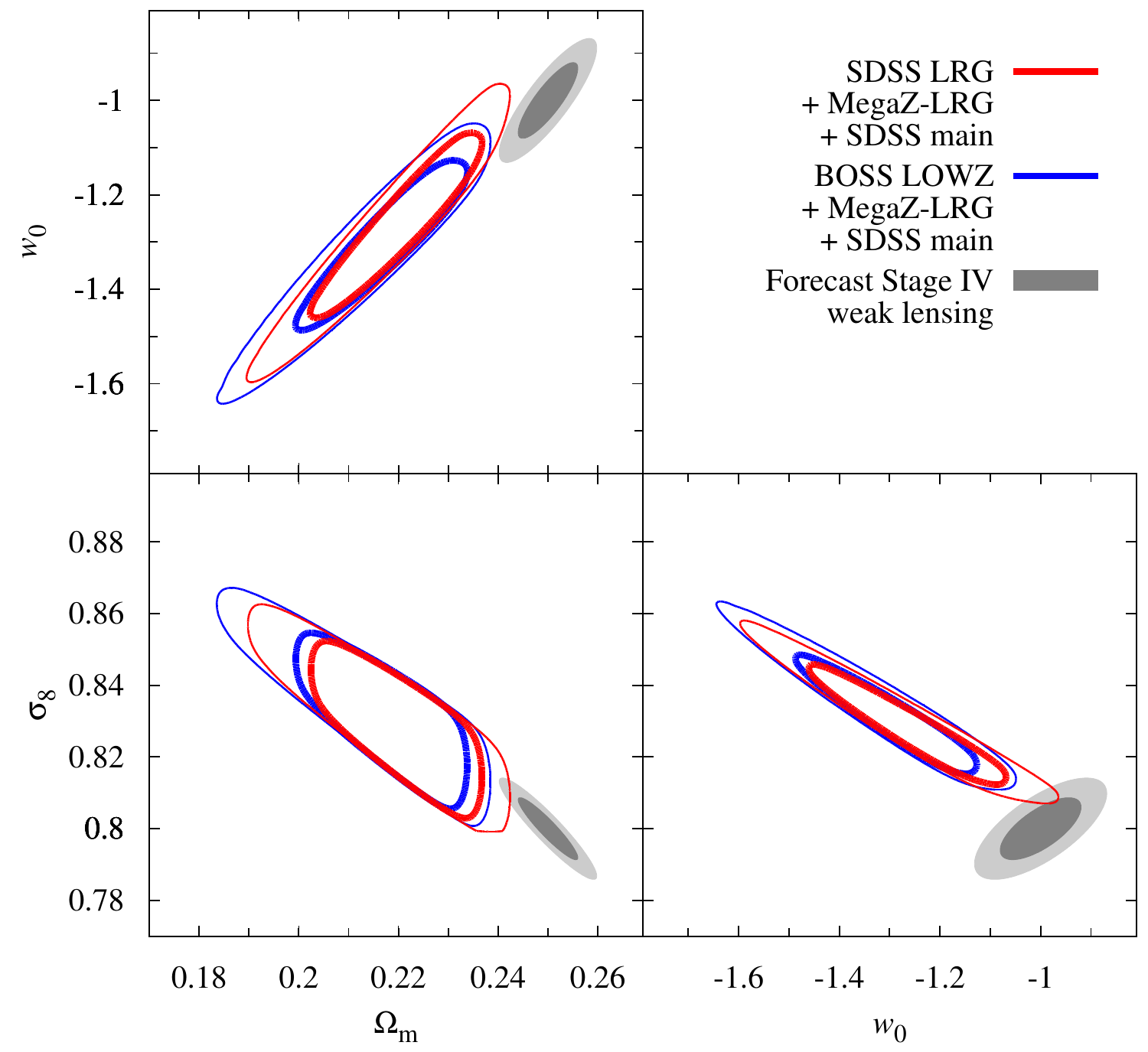}
\caption{Bias on cosmological parameters due to unmitigated intrinsic galaxy alignments for a next-generation (\lq Stage IV\rq, Euclid-like) weak lensing survey. A six-parameter flat $\Lambda$CDM cosmology was considered, with marginal constraints on the matter density parameter, $\Omega_{\rm m}$, the normalisation of density fluctuations, $\sigma_8$, and the dark energy equation of state parameter, $w_0$. Contour lines encompass the regions in which $99\%$ of the possible biases on these parameters are located when we sample from the $1\sigma$ credible region (thick lines) and the $2\sigma$ credible region (thin lines) of the three parameters in the intrinsic alignment model constrained in \citet{JMA+11}. The set of red contours is obtained for the posterior of the combined samples (6 SDSS LRG samples, 2 MegaZ-LRG samples, and 2 SDSS main survey samples) analysed in \citet{JMA+11}, while for the blue set the SDSS LRG samples have been replaced by four subsamples of the BOSS LOWZ measurements presented in \citet{SMM14}, split according to luminosity. The grey regions correspond to Fisher matrix forecasts of $1\sigma$ and $2\sigma$ constraints from a tomographic cosmic shear analysis. For details of the modelling see \citet{JMA+11}. Note in particular that it was optimistically assumed that blue galaxies have zero intrinsic alignments, and that this is known with zero uncertainty.}
\label{fig:iabias}
\end{figure}

\Cref{fig:iabias} shows the ranges of possible biases on key cosmological parameters for a tomographic weak lensing survey design similar to Euclid \citep{LAA+11}, which are derived from a three-parameter intrinsic alignment model (representing the overall amplitude of the correlations, as well as the indices of two power-laws encoding a redshift and luminosity dependence) constrained by a collection of recent observations of early-type galaxy samples in SDSS \citep[see][]{JMA+11}. The extent of the regime of possible biases beyond the credible regions of the weak lensing constraints demonstrates that ignoring intrinsic alignments would lead to significant misestimates of dark matter and dark energy parameters, while the size of the regions of possible biases, which is substantially larger than the corresponding credible regions, indicates that marginalising over this intrinsic alignment model would significantly weaken the cosmological parameter constraints. Note that we have optimistically assumed that blue galaxies do not have intrinsic alignment and, more importantly, that there is zero uncertainty on this statement. In reality, quite the opposite holds true: constraints on the intrinsic alignment signal of disc galaxies at typical redshifts and luminosities for weak lensing surveys are very poor; see \Cref{sec:latetype}. It is likewise optimistic to assume that early-type intrinsic alignments will be exhaustively described by just these three parameters for all relevant redshifts, luminosities, and spatial scales.

It is clear that adding additional parameters to model intrinsic alignments can increase the error bars on cosmological 
parameters. This was first shown empirically by \citet{HMI+07}, and in a predictive capacity by \citet{BK07} who demonstrated that the 
dark energy figure of merit \citep{detf} could be affected by several tens of percent. Subsequent papers cited in the previous paragraph 
showed similar potential increases cosmological parameter error bars, however the exact predictions of each depend on the details of 
modelling assumed. 

\begin{figure}[t]
\centering
\includegraphics[scale=0.5]{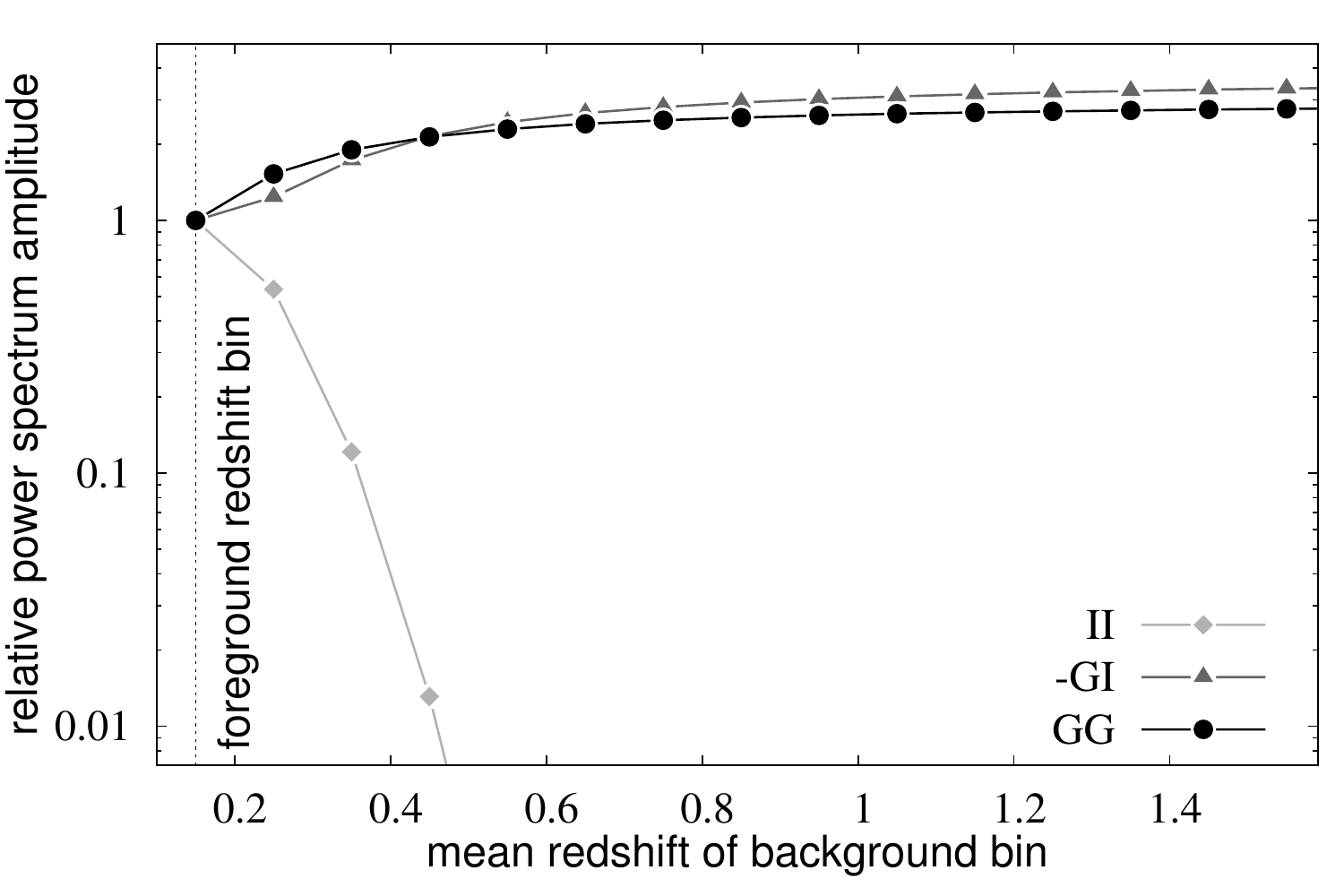}
\caption{Redshift scaling of the lensing and intrinsic alignment power spectra. Shown is the
amplitude of the tomographic GG, II, and the negative of the GI power spectra at $\ell=1000$ as a function of the
mean redshift of the background bin. The amplitude is normalised to the autocorrelation in a bin
centred on $z=0.15$. The photometric redshift scatter was assumed to be Gaussian with a width of
$0.05(1+z)$, without bias or outliers due to catastrophic failures. Note the very similar scalings of the GG and GI signals.}
\label{fig:ps_redshift_dependence}
\end{figure}

The majority of the bias in cosmological parameter inference is expected to be caused by incorrect modelling of the GI correlation. 
The II signal is generated by galaxy pairs which have physically interacted with the same matter structures. Therefore the II correlation is expected 
to quickly decrease in amplitude once galaxies are separated by several megaparsecs, which is readily achieved by cross-correlating galaxy samples 
with disjoint redshift distributions \citep{KS03}. Well-separated galaxy samples will still yield a GG signal as they share the gravitational shear exerted 
by the matter between the observer and the foreground sample. From a more formal perspective, consider the kernels in the line-of-sight projections of the GG, GI, and II signals, \Cref{eq:limberlensing} and \Cref{eq:limberia}. 
Lensing efficiencies, as given by \Cref{eq:weightlensing}, are smooth functions of redshift, non-zero between $z=0$ and the source redshifts, whereas the intrinsic alignment 
signals are local effects with the redshift distribution as a kernel, which can be quite compact if adequate redshift information is available.
This is illustrated in \Cref{fig:ps_redshift_dependence} which shows the redshift scaling of tomographic lensing and intrinsic alignment power spectra, 
keeping the foreground redshift bin fixed. The GG signal displays a slow increase with redshift while the II signal quickly drops with increasing 
separation of the redshift distributions of the galaxy samples correlated. It is also clear from the figure that removing the GI term is 
more challenging: this term contains one lensing efficiency in its kernel and thus has a redshift scaling that is slightly steeper but 
otherwise very similar to the lensing signal.

Instead of modelling the intrinsic alignment signal, the data vector itself can be changed to remove any contaminating signal. In reference to \Cref{like} this is the act of modifying $\vek{D}$, as opposed to $\vek{M}$. This technique is known as `nulling', and takes combinations of data points 
that have small or zero contamination from the intrinsic alignment signal; in practice this amounts to taking linear combinations of the data points in an analogous way to 
principal component analysis. This approach was first suggested for weak lensing, in the context of removing small-scale information, by \cite{HW05}. For intrinsic 
alignments this was revived by \citet{JS08} and \cite{JS09}, relying on accurate and moderately precise redshift estimates to separate the cosmic shear and intrinsic alignment signals via the different scaling with redshift as shown in \Cref{fig:ps_redshift_dependence}. When nulling the intrinsic alignment signal from the data vector, information is necessarily removed 
(in the principal component analysis this is equivalent to ignoring, or removing, these contaminated modes) and as a result the constraints on the parameters of 
interest again have larger error bars, although the maximum likelihood should be unbiased if the remaining part of the data vector can be modelled well; see \Cref{fig:bias_diagram}. 
An analogous null test can be constructed from decomposing the shear field into a divergence-free (B-mode) and a curl-free (E-mode) part. To very good approximation, lensing only generates E-modes \citep{HHW+09}, so that a B-mode correlation signal would indicate the presence of systematic effects. Some intrinsic alignment models predict significant B-modes \citep{GS14} although measurements from mock catalogues suggest these are also negligible \citep{JSH+13} while the E-mode contribution may be significant.

The final approach that can be taken, in reference to \Cref{like}, is to modify the covariance matrix $\vek{C}$, effectively increasing the error bars to account 
for a marginalisation over non-parameter functional behaviour that cannot be modelled. This `path integral' approach was derived by \cite{KT11} (and a similar 
non-parametric approach taken in \citealt{TDK09}) and applied to intrinsic alignments. The authors found that the scale and redshift dependence of the functional form of the intrinsic alignment signal needs to be known 
to better than $10\,\%$ for the cosmological error bars to be unaffected. 

It is possible to include additional data in the likelihood analysis, which, due to a different sensitivity  to weak lensing and intrinsic alignment signals, calibrates the nuisance signal internally. Weak lensing catalogues also contain galaxy positions, so that clustering signals and cross-correlations between galaxy position and galaxy ellipticity (which on small scales corresponds to galaxy-galaxy lensing) can be added to $\vek{D}$ without the need for extra data. This was first proposed by \citet{Bernstein09} and shown by \citet{JB10} to self-calibrate flexible models of intrinsic alignments with minimal assumptions such that cosmological parameter constraints are not degraded by more than factors of a few compared to pure weak lensing analysis without intrinsic alignments or extra data \citep[see also][]{Zhang2010}. Sub-dividing the data into tomographic redshift bins is crucial for this purpose, and the self-calibration performance may be further improved by a split between blue and red galaxies which are known to have largely different intrinsic alignment signals (see \Cref{sec:galaxyalign}). The combination of weak lensing using galaxies as light sources with lensing statistics derived from the cosmic microwave background (CMB), which are not affected by intrinsic alignment effects (note that a cross-correlation between galaxy samples and CMB lensing \emph{would} be affected by a GI term; \citealp{HT14,TI14}), also seems a promising route \citep{KHD14}. See \citet{Paper3} for a more quantitative comparison of some intrinsic alignment mitigation techniques.

In principle there is also cosmological parameter information in the intrinsic alignment signals, e.g. in the distance ratios that govern the redshift scaling of the signals \citep{KT11,SJ12,CD13}. However, it is currently unclear if the sensitivity to cosmology will ever outweigh the statistical errors and the uncertainty in the models. Intrinsic alignments also effect the three-point statistics of the ellipticity field, analogously to their impact on two-point statistics. \citet{SHW+08} found stronger contamination for three-point statistics in simulated data than at the two-point level, which could be exploited to self-calibrate intrinsic alignments by combining these statistics. However the modelling is commensurately more difficult at the three-point level. Maps of weak lensing convergence have been proposed as a starting point to extract more advanced statistics of  the shear pattern on the sky. The impact of intrinsic alignments on the map-making process and the derived statistics clearly warrants further investigation (see \citealp{F07} for an early investigation into the impact of intrinsic alignments on peak counts in such maps). Moreover, if one attempts to measure lensing using galaxies in the background, but has imperfect redshift information and thus contamination from galaxies at the same redshift as the lens, then intrinsic alignments can also contaminate galaxy-galaxy lensing \citep{BMS+12} and need to be mitigated, perhaps using forward modelling approaches analogous to those used for cosmic shear.

Because the intrinsic alignment of galaxies with matter densities is a local effect, photometric surveys that infer the distance information of galaxies using broad-band colours do not provide a sufficiently precise measurement, and as a result the intrinsic alignment and lensing signals are mixed up in the ellipticity measurements. Hence, to provide calibration information that yields priors on parameters of an intrinsic alignment model at the likelihood stage, data with precise redshift information is required. In the following section we will return to the question of what kind of survey will be able to provide this critically important extra information for the galaxy samples observed with forthcoming weak lensing surveys.

\section{Quo vadis?}
\label{sec:outlook}

What are the challenges and opportunities for galaxy alignment studies in the nascent era of precision cosmology with deep and large galaxy surveys? Increasingly large datasets with high-quality imaging superior to what the SDSS provided will be produced by on-going surveys\footnote{including the Kilo Degree Survey, \texttt{http://kids.strw.leidenuniv.nl}; the Dark Energy Survey, \texttt{http://www.darkenergysurvey.org}; and the Hyper Suprime-Cam Survey, \texttt{http://www.naoj.org/Projects/HSC}}, and by the even larger projects of the 2020s\footnote{including the Large Synoptic Survey Telescope \citep{LSST09}, \texttt{http://www.lsst.org/lsst}; the ESA Euclid satellite \citep{LAA+11}, \texttt{http://sci.esa.int/euclid} and \texttt{http://www.euclid-ec.org};  and the NASA Wide-Field Infrared Survey Telescope (WFIRST, \citealp{SGB+13}), \texttt{http://wfirst.gsfc.nasa.gov}}. Not only will these surveys allow for galaxy alignment measurements with improved precision and accuracy, they will also feature weak lensing as a key cosmological probe and thus raise the bar for the performance of intrinsic alignment mitigation. They will be complemented by new spectroscopic redshift surveys\footnote{for instance with the Dark Energy Spectroscopic Instrument, \texttt{http://desi.lbl.gov/cdr}; the Subaru Prime Focus Spectrograph \texttt{http://sumire.ipmu.jp/pfs}; as well as Euclid and WFIRST}, which will enable alignment studies (which usually require precise three-dimensional positions of galaxies) for the samples of early-type and emission-line galaxies that are targeted. The advancements on the observational side will be matched by a continued development of algorithms for simulations of structure formation, paired with the rapid evolution of available computational power.

Based on this framework, it is reasonable to expect that the currently good understanding of early-type alignments is going to be consolidated using fainter galaxy samples at higher redshift. Galaxy groups and clusters can be viewed as a high-mass extension to the range of objects for which the tidal stretching paradigm should constitute an accurate description of alignment processes. Also for these comparatively rare objects we anticipate better constraints as data and simulations provide denser sampling of large cosmological volumes, which allows us to test models over a continuous mass range of about three decades from Milky Way-size haloes to massive clusters.

While the linear alignment model works remarkably well on large scales, its extensions to non-linear
scales will require further scrutiny. A halo model of alignments as proposed by \citet{SB10} is a
promising framework, but it is yet to be tested if the assumption that all matter is bound in haloes
breaks down on intermediate scales \citep{DS15}, where filamentary structure may play a decisive role. Once a
successful model for describing intrinsic alignments down to megaparsec scales exists, it will be desirable to measure higher-order alignment statistics, such as three-point functions. They have to be self-consistently predicted by the model and may in addition help to disentangle intrinsic alignment from weak lensing signals \citep{SHW+08}. The current and future high signal-to-noise alignment detections for early-type galaxies, groups, and clusters could in principle be interesting probes of the properties of dark matter and gravity \citep[e.g.][]{CD13}. However, signals of interest would have to manifest themselves in the redshift dependence or the scale dependence on moderately large scales, whereas the non-linear regime and the overall amplitude of the alignment signals is likely to always be dominated by the effects of highly non-linear physics, baryons, and stochastic processes.

Only in the last few years have hydrodynamic simulations begun to cover cosmological volumes with high enough resolution to allow for measurements of shapes and their alignments (see \citealp{Paper2} for details). They will be key to unravelling the link between the alignments of dark matter structures and the visible distribution of stars, as well as elucidating correlations with other observables, such as colour, size, and dynamical state. A work plan for the near future has to include tests of the sensitivity of simulated galaxy shapes and alignments to the choice of simulation code, and to what is referred to as \lq sub-grid physics\rq, i.e. effective descriptions of small-scale processes below the simulation resolution like the impact of supernova explosions and active galactic nuclei on the temperature, distribution, and chemical composition of gas within the galaxy. The implementation of the physics behind these processes will have to mature to a degree that hydrodynamic simulations can simultaneously predict galaxy alignments and basic observables such as the number density of galaxies as a function of their luminosity and the mass-size relation \citep[perhaps via calibration techniques; see e.g.][]{schaye15,crain15}. Recent results have highlighted the sensitivity of alignment signals to implementation details of the simulations \citep{VCS+15}, but have also yielded promising results in that they show quantitative agreement between observations and simulations \citep{TMD+14b}.
By being able to predict the correct amplitude of alignments, one can begin to answer key issues, such as if alignments are frozen in at the time of galaxy formation, or generated or reset in major merger events, and how these alignments are transferred from the dark matter to the stellar distribution.

For the foreseeable future hydrodynamic simulations will be too computationally expensive to be run in boxes that cover full surveys, and with many realisations. Hence we anticipate that a substantial effort will go into developing statistical or analytic prescriptions \citep{HWH+06,JSB+13,JSH+13} for galaxy alignments in order to paste galaxy properties into pure dark mater simulations, informed by results from smaller hydrodynamic simulations. This is analogous to what is routinely done to include photometric properties of galaxies in mock catalogues based on $N$-body simulations \citep{Baugh06}. Since galaxy morphology and colours are strongly correlated, one should expect that incorporating galaxy shapes into the formalism will improve the overall model, which can additionally be tested against observed galaxy shape distributions and alignments. See \citet{Paper2} for a detailed discussion of requirements on future simulations in relation to galaxy alignment studies.

While the observational prospects for elliptical galaxies are good, the situation for spiral galaxies is much more uncertain. To date, there are few (if any) convincing, highly significant detections for any kind of alignment involving the shape of disc galaxies or their spin. Simulations do see signatures of tidal torque theory, or intriguing alternative models \citep[e.g.][]{LHF+13}, but these are largely eradicated in observational data by a combination of projection effects and stochastic misalignments between dark matter and stars \citep{HWH+06}. Nor is it entirely clear whether elements of large-scale structure like voids can be identified from future high-redshift spectroscopic galaxy samples in a sufficiently clean manner to yield constraints on spin alignments that are readily interpretable.

Since the typical galaxy samples used for cosmic shear surveys are dominated by late-type galaxies, the lack of evidence for intrinsic alignments among them may suggest that a straightforward cosmological analysis is possible without invoking complicated mitigation schemes. However, the decisive quantity in this case is not the signal amplitude, but the level of uncertainty on this signal which, for blue disc galaxies above $z \sim 0.25$, is very large. There is currently no clear avenue to change this situation as the forthcoming spectroscopic surveys will target other galaxy types which have more obvious spectral features that facilitate the determination of a redshift. These datasets will only fill in the current gaps for red galaxies, above $z\sim 0.6$ and for low luminosities down to Milky Way brightness. This will be different for future observational campaigns required to calibrate photometric redshifts. They will generate spectroscopic samples that are more representative of those found in weak lensing surveys, but are optimised to cover small areas in different parts of the sky to beat down sample variance, which prohibits the measurement of spatial correlations on the scales used for cosmological weak lensing measurements.

\begin{figure}[t]
\centering
\includegraphics[scale=0.6]{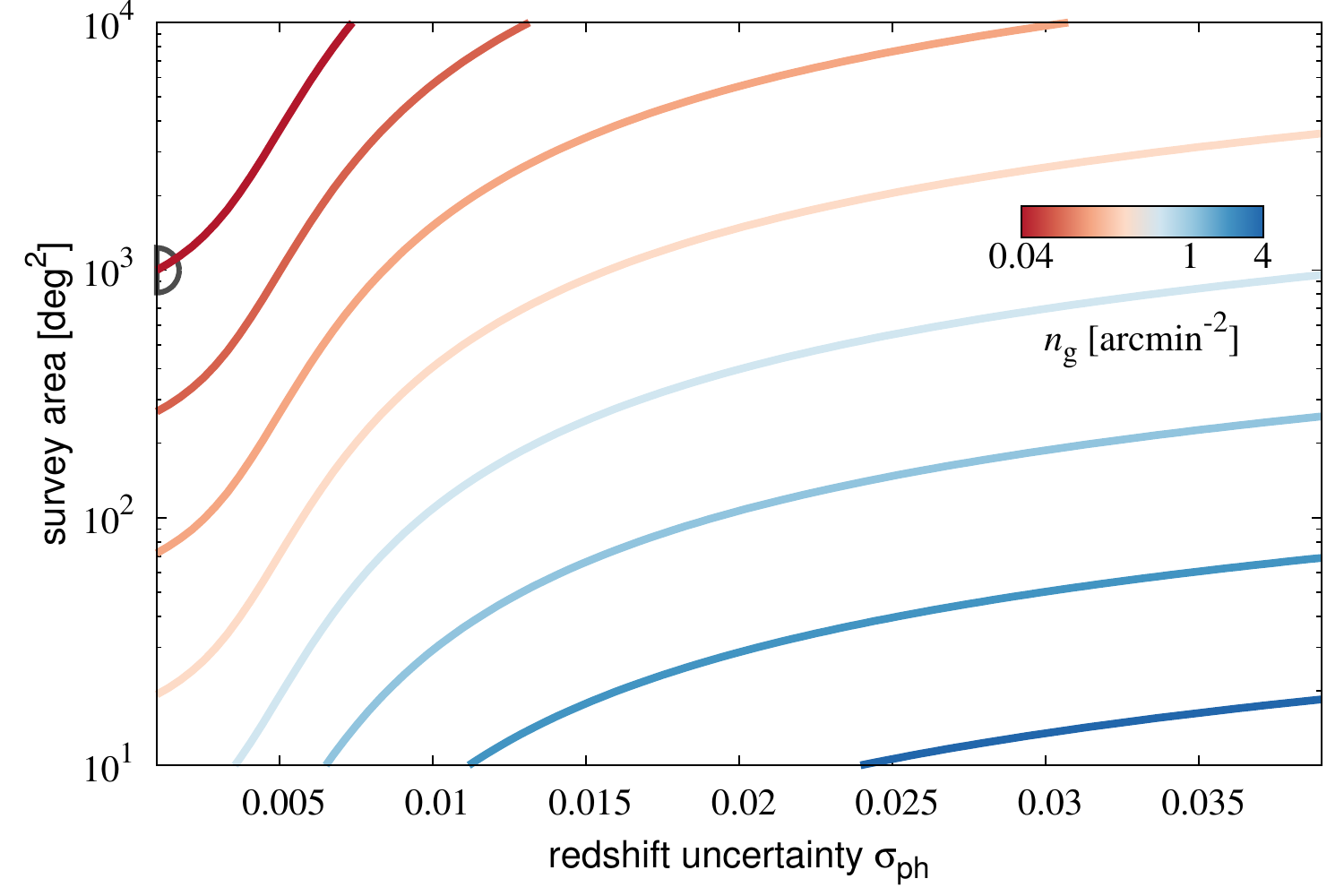}
\caption{Relation between survey parameters and signal-to-noise of a galaxy alignment correlation function as in \Cref{eq:wgplus}. A simple analytic estimate neglecting sample variance was employed. While the absolute value of signal-to-noise determined by this method is not very realistic, the relative values used in this graph are expected to be a good approximation on small and intermediate scales. Lines indicate the survey area required to attain the same signal-to-noise as from a spectroscopic galaxy sample akin to BOSS CMASS (in a redshift range of $\bb{0.45;0.65}$; position in plot marked by the grey semi-circle) over $1000\,{\rm deg}^2$. This is shown as a function of the statistical uncertainty in the redshift determination, given by a Gaussian of width $\sigma_{\rm ph}(1+z)$, and of the angular number density, $n_{\rm g}$, of galaxies with shape measurements of sufficient quantity in the corresponding redshift range. Note the steep rise of the curves above $\sigma_{\rm ph} \approx 0.003$, which is due to the rapid dilution of information in the alignment statistic because of the increased scatter in redshifts.}
\label{fig:surveyreq}
\end{figure}

Although the data obtained from a weak lensing survey offers the ability to self-calibrate intrinsic alignments \citep{JB10}, it is still highly desirable to have a direct measurement for blue galaxies over relevant redshift and luminosity ranges to either put priors on the intrinsic alignment contamination in the self-calibration analysis or verify that model choices when marginalising over nuisance parameters related to intrinsic alignments are justified. The redshift survey required for this purpose would have to go to similar depths as the weak lensing data, cover a sufficiently large contiguous area to sample galaxy pair separations of tens of megaparsecs with small errors, and obtain an average galaxy number density of the same order as the weak lensing survey to avoid being limited by shot noise on small scales. Realistically, such datasets could only be obtained in the near future by low-resolution spectroscopic surveys or narrow-band photometric surveys\footnote{such as PRIMUS, \texttt{http://primus.ucsd.edu}; PAU, \texttt{http://www.ieec.cat/project/pau-physics-of-the-accelerating-universe}; J-PAS, \texttt{http://j-pas.org}; and (limited to bright galaxies) the proposed SPHEREx mission, \texttt{http://spherex.caltech.edu}, which employs filters with spatially varying response}. These produce redshift estimates with a low catastrophic failure rate and scatter that is an order of magnitude or more smaller than for broad-band photometric surveys, which preserves most of the large-scale alignment signals and avoids confusion with the lensing signal. \Cref{fig:surveyreq} illustrates how the key survey parameters -- area, galaxy number density, and redshift scatter -- influence the signal-to-noise achievable for intrinsic alignment correlation functions of the form given in \Cref{eq:wgplus}.

The Square Kilometre Array (SKA)\footnote{\texttt{https://www.skatelescope.org}} will extend galaxy
survey astrophysics to radio wavelengths and mark a transition into a new era for large-scale structure cosmology, including the study of galaxy
alignments. Galaxy shapes will be determined by the distribution of neutral gas rather than stellar
light. The former extends to much larger radii and may therefore not be perfectly correlated with the
optical shape, and be subject to a different alignment strength. Both effects can be exploited to
separate the lensing and alignment signals in a joint analysis. Moreover, the SKA will produce maps
of polarisation and radial velocities across a galaxy which may be used as independent tracers of
the gravitational lensing effect (see \citealp{Paper3} for details). Finally, the full SKA will
deliver precise redshift estimates for up to $10\,$galaxies$/{\rm arcmin}^{2}$ over the whole
extragalactic sky and will thereby provide exquisite data for alignment measurements among the
star-forming disc galaxies to which it is most sensitive.

In any case, the subtle observational signatures of galaxy alignments will remain challenging to measure, and the underlying highly non-linear, baryonic physics-dependent processes challenging to model. However, being at the interface between galaxy formation and evolution on the one side and fundamental cosmology on the other, with potentially considerable impact on both, galaxy alignments are expected to feature prominently in this new era of precision cosmology.

\section*{Acknowledgements}
\label{sec:acknowledgements}

We acknowledge the support of the International Space Science Institute Bern for two workshops at which this work was conceived.
We thank E. Brunnstrom for an investigation into alignments in Palomar Sky Survey catalogues, and our referee, J. Blazek, for many helpful comments and stimulating discussions. We are grateful to B. Binggeli, C. Heymans, S. Singh, A. Slosar, A. Tenneti, and I. Trujillo for sharing their data.

BJ acknowledges support by an STFC Ernest Rutherford Fellowship, grant reference ST/J004421/1. 
MC was supported by the Netherlands organisation for Scientific Research (NWO) Vidi grant 639.042.814.
TDK is supported by a Royal Society URF.
AL acknowledges  the  support of  the  European  Union Seventh Framework Programme (FP7/2007-2013) under  grant  agreement  number 624151.
RM acknowledges the support of NASA ROSES 12-EUCLID12-0004.
CS and HH acknowledge support from the European Research Council under FP7 grant number 279396.
AK was supported in part by JPL, run under a contract by Caltech for NASA. AK was also supported in part by NASA ROSES 13-ATP13-0019 and NASA ROSES 12-EUCLID12-0004.


\bibliographystyle{apj_title}

{\small
\bibliography{bibliography} 
}

\end{document}